\documentclass[traditabstract]{aa}

\usepackage{graphicx}
\usepackage{txfonts}
\usepackage{natbib}
\usepackage{adjustbox}
\usepackage[colorlinks=true,citecolor=blue]{hyperref}
\usepackage{subfigure}
\usepackage{xspace}
\usepackage{amsmath,amssymb}
\usepackage{tablefootnote}
\usepackage{multirow}

\bibpunct{(}{)}{;}{a}{}{,}

\newcommand{\Teff}{\ensuremath{T_\mathrm{eff}}\xspace}
\newcommand{\logg}{\ensuremath{\log~g}\xspace}
\newcommand{\mic}{\ensuremath{\mu\mathrm{m}}\xspace}
\newcommand{\as}{\hbox{$^{\prime\prime}$}\xspace}

\begin{document}

\title{Direct characterization of young giant exoplanets at high spectral resolution by coupling SPHERE and CRIRES+}
\titlerunning{Coupling SPHERE and CRIRES+}
\author{
    G.~P.~P.~L.~Otten\inst{\ref{lam}} \and
    A.~Vigan\inst{\ref{lam}} \and
    E.~Muslimov\inst{\ref{lam},}\inst{\ref{kazan}} \and
    M.~N'Diaye\inst{\ref{oca}} \and
    E.~Choquet\inst{\ref{lam}} \and
    U.~Seemann\inst{\ref{gottingen}} \and
    K.~Dohlen\inst{\ref{lam}} \and
    M.~Houll\'{e}\inst{\ref{lam}} \and
    P.~Cristofari\inst{\ref{lam}} \and
    M.~W.~Phillips\inst{\ref{exeter}} \and
    Y.~Charles\inst{\ref{lam}} \and
    I.~Baraffe\inst{\ref{exeter}} \and
    J.-L.~Beuzit\inst{\ref{lam}} \and
    A. Costille\inst{\ref{lam}} \and
    R.~Dorn\inst{\ref{eso_garching}} \and
    M.~El~Morsy\inst{\ref{lam}} \and
    M.~Kasper\inst{\ref{eso_garching}} \and
    M.~Lopez\inst{\ref{lam}} \and
    C.~Mordasini\inst{\ref{bern}} \and
    R.~Pourcelot\inst{\ref{lam}} \and
    A. Reiners\inst{\ref{gottingen}} \and
    J.-F.~Sauvage\inst{\ref{lam}}
}

\institute{
    Aix Marseille Univ, CNRS, CNES, LAM, Marseille, France \label{lam} \\    \email{\href{mailto:gilles.otten@lam.fr}{gilles.otten@lam.fr}}
    \and
    Kazan National Research Technical University named after A.N. Tupolev KAI, 10 K. Marx, Kazan, Russia, 420111 \label{kazan}
    \and
    Universit\'{e} C\^{o}te d'Azur, Observatoire de la C\^{o}te d'Azur, CNRS, Laboratoire Lagrange, Bd de l'Observatoire, CS 34229, 063404 Nice Cedex 4, France \label{oca}
    \and
    Institute for Astrophysics, Georg-August University, Friedrich-Hund-Platz 1, 37077, G\"{o}ttingen, Germany \label{gottingen}
    \and
    School of Physics and Astronomy, University of Exeter, Exeter, EX4 4QL, UK \label{exeter}
    \and
    Physikalisches Institut, University of Bern, Sidlerstrasse 5, 3012 Bern, Switzerland \label{bern}
    \and
    European Southern Observatory (ESO), Karl-Schwarzschild-Str. 2, 85748 Garching, Germany \label{eso_garching}
}

\date{Received 28 May 2020; accepted 27 August 2020}

\abstract{Studies of atmospheres of directly imaged extrasolar planets with high-resolution spectrographs have shown that their characterization is predominantly limited by noise on the stellar halo at the location of the studied exoplanet. An instrumental combination of high-contrast imaging and high spectral resolution that suppresses this noise and resolves the spectral lines can therefore yield higher quality spectra. We study the performance of the proposed HiRISE fiber coupling between the direct imager SPHERE and the spectrograph CRIRES+ at the Very Large Telescope for spectral characterization of directly imaged planets. Using end-to-end simulations of HiRISE we determine the signal-to-noise ratio (S/N) of the detection of molecular species for known extrasolar planets in $H$ and $K$ bands, and compare them to CRIRES+. We investigate the ultimate detection limits of HiRISE as a function of stellar magnitude, and we quantify the impact of different coronagraphs and of the system transmission. We find that HiRISE largely outperforms CRIRES+ for companions around bright hosts like $\beta$\,Pictoris or 51\,Eridani. For an $H=3.5$ host, we observe a gain of a factor of up to 16 in observing time with HiRISE to reach the same S/N on a companion at 200\,mas. More generally, HiRISE provides better performance than CRIRES+ in two-hour integration times between 50--350\,mas for hosts with $H<8.5$ and between 50--700\,mas for $H<7$. For fainter hosts like PDS\,70 and HIP\,65426, no significant improvements are observed. We find that using no coronagraph yields the best S/N when characterizing known exoplanets due to higher transmission and fiber-based starlight suppression. We demonstrate that the overall transmission of the system is in fact the main driver of performance. Finally, we show that HiRISE outperforms the best detection limits of SPHERE for bright stars, opening major possibilities for the characterization of future planetary companions detected by other techniques.}


\maketitle

\section{Introduction} 
\label{sec:intro}

In contrast to indirect methods, direct imaging permits us to spatially separate and directly measure radiation from an exoplanet, which allows us to spectrally analyze its atmosphere with minimized impact from the host star. Direct imagers such as SPHERE \citep{Beuzit:2019}, GPI \citep{Macintosh:2014}, and SCExAO \citep{Jovanovic:2015} are designed to find and detect young planets around nearby stars \citep{Chauvin:2005, Marois:2008, Lagrange:2009, Chauvin:2017a,Keppler:2018}, but their ability to characterize them is limited by their spectral resolution of $R=\lambda/\Delta\lambda=400$ at most \citep{Vigan:2008}. 

High-dispersion spectrographs (HDS) have detected the thermal radiation of both transiting and non-transiting planets  \citep[e.g.,][]{Snellen:2010,Brogi:2012,Birkby:2013}, and have in theory demonstrated to be a very promising trajectory for the detection of biomarkers around Earth-twins with the ELT \citep{Snellen:2013}. As  starlight is the greatest  contributor of noise, it is highly beneficial to first spatially separate the stellar and planetary point spread functions (PSFs) through high angular resolution imaging techniques like adaptive optics \citep{Sparks:2002,Riaud:2007}, and then implement mid- to high-resolution spectroscopy \citep[$R=5\,000-100\,000$;][]{Thatte:2007,Konopacky:2013,Barman:2015}. 

This combination of the spatial separation of planet and star and mid- to high-resolution spectroscopy has been given a clear demonstration through the detection of the atmosphere of the directly imaged planet $\beta$\,Pictoris\,b \citep{Snellen:2014} using the CRIRES instrument \citep{Kaeufl:2004} with its MACAO adaptive optics system \citep{Arsenault:2003} on the ESO Very Large Telescope (VLT). This measurement not only allowed us to determine the planet's orbital velocity, which was used to better constrain its orbital parameters, but also to determine the planet's rotational period and probable length of day, which were derived through the broadening of the CO and H$_2$O lines. This method has since provided more atmospheric detections and rotation speeds of young directly imaged companions \citep{Schwarz:2016,Hoeijmakers:2018,Bryan:2018}, providing unique insight into the properties of this population of objects \citep{Bryan:2018}. 

Measurements on young companions have been obtained with low-order adaptive optics systems like MACAO for the $\beta$\,Pictoris\,b result. With such systems, the atmospheric turbulence correction   concentrates a moderate fraction of the energy in the PSF core (typically 50--60\% in $K$ band; \citealt{Paufique:2006}), but the level of the uncorrected halo is high. This means that the strongest limiting factor of the signal-to-noise ratio (S/N) of the planet signal close to the star remains the noise contributed by the uncorrected stellar halo. To significantly decrease the halo, it is necessary to rely on high-order adaptive optics known as extreme adaptive optics (ExAO), which can provide diffraction-limited images in the near-infrared (NIR) and therefore decrease the level of stellar halo (and noise) at the location of the planet \citep{Kawahara:2014a,Snellen:2015,Wang:2017,Mawet:2017,Vigan:2018}. Diffraction-suppressing coronagraphs can be used to further improve the contrast and decrease the level of stellar residuals at the location of the planet.

The implementation of the spectrograph downstream of the ExAO system and the coronagraph can either rely on traditional integral field spectrographs \citep[e.g.,][]{Antichi:2009} or on fiber-fed spectrographs \citep{Jovanovic:2017}. While the former offer a full spatial and spectral information but can prove costly in terms of pixels for high numbers of resolution elements, the latter offer significant advantages when based on single-mode fibers (SMFs). Indeed, SMFs provide a positionally and spectrally stable source at the entrance of the spectrograph \citep{Ge:1998,Jovanovic:2017}, which neutralizes the issue of modal noise seen in multi-mode fibers commonly used for seeing-limited telescopes \citep{Baudrand:2001,Plavchan:2013}. Moreover, because SMFs only support the propagation of a single quasi-Gaussian mode, they offer an additional level of spatial filtering of the stellar halo, which can further decrease the contribution of the stellar halo at the location of the planet \citep{Mawet:2017}. However, a drawback is that the planet's PSF will also couple less efficiently into the SMF for typical telescope apertures with central obscuration and spiders \citep{Ruilier:98,Jovanovic:2017}. In addition, the telescope PSF has to be accurately centered on the fiber (on the order of 10\% of a $\lambda/D$, where $\lambda$ is wavelength and $D$ is effective diameter) to have minimal coupling losses.

In recent years several projects have proposed  combining ExAO systems with existing mid- or high-resolution spectrographs using SMF either in the NIR, for example  between NIRC2 and NIRSPEC at Keck \citep[KPIC;][]{Mawet:2016,Mawet:2017} or between SCExAO and IRD at Subaru \citep[REACH;][]{Kawahara:2014,Kawahara:2014a,Kotani:2018}, or in the visible, for example  between SPHERE and ESPRESSO at the VLT \citep{Lovis:2017}. For the NIR, the VLT offers a unique opportunity to achieve a similar feat by coupling the high-contrast imager (HCI) SPHERE \citep{Beuzit:2019} with the high-resolution spectrograph CRIRES+ \citep{Dorn:2016}, which will both be available at the same unit telescope (UT3) in   2020. The coupling between these two flagship instruments has been proposed as the High-Resolution Imaging and Spectroscopy of Exoplanets project \citep[HiRISE;][]{Vigan:2018}.

SPHERE offers a unique ExAO system \citep[called SAXO;][]{Fusco:2006} that has demonstrated exquisite performance on-sky \citep{Sauvage:2014,Petit:2014,Milli:2017} and efficient coronagraphs \citep{Carbillet:2011,Guerri:2011};  its infrared arm covers the $Y$, $J$, $H$, and $K_s$ bands. CRIRES+ is the fully refurbished and upgraded CRIRES spectrograph \citep{Kaeufl:2004}, operating in the $Y$, $J$, $H$, $K$, $L$, and $M$ bands (0.9-5.3 um) at $R=100\,000$ or $R=50\,000$ resolving power (for the 0.2\as and 0.4\as slits, respectively). It features three Hawaii-2RG 2k$\times$2k detectors and a cross-disperser for an  increase in its simultaneous spectral coverage of up to
tenfold. 

The large overlap in terms of spectral coverage between SPHERE and CRIRES+ is a key advantage in particular in the $H$ and $K$ bands, which contain strong molecular features for CO, CH$_4$, H$_2$O, or NH$_3$. The high angular resolution and high-contrast capabilities of SPHERE at these wavelengths, combined with the high spectral resolution of CRIRES+, will provide a unique system capable of characterizing known directly imaged companions.

In this paper we present the expected performance of the HiRISE system. In Sect.~\ref{sec:modelling} we present the full simulation model that we have developed to investigate the performance of the system. In Sect.~\ref{sec:perf_known_targets} we quantify the S/N that can be expected on individual molecules for known planetary targets, and then in Sect.~\ref{sec:discovery_potential} we investigate the discovery potential of HiRISE for the detection of companions that are below the detection threshold of current high-contrast instruments like SPHERE. In particular, we compare the expected performance between CRIRES+ in standalone mode with the combination SPHERE/HiRISE/CRIRES+ system. We further study the impact of different coronagraphic modes and the impact of measures that increase  transmission. Finally, we present our conclusions  and   perspectives in Sect.~\ref{sec:conclusion}.

\section{Modeling the combination of HCI and HDS}
\label{sec:modelling}

To model the combination of a high-resolution imager and high spectral resolution spectrograph we parameterize and quantify the properties of the stellar and planetary sources, atmosphere, telescope, imager, spectrograph, and coupling system, which are all described in the following subsections. By using such a quantified approach with multiplicative transmission components we are able to easily study different configurations from the early design phase to the nearly finalized design. Generally, conservative values are taken when estimates are required. The full model is described in Sects.~\ref{sec:stellar_source} to \ref{sec:background}. At the end (Sect.~\ref{sec:combiningmodel}), the different noise contributions are injected into the simulated planetary spectra, which are used in subsequent sections for performance estimation based on S/N calculations. The same procedure is used to describe CRIRES+ in standalone mode by not including the effect of SPHERE and HiRISE.

\subsection{Stellar source}
\label{sec:stellar_source}

We model the star using a PHOENIX \citep{Husser:2013} model with a given effective temperature (\Teff) and surface gravity (\logg), metallicity [Fe/H] equal to solar metallicity, and no alpha enhancement (overabundance of He with respect to metallicity, [$\alpha$/Fe]). First, we determine a flux scaling factor to rescale the model to the observed values for known stars so that we can accurately evaluate the number of photons per spectral channel for the planetary hosts. To determine this scaling factor, the model (which is sampled at $R=500\,000$ between 300--2500\,nm) is interpolated on a regular grid with a wavelength step size of 1\,\AA\ from 0.6\,\mic to 30\,\mic. Photometric spectral energy distribution (SED) data points for the host star are obtained from the VOSA tool \citep{Bayo:2008,Skrutskie:2006} in the 2MASS $J$, $H$, and $K_s$ bands. After integrating the flux of the stellar model over the band pass of each of these filters using their response curve, we perform a least-squares fit to obtain the flux scaling factor. This flux rescaling factor is of the same order of magnitude as the geometric flux scaling factor $(d_\mathrm{radius}/d_\mathrm{distance})^2$, derived from the stellar radius $d_\mathrm{radius}$ and distance $d_\mathrm{distance}$. The former solution was chosen over the latter to be able to easily simulate targets of known observed magnitudes rather than known physical radius and distance. The observed spectrum of the star is finally obtained by interpolating the original spectrum to a spectrum resolution of $R$, converting it to photons per units of time and rescaling it with the flux scaling parameter.

\subsection{Planetary source}

For the planet we use models generated using the one-dimensional radiative-convective equilibrium code \texttt{ATMO} 2020 \citep{Phillips:2020}. The models are computed on a grid of self-consistent pressure--temperature profiles and chemical equilibrium abundances for a range of effective temperatures (800 to 2000\,K) and gravities (3.5-5.5\,dex). The line-by-line radiative transfer calculation in ATMO is then performed using these profiles and abundances as inputs to generate a high-resolution thermal emission spectrum. Separate spectral templates isolating the contribution of the specific molecules ($\mathrm{CH_4}$,  $\mathrm{H_2O}$, $\mathrm{NH_3}$, CO, $\mathrm{CO_2}$) are also generated using a  method similar to that described in \citet{Wang:2017}. The template for a given molecular absorber is the emission spectrum calculated by removing all opacities in the model atmosphere apart from the respective absorber and the absorption from $\mathrm{H_2}$-$\mathrm{H_2}$ and $\mathrm{H_2}$-$\mathrm{He}$ collisions. This allows the contributions to the thermal emission of dominant molecular absorbers such as $\mathrm{H_2O}$, $\mathrm{CO,}$ and $\mathrm{CH_4}$ to be isolated. In total, \texttt{ATMO} 2020 contains 22 molecular and atomic opacity sources primarily originating from the ExoMol database \citep{Tennyson:2016}, and the full model is described in \citet{Goyal:2018} and \cite{Phillips:2020}. In this paper the non-equilibrium models produced with this code are used.

Similarly to the stellar model, the planetary model is first interpolated on a regular grid in wavelength and its flux is then integrated over the 2MASS filter band passes. We then use the known delta magnitude $\Delta m$ of the planet with respect to the star in a certain band ($H$ or $K_s$) to define a scaling factor that will rescale the planetary spectrum to be $10^{\Delta m/2.5}$ times fainter than the star in the chosen band. To obtain the final spectrum of the planet we again interpolate the original planetary spectrum to the spectral resolution, convert it to photons per units of time and rescale it with this contrast scaling factor. We set the rotational velocity (i.e.,~$v\sin{i}$, with $v$ the velocity and $i$ the inclination) of the planet to zero, as if  seen pole-on. We ran dedicated simulations following the approach of the current and following section to investigate the effect of up to 15 km/s rotational velocity on the S/N. For a 1200 K planet around a $\beta$\,Pic b-like host star with a contrast $\Delta m = 10$ and a separation of 300 mas in $H$ band we see a reduction in the total S/N by a factor of $1.46$, so more than two thirds of the S/N is retained with a typical edge-on rotation speed. This reduction in S/N can be compensated for by approximately doubling the exposure times.

\subsection{Atmospheric transmission and emission}

For the transmission and emission components of Earth's atmosphere we use the ESO SkyCalc models\footnote{\url{https://www.eso.org/observing/etc/skycalc/}} \citep{Noll:2012,Jones:2013}. By default we use a precipitable water vapor (PWV) content of 2.5\,mm and a seeing of 0.8\as, which both represent median conditions at the Paranal observatory. We include all available transmission and emission terms except for instrumental radiation, which are handled separately in our model (see Sect.~\ref{sec:background}). We do not consider any variability of the sky lines over the course of the science exposure.

\subsection{Telescope, ExAO \& high-contrast imager: SPHERE}
\label{sec:tel_sphere}

The transmission of the telescope and the SPHERE instrument up to the installation point of HiRISE in the IFS branch is taken from tabulated values available from the Phase B study of SPHERE, which was cross-checked on-sky \citep{Dohlen:2016,Beuzit:2019}. In that study, measurements of individual optical components were combined together with reasonable assumptions, when no measurements were available, to produce a wavelength-dependent transmission model. The telescope measurements were derived from reflection measurements for a single VLT mirror, which has a coating material made of aluminum. The study considers reflections on the telescope mirrors M1, M2, and M3 before entering SPHERE, so the single mirror reflection measurements are cubed and combined to have a nearly  gray reflectance of $\sim$80\%. In addition to the reflectance of the mirrors, a flat 86\% transmission for the dust on the primary is assumed (again following the Phase B study of SPHERE's transmission).

Within SPHERE, the IFS arm where HiRISE picks up the planetary light can be fed using the two dedicated IRDIFS dichroics: the \texttt{DICH-H} that reflects $H$ band into IRDIS and transmits the $Y$ and $J$ bands into IFS as well as some of the $K$ band, and the \texttt{DICH-K} that reflects $K$ band into IRDIS and transmits the $Y$, $J$, and $H$ bands into IFS. The transmission values for both dichroics were measured and provided by the manufacturer (CILAS, France) for the $Y$, $J$, and $H$ bands. However, no measurements were provided in $K$ band because the SPHERE IFS is not designed to observe in this band. Since the $YJH$ measurements are a very good match with the theoretical curves of the coatings provided by the manufacturer CILAS, we use the theoretical transmission of the coatings in $K$ band to supplement the $H$-band measurements. For HiRISE we use the existing SPHERE dichroics for the science observation, \texttt{DICH-K} for $H$-band observations, and \texttt{DICH-H} for $K$-band observations. The average common path instrument (CPI) transmission for the \texttt{DICH-K} dichro in $H$ band is 47.3\%, while the average transmission of CPI using the \texttt{DICH-H} dichro in $K$ band is 36.8\%. 

\subsection{HiRISE}

HiRISE consists of a NIR single-mode fiber bundle with at least four science fibers, and injection and extraction optics that provide efficient coupling between SPHERE and CRIRES+ using a tracking system. Details of the proposed implementation are provided in other publications \citep{Vigan:2018,Vigan:2019a}.

\subsubsection{Fiber injection}
\label{sec:fibereff}

The optics in the Fiber Injection Module (FIM) cause a reduction in transmission, predominantly due to Fresnel reflection on the surfaces. Here we assume a 2\% loss per optical surface of the lenses and a 5\% loss per optical surface for the dichroic following the design of the anti-reflection and dichroic coatings (Fresnel Institute, France). For six lenses, one mirror, and one dichroic in the injection optics, this results in a total transmission factor of $0.98^{13}\times0.95^{2} = 0.71$.

The fiber injection efficiency, or the amount of planet light that couples and propagates into the fiber, is dependent on the complex field of the PSF of the telescope including all upstream (amplitude and phase) aberrations. We evaluate the impact of the different aberrations on the coupling efficiency.

First, we use the \texttt{Coronagraphs} Python toolkit (N'Diaye, private communication) to simulate the complex focal plane electric field ($E_1$) of the SPHERE PSF out to a radius of 2.45\as, and with a pixel resolution of 12.25 mas, which corresponds to the Nyquist sampling of SPHERE at 0.95\,\mic. The simulation implements a realistic model of the SPHERE instrument based on several inputs: VLT pupil, ExAO residuals, non-common path aberrations (NCPA), amplitude aberrations, coronagraphic masks, and wavefront errors of the HiRISE injection optics. These values were derived from calibration measurements in SPHERE: imaging of the pupil for the amplitude maps, the Lyot stops,   the coronagraphic masks, and ZELDA wavefront sensor measurements for the NCPA \citep{Vigan:2019}. For the FIM, wavefront errors derived by the optical design software \texttt{ZEMAX} were used.  For the residual atmospheric wavefront error we generated a series of 20 uncorrelated residual atmospheric phase screens using a Fourier-based AO simulation code \citep{Fusco:2006} with parameters representative of the SPHERE ExAO system, median observing conditions, and a moderately bright AO star (R=5). The three coronagraphic scenarios that we consider are the apodized-pupil Lyot Coronagraph \citep[APLC;][]{Soummer:2005,Carbillet:2011,Guerri:2011}; the classical Lyot Coronagraph (CLC), implementable by not selecting the apodized pupil of the APLC; and a no-coronagraph option. All three of these options are available in SPHERE without hardware interventions and are therefore the only coronagraphs that are evaluated in this paper.

We then define the focal plane electric field for the fiber as the Gaussian $E_2=e^{-\frac{1}{2}\left(\frac{r}{\sigma}\right)^2}$, with radius $r$ and $\sigma$ as the standard deviation of the radius. In practice, fibers are often defined with a mode field diameter equal to $\mathrm{MFD}=2\omega$, with $E_2=e^{-\left(\frac{r}{\omega}\right)^2}$. This means $\omega=\sqrt{2}\sigma$ and $\mathrm{MFD} = 2\sqrt{2}\sigma$. The corresponding intensity is given by $I=\left| E_2 \right|^2$.

The fiber coupling efficiency represents how well the incoming wavefront projects onto the fundamental mode of the MFD. This can be calculated through an overlap integral with the two previous computed electric fields using the equation
\begin{equation}
    \eta=\frac{\left| \int E^*_1 E_2 dA \right|^2}{\int \left| E_1 \right|^2 dA\int \left| E_2 \right|^2 dA}, 
    \label{eq:fibereff}
\end{equation}
where $\eta$ is the coupling efficiency, $E_1$ and $E_2$ are the complex electric fields, and the integral is taken over all pixels of the field $A$. This equation is identical to the coupling efficiency equations in \citet{Wagner:1982} and \citet{Jovanovic:2017}.

The sigma value of the Gaussian needs to be critically sized in terms of the scale of the PSF (in $\lambda/D$) to obtain an optimal coupling efficiency. We optimize the coupling efficiency for the parameter $\sigma$ to derive the optimal size of the instrumental PSF, and therefore focal ratio $F$ of the FIM at the fiber entrance plane. For a SPHERE-like pupil (14\% secondary obscuration) we calculate, in a case without a coronagraph, that the optimally matching Gaussian has a $\sigma=0.504\,\lambda/D$. Using the APLC coronagraph this changes to the slightly smaller $\sigma=0.493\,\lambda/D$. We settle for an intermediate relation where $\sigma = 0.5\,\lambda/D$ to cover both scenarios. While taking the relation between MFD and $\sigma$ into account, we calculate that to achieve the best average efficiency from 1.5 to 2.5\,\mic, considering the optimal MFD of the PSF and the factory-given MFD of the fiber, we need to provide a focal ratio $F=3.3$ at the fiber. Details of the selected fibers are provided in Sect.~\ref{sec:fibertrans}.

The $F=3.3$ design, while optimal for injection into the fiber, does not lead to a practical opto-mechanical design, so we explored the impact of using slightly higher values of $F$. We measure that the relative loss in coupling efficiency for $F=3.4$ and $F=3.5$ is respectively, $0.1-0.8$ and $0.6-1.8$ percentage points, depending on the coronagraphic mode, which we consider acceptable. We decided to settle on a final design with $F=3.5$.

For the fraction of planet light that couples into the fiber, we calculate the coupling efficiency for SPHERE (APLC, CLC, and no coronagraph) from 1.5 to 2.5\,\mic with 20 decorrelated realizations of ExAO residuals generated with the AO code mentioned previously.  To mimic an off-axis source (i.e.,~a planet) we use the complex PSF where the focal plane mask is not present. We extract the mean coupling efficiency and the average radial profile over the 20 realizations as a function of wavelength and use them for our simulator. 

\begin{figure}
  \centering
    \includegraphics[width=1.0\columnwidth]{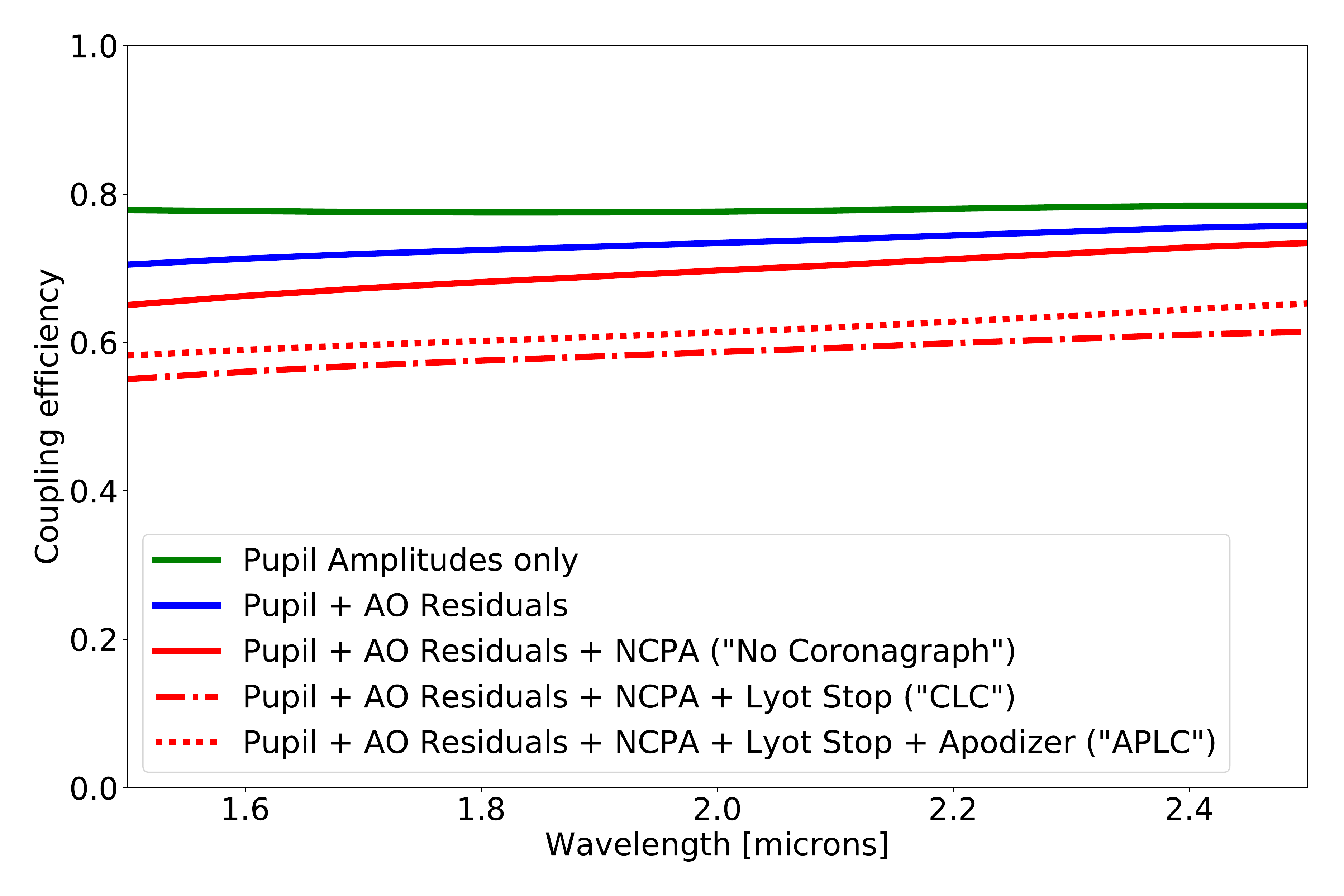}
    \caption{Coupling efficiency as a function of wavelength for an off-axis companion to the host star. For a well-aligned fiber on top of the companion this shows the fraction of planet light that couples and propagates into the fiber. Contributions are added in one at a time to show their impact; highlighted in red are the three coronagraphic configurations considered in this paper.}
    \label{fig:coupling_planet}
\end{figure}

Figure \ref{fig:coupling_planet} shows the coupling efficiency as a function of wavelength for the planet.  The green line shows the coupling efficiency only using the SPHERE pupil amplitude, which includes amplitude aberrations measured directly by SPHERE. The blue line shows the coupling efficiency with simulated AO residuals added in. The lines in red additionally include the NCPA and HiRISE FIM wavefront error, and the effect of the coronagraphs. Due to the oversized spiders in the coronagraphic pupil stops, the coupling efficiency is significantly lower for the coronagraphic cases (CLC and APLC) than for the no-coronagraph case. The APLC mode gets higher coupling efficiency than the classical Lyot Coronagraph due to its apodized pupil optic that makes the PSF more Gaussian, which helps couple the light slightly more efficiently. Unfortunately, this increase in coupling efficiency is created by an amplitude mask in the pupil (apodized pupil) that blocks 57\% of the incoming light, and therefore creates a major loss of photons that is not compensated by the increased coupling efficiency.

In Appendix \ref{app:fiberinjection} the dependence of the fiber injection efficiency on tip/tilt and lateral displacement in the focal plane is shown. To remain at 99\% and 95\% of peak coupling efficiency a pointing accuracy of respectively $0.030\,\lambda/D$ and $0.097\,\lambda/D$ and a tilt accuracy of respectively $0.041\,\lambda$ and $0.11\,\lambda$ per $\lambda/D$ (0.66 and 1.68 degrees from normal incidence) is required for the non-coronagraphic case, which has the tightest tolerance.

\begin{figure}
  \centering
    \includegraphics[width=1.0\columnwidth]{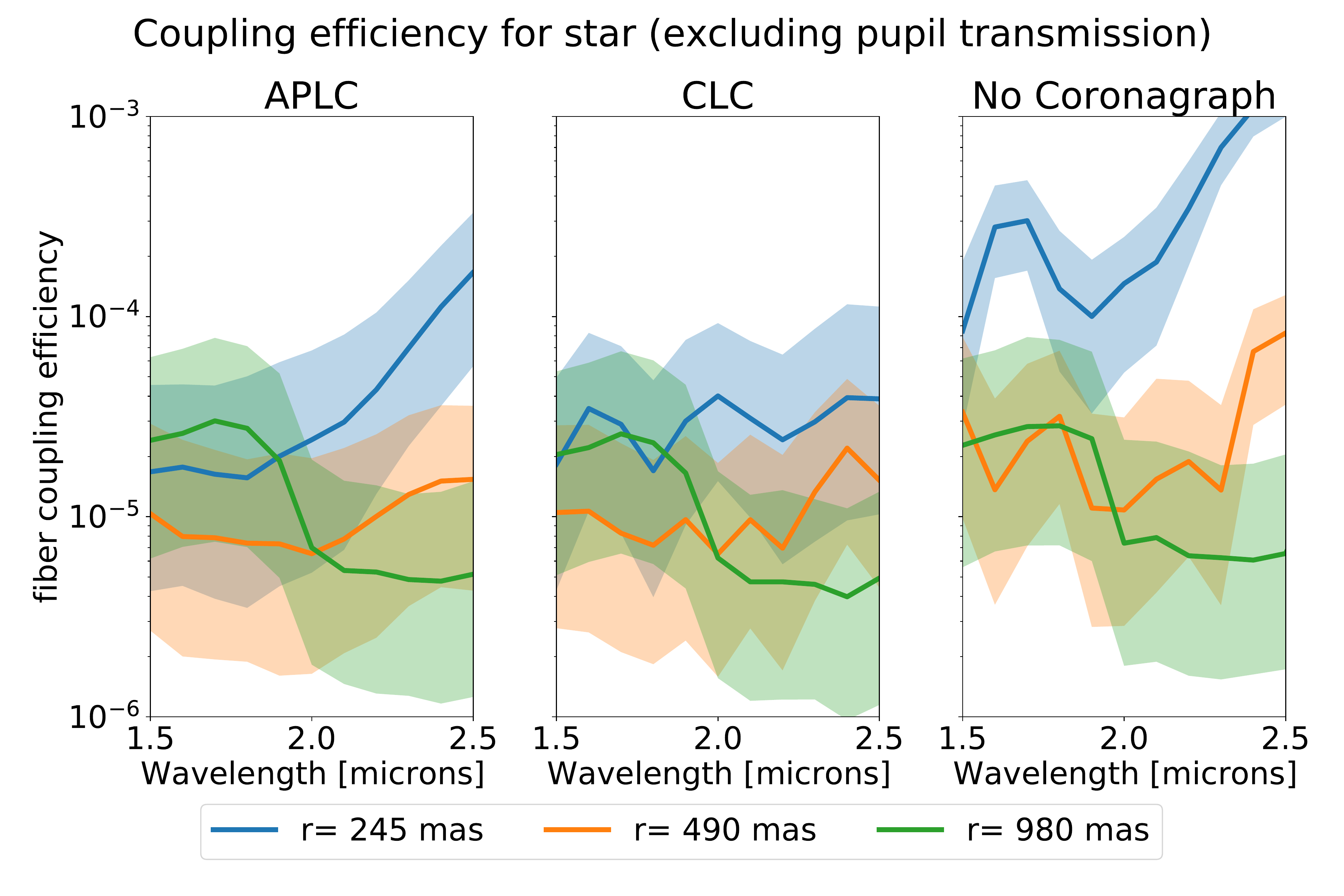}
    \caption{Fraction of starlight coupling into the fiber as a function of wavelength at angular separations of 245, 490, and 980 mas. The curve and shaded regions show the average of the mean and 1$\sigma$ boundary of the azimuthal statistics of 20 independent realizations of ExAO residuals.}
    \label{fig:coupling_star}
\end{figure}

\begin{figure}
  \centering
    \includegraphics[width=1.0\columnwidth]{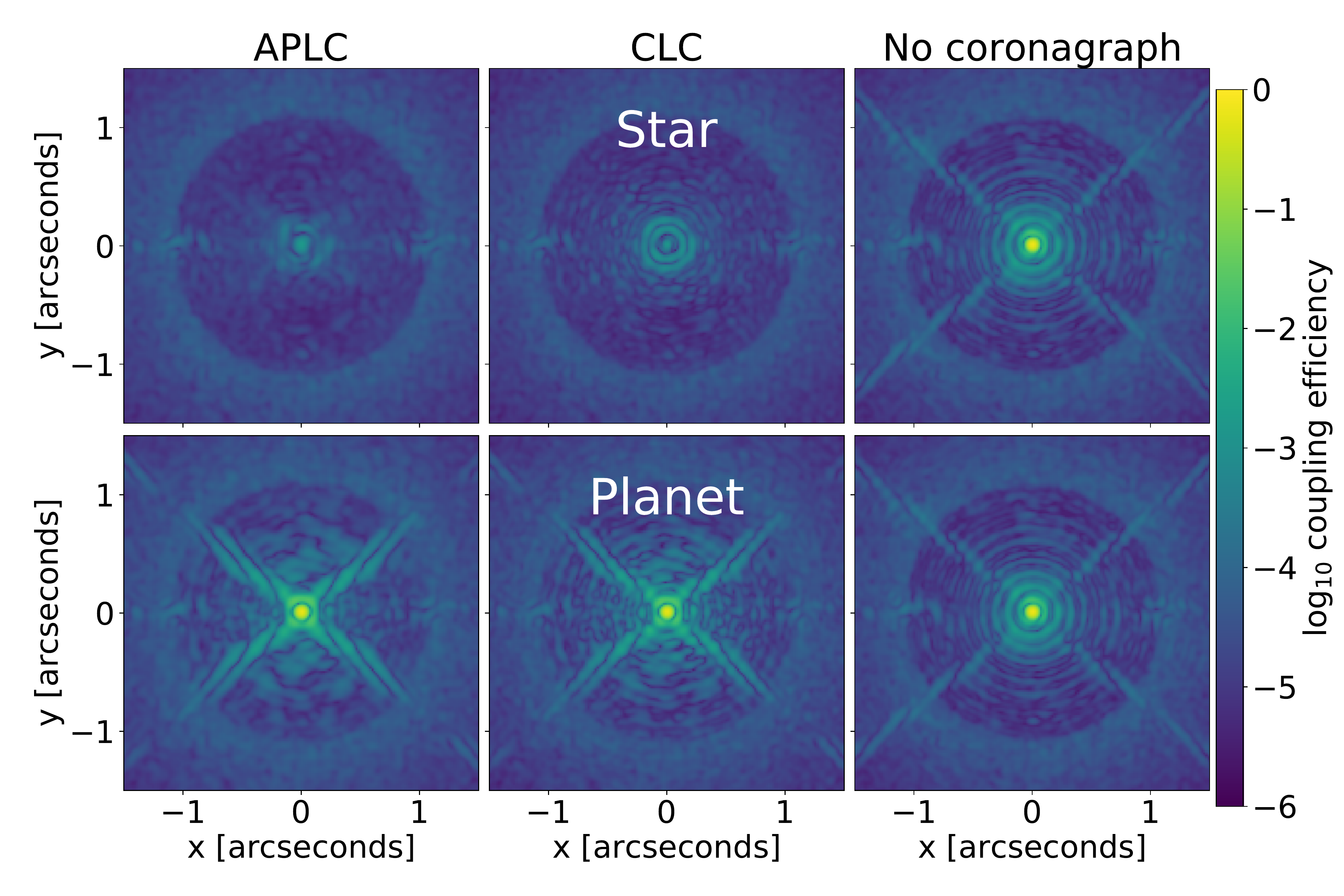}
    \caption{Two-dimensional maps of the coupling efficiency as a  function of separation, simulated at a wavelength of 1.6\,\mic and averaged over 20 realizations of atmospheric residuals. For the coronagraphic efficiencies of the star the inner $\sim$100\,mas is additionally impacted by the transmission loss caused by the focal plane mask. The coupling efficiencies for the star have been corrected to account for the light blocked by the focal plane mask as described in the main text.}
    \label{fig:coupling_2d}
\end{figure}

As the planet lies in the speckle field surrounding the star, part of the starlight will also couple into the fiber at the location of the planet. To estimate this contamination, we recompute the overlap integral between the complex amplitude computed for the star in the two considered coronagraphic cases and non-coronagraphic case ($E_2$) and a Gaussian placed at each location in the simulated field of view ($E_1$). With this procedure we obtain a map that represents for each location in the field the fraction of starlight coupling into the fiber. For each PSF we also store the total flux with and without the coronagraphic mask and renormalize the coupling efficiency of the starlight with this ratio to get the coupling efficiency with respect to the total of incoming light (i.e.,~relative to the flux before the focal plane mask) instead of the total remaining light in the focal plane. The results are summarized in Figs.~\ref{fig:coupling_star} and \ref{fig:coupling_2d}.

Figure \ref{fig:coupling_star} shows the fraction of starlight from the halo that couples into the fiber as a function of wavelength and for three selected angular separations (i.e.,~0.245\as, close to the inner working angle of the coronagraph where the stellar halo should dominate; 0.49\as, an intermediate point; and 0.98\as, where we expect the total flux to be dominated by the background). We see remarkably similar coupling of the stellar light at a level of about $10^{-5}$ for all three options. Only very close to the star with the non-coronagraphic option do we see an increase in coupling efficiency, and therefore additional contamination from starlight with respect to the two coronagraphic options. In radial contrast profiles there is a more pronounced difference in the stellar suppression between coronagraph types (more than an order of magnitude difference at a few diffraction widths away from the star; see also Appendix~\ref{sec:rawcontrast} for a comparison between raw contrast and the contrast achieved when sampling the PSF with a fiber). In Fig.~\ref{fig:coupling_star}, where the focal plane is sampled by a single-mode fiber, this difference in contrast between coronagraphic modes is less apparent. This shows that the halo features for all three coronagraph options couple almost equally well and therefore resemble a Gaussian mode to the same extent. This effect has been explored in more detail by the work of \citet{Por:2018} and \citet{Haffert:2018} who   propose and optimize a combination of coronagraph and single-mode fibers that minimizes the amount of starlight coupled into the fiber. The difference in transmission and the inner working angles of the three coronagraphs are therefore important aspects to consider. As we   show in Sect.~\ref{sec:effect_transmission}, performance is strongly driven by the total transmission of the system, so gaining in coupling efficiency is less critical than maximizing photons in the first place. 

For completeness we show in Fig.~\ref{fig:coupling_2d} two-dimensional maps of the coupling efficiency for both the star and the planet. The stellar map assumes an aligned coronagraphic focal plane mask and can be used to read off the amount of starlight that ends up in the fiber, which is valid beyond 100 mas from the center. The planetary map has no focal plane mask applied and can be used to read off the amount of planet light that ends up in the fiber. For example, the top plots are read off at a certain separation ($\sim300$ mas for $\beta$\,Pictoris b) to get the fraction of starlight coupling into the fiber (see Fig. \ref{fig:coupling_star} for more exact values). Instead,  for the bottom plots a fiber that is well aligned on the planet will couple with the efficiency that is visible at position $x=0$, $y=0$ (see Fig. \ref{fig:coupling_planet} for more exact values).

Most noticeable, spatially, is the widened diffraction spikes from the spiders of the coronagraphic masks, which has a strong impact on the coupling efficiency for the planet. 

\subsubsection{Fiber transmission and extraction}
\label{sec:fibertrans}

To carry the light from SPHERE to CRIRES+ with a clean PSF and therefore line-spread function on the spectrograph, we use single-mode fibers. Single-mode ZBLAN fibers manufactured by Le Verre Fluor\'e (LVF) are ideally suited to cover both $H$ and $K$ bands with the highest possible transmission. Best matches within their catalog are the fibers with a core diameter of 6.5\,\mic, which have a single-mode cutoff at 1.48\,\mic, and a MFD of 7.2\,\mic at $\lambda=1.5\,\mic$ and 12.67\,\mic at $\lambda = 2.5\,\mic$. While silica fibers designed for telecommunications could also  satisfy the requirement of high transmission in $H$ band very well, the ZBLAN fiber additionally enables observations in $K$ band.

The fiber transmission of HiRISE is calculated from attenuation values provided by LVF for the selected SMF design and for a fiber length of 55 m. This distance is approximately equal to that spanned by the UVES-FLAMES fiber connection on the VLT-UT2 (\citealt{Pasquini:2002}; FLAMES P103 user manual\footnote{\url{https://www.eso.org/sci/facilities/paranal/instruments/flames/doc/VLT-MAN-ESO-13700-2994\_p103.pdf}}), which is similar to the setup that will be adopted for HiRISE.

To reduce manufacturing risks and to facilitate the installation of the fibers, the complete fiber assembly consists  of three pieces joined with connectors, with losses of approximately 0.15\,dB per connector. We assume a 2\% loss for reflection on the input and output of the fiber.

The current design of the fiber extraction module (FEM) is based on four lenses to re-image the fiber onto the CRIRES+ focal plane and to match the $F$-number expected by CRIRES+ at the entrance of the spectrograph (i.e.,~$F=15$) in a broad wavelength range from $V$ to $K$ band. This allows  the operation of MACAO to keep the deformable mirror flat and provides high  efficiency at the slit. These additional optical surfaces, together with the absorption in the medium of a heavy glass type required for one lens, combine to give a transmission factor of $0.80$.  

\subsection{CRIRES+ spectrograph}
\label{sec:crires+}

Our baseline is that the high-resolution spectrograph that is used in combination with SPHERE is CRIRES+. CRIRES+ was installed in UT3 in early 2020, upgraded with a cross-disperser to increase the bandwidth, and with new low-noise detectors \citep{Dorn:2016}. It provides a spectral resolution up to $R=100\,000$ in six different bands ($Y$ to $M$), although not all bands are covered simultaneously. The bands of interest to HiRISE, $H$ and $K$, are covered by four grating settings each to provide 100\% wavelength coverage of the bands. In practice, a single setting will be used that covers approximately $68\%$ of $H$ band and 52\% of $K$ band, with gaps in the spectrum due to the way the three detectors cover the cross-dispersed orders.

The transmission of CRIRES+ was measured by its instrument team during the integration phase in Europe. The slit-independent transmission, but including the quantum efficiency of the detectors, is approximately 20\% in $H$ and $K$ band (Seemann, private communications). These values are approximate and subject to change;   final values will only be available when the instrument has been tested on-sky. Since they are currently the best available estimation, we use them for our simulations.

The 120\,\mic 0.2\as-wide, 10\as-long slit spans 3.1 pixels in width projected on the H2RG detectors. The read noise per pixel is approximately 7 electrons per readout and the dark current is 0.028 electrons per second\footnote{CRIRES+ forward simulator (\texttt{CRIFORS}): \url{https://github.com/ivh/crifors}}, with a gain that is 2.1 electrons per count. The dark current is scaled with the integration time and converted to dark noise by taking the square root. We assume that the light hits a $3.1 \times 3.1 $\,pixel area on the detector, and therefore both the read noise and dark noise are scaled by a factor of 3.1 to give a conservative estimate of the noise per resolution element.

\begin{figure}
  \centering
    \includegraphics[width=1.0\columnwidth]{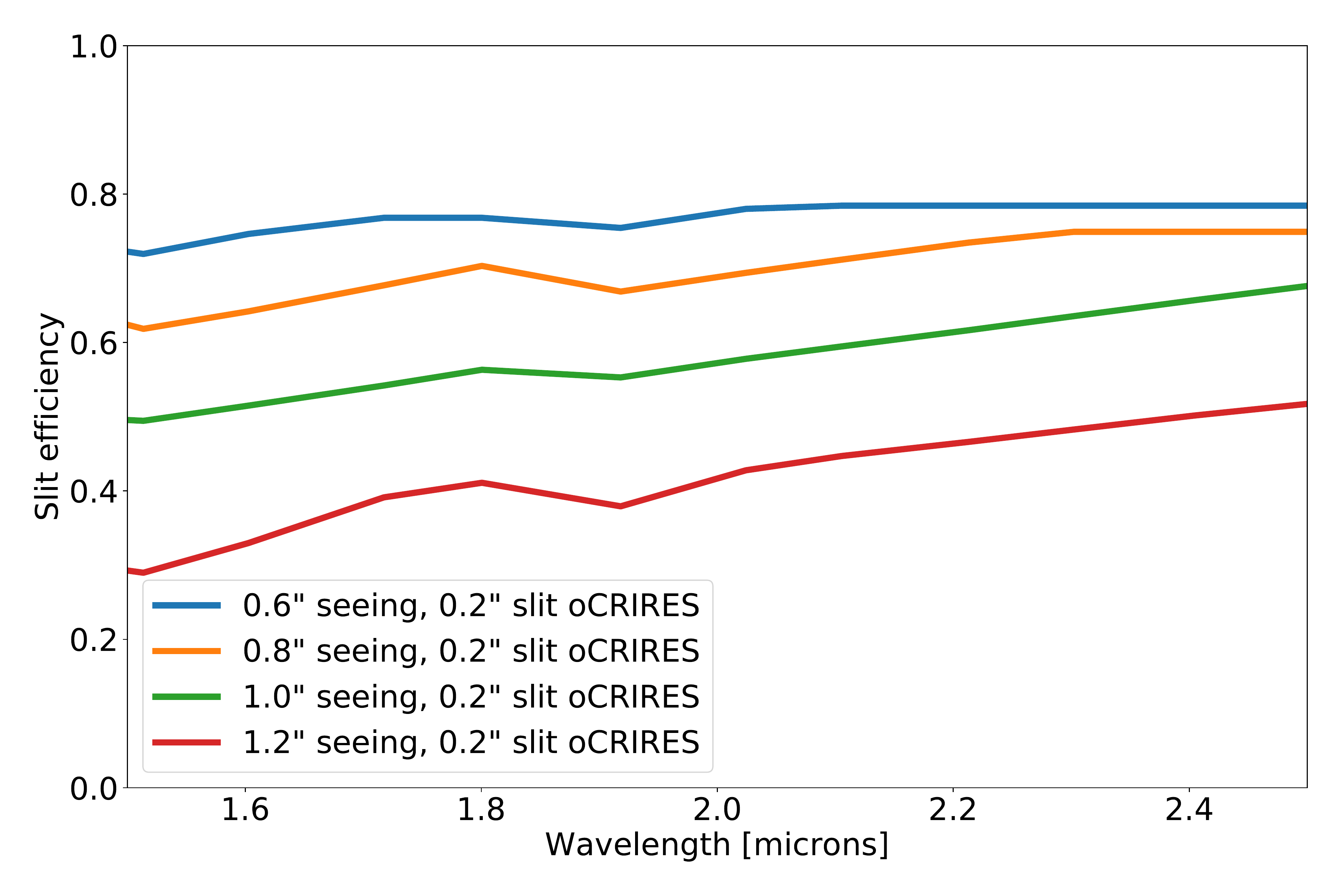}
    \caption{Slit efficiency as a function of wavelength for seeing values of 0.6\as, 0.8\as, 1.0\as, and 1.2\as, extracted from the P95 CRIRES ETC. These values are not expected to change for CRIRES+.}
    \label{fig:sliteff}
\end{figure}

As we want to compare the combined HiRISE system with CRIRES+ standalone (i.e.,~where the slit is placed across star and planet, similar to \citealt{Snellen:2014}) we determine the slit efficiency for both instrument configurations. Slit efficiencies were not included in the transmission measurement of CRIRES+ as the full slit was illuminated to be insensitive to these effects for the measurement of the transmission. The slit losses are small and constant with wavelength for HiRISE: the enclosed energy diagrams of the optical design of the FEM indicate that more than 93\% of the incoming light goes through the slit. Any AO degradation for HiRISE would translate into reduced fiber coupling efficiencies with unchanged slit losses. 

This is in contrast to CRIRES+ used in standalone, where slit losses are much higher and variable with seeing due to the lower order AO system MACAO. Figure~\ref{fig:sliteff} shows the fraction of enslitted light inside of the 0.2\as slit, referred to as slit efficiency, as a function of wavelength and for seeing values from 0.6\as up to 1.2\as, extracted from the P95 CRIRES exposure time calculator (ETC). We see that the slit efficiencies drop sharply with increasing seeing. At a seeing of 1.2\as the slit efficiency is almost halved compared to 0.8\as. If we look at a similar seeing difference for SPHERE we see a drop in Strehl, and therefore coupling efficiency, of about 15\%. As a cross-check for the CRIRES ETC slit efficiencies we extracted the spatial profile at 2.3\,\mic from the data of \citet{Snellen:2014} (program ID 292.C-5017) and rotated this profile to get an artificial PSF. The enslitted energy within a 0.2\as artificial slit corresponds within a few percent to the slit efficiency of the ETC at the same wavelength and DIMM seeing. As the MACAO system of CRIRES+ has not been significantly upgraded (and performance has to be revalidated on-sky) we assume that the slit losses of CRIRES can be directly used for our calculations of CRIRES+. In the remainder of the present work we use the average seeing of 0.8\as for CRIRES+ in standalone, as for the SPHERE ExAO simulations presented in Sect.~\ref{sec:fibereff}.

Lastly, the PSF profiles of CRIRES+ are taken from the ETC to simulate the strength of the stellar halo at the location of the planet. The previously mentioned slit efficiencies (enslitted energy) determine which fraction of this stellar light falls into the slit.

\subsection{End-to-end transmission}

\begin{figure*}
    \centering
    \includegraphics[width=0.9\textwidth]{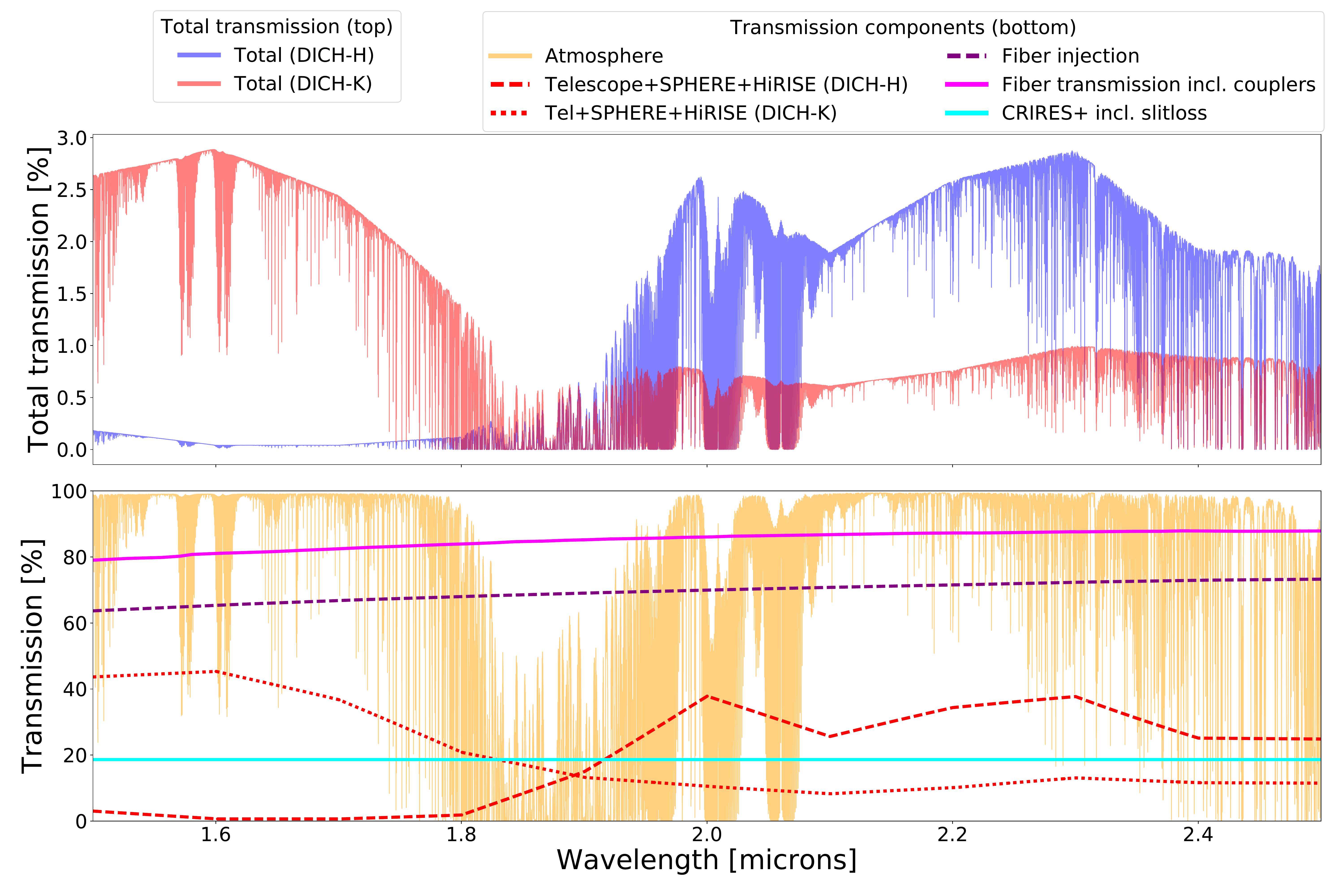}
    \caption{Upper panel: Total transmission for \texttt{DICH-H} and \texttt{DICH-K} (optimized  respectively for $K$ and $H$ band). Lower panel: Breakdown of transmission into components. Fiber injection and transmission based on the no-coronagraph case.}
    \label{fig:trans_breakdown}
\end{figure*}

Each transmission component is collected and multiplied to give the total transmission of the combination of instruments. Averaging over $H$- and $K$-band grating wavelength ranges we get the values in Table \ref{tab:trans}. The total transmission in $H$ and $K$ band are 1.90\% and 1.41\%, respectively, for the non-coronagraphic case. The corresponding breakdown of the transmission into multiple components as a function of wavelength can be seen in Fig. \ref{fig:trans_breakdown}. In this figure we also see the impact of using two different dichroics and the regimes over which they are optimal. The highest transmission is achieved by  using CRIRES+ alone, which has an end-to-end transmission of 9.56\% and 8.82\% in $H$ and $K$. Focusing on HiRISE the highest transmission is seen in the non-coronagraphic mode, then the CLC mode which only has a pupil stop and focal plane mask, and the poorest transmission is seen for the APLC, which includes an additional apodizing pupil mask. The transmission differences are purely driven by the change in injection efficiency and the additional light blocking of the coronagraphic masks. Swapping the existing dichroic for a clear plate that conserves the beam path and optical path length will give 24\% in $H$ and 58\% in $K$ band of additional light and is broken down in the upgrade column.

\begin{table*}
    \centering
    \caption{Average end-to-end transmission in $H$ and $K$ bands broken down into individual components. The baseline case uses current SPHERE dichroics. The upgrade case assumes a clear plate instead of less efficient SPHERE dichroics. For coronagraphic and fiber coupling transmission factors only one of each applies.}
    \label{tab:trans}
    \begin{tabular}{lcccccc}
    
    \hline\hline
                    & \multicolumn{2}{c}{HiRISE $H$}& CRIRES+ $H$  & \multicolumn{2}{c}{HiRISE $K$}& CRIRES+ $K$     \\
                    & Baseline& Upgrade&  Standalone  & Baseline& Upgrade& Standalone    \\
    \hline
    Atmosphere      & \multicolumn{2}{c}{91.2\%} & 91.2\%    & \multicolumn{2}{c}{75.4\%} & 75.4\%    \\ 
    \hline
    Telescope       &\multicolumn{2}{c}{79.8\%}&79.8\%&\multicolumn{2}{c}{80.6\%}&80.6\% \\ 
    \hline
    SPHERE CPI      &47.3\% & 58.8\% & -  & 36.8\% & 58.1\%   &- \\
    Coronagraph (APLC)      &\multicolumn{2}{c}{48.6\%} & -  & \multicolumn{2}{c}{48.6\%}   &- \\ 
    Coronagraph (CLC)      &\multicolumn{2}{c}{84.6\%} & -  & \multicolumn{2}{c}{84.6\%}   &- \\  \hline
    FIM optics      &\multicolumn{2}{c}{71.0\%}& - & \multicolumn{2}{c}{71.0\%}&-\\
    Fiber coupling (no coro.) &\multicolumn{2}{c}{66.7\%}& - & \multicolumn{2}{c}{71.5\%}&-\\ 
    Fiber coupling (APLC)  &\multicolumn{2}{c}{59.2\%}& - & \multicolumn{2}{c}{63.2\%}&-\\ 
    Fiber coupling (CLC)  &\multicolumn{2}{c}{56.3\%}& - & \multicolumn{2}{c}{60.1\%}&-\\ 
    Fiber transmission  & \multicolumn{2}{c}{78.2\%} & - &    \multicolumn{2}{c}{83.6\%}&-  \\
    FEM optics      &\multicolumn{2}{c}{80.0\%}&-  &\multicolumn{2}{c}{80.0\%}& -\\
     \hline
    Slit efficiency     &\multicolumn{2}{c}{93.0\%}&65.7\%  &\multicolumn{2}{c}{93.0\%}&72.6\% \\
    CRIRES+         &\multicolumn{2}{c}{20.0\%}& 20.0\%  &\multicolumn{2}{c}{20.0\%} &20.0\%\\ 
    \hline
    Total (no coro.) & 1.90\%& 2.36\% & 9.56\% & 1.41\%& 2.23\%& 8.82\%\\ 
    Total (APLC) & 0.82\%& 1.02\% & - & 0.61\%& 0.96\%& -\\ 
    Total (CLC) & 1.35\%& 1.68\% & - & 1.00\%& 1.59\%& -\\ 
    \hline
    \end{tabular} 
\end{table*}

\subsection{Background thermal radiation}
\label{sec:background}

The majority of the path of the full instrument combination is at an ambient temperature of about 14 degrees Celsius. Only inside of CRIRES+ is there a cryogenic vessel with the slit acting as a cold field stop. The slit sees the upstream thermal radiation, which comes from the re-imaging optics for the fiber extraction unit, the front surface of the fibers, and the thermal radiation transmitted from the SPHERE side. In the case of mirrors, the light that is not reflected is mostly absorbed so this directly gives us the emissivity of the mirror. In the case of lenses or dichroics, the light that is not transmitted is mostly reflected. The reflected paths look into the instrument, which is at ambient temperature, and so the emissivity is $(1-\mathrm{transmission})$ for these optical elements. The dominant thermal components will therefore be the poorest transmitting or reflecting optics that are at room temperature. 

In the case of SPHERE and HiRISE this is the IRDIFS dichroic and the coronagraph optics. As the Lyot apodizer and the Lyot stop have a chromium mask, the fraction of light that is not transmitted is reflected and therefore it sees the surrounding warm surfaces. We assume that only the light that geometrically fits into the 6.5\,\mic core of the fiber and in the angular cone that is spanned by the numerical aperture can pass through the fiber. 

On the CRIRES+ side, most of the optical path is cooled, but the fiber face can in the worst case be considered a half-open cylindrical cavity with emissive walls, and can therefore be considered to have strong emissivity and to be the dominant background source on the CRIRES+ side as seen from the cold slit. In our simulation we therefore implement SPHERE at an emissivity of 0.7, approximating the amount of light lost in the upstream branch, and the fiber-end at a very conservative emissivity of 1.0. For the SPHERE side we integrate background light over a circular surface with a diameter of 6.5\,\mic and a cone with a half-angle of $\arcsin(0.17)$, derived from the $F$-number. For CRIRES+, a larger part of the face of the fiber is visible from the slit than just the core. The 120\,\mic slit width back-projected onto the fiber gives a width of 28\,\mic, which is used as the emitting surface. We also use a cone with a half-angle of $\arcsin(0.17)$ as the emissive angular area. Using Helmholtz reciprocity, the surface area and the area in the cone are multiplied to give us the amount of light arriving at the slit, which corresponds to the thermal background radiation considered in the simulation.

\subsection{Simulation of the final spectra}
\label{sec:combiningmodel}

We use the source spectra scaled by the simulated detector integration time (DIT) and multiply them by all transmission factors down to the CRIRES+ detector, while the background light from the sky, SPHERE, and fiber are transmission-corrected from their respective locations of origin. The starlight, planet light, and atmospheric radiance are multiplied by the surface area of the telescope, which is $49.28\,\mathrm{m}^2$, considering the 8 m telescope and 14\% secondary obscuration and the width of each wavelength bin.

A science signal is made from the sum of the stellar halo, planet light, Earth's atmospheric radiance, dark current, background from fiber and background from the SPHERE instrument as measured at the CRIRES+ detectors over each $3.1 \times 3.1$ pixel area. We also construct a reference signal that contains all the components except for the planet light, which in practice will be obtained using multiple reference fibers (4+) that sample the PSF, background, and sky at different locations in the focal plane. While speckles have a chromatic dependence, it is a low-order effect, which is suppressed by our high-pass filtering. The fibers sampling the PSF provide multiple references for the starlight. Additionally, a reference spectrum can be taken by putting the star on the science fiber. The detailed data analysis of the HiRISE data will be studies in future work. For each spectral bin of both the science and reference signal we draw from a Poisson distribution with its parameter $\lambda$ as the total flux, and sum it with read noise drawn from a normal distribution. The final planetary spectrum to be analyzed is the difference between the science signal and the reference signal. For CRIRES+ standalone the impact of SPHERE and HiRISE is excluded from these calculations.

\section{Performance for known targets}
\label{sec:perf_known_targets}

\subsection{Simulations and signal-to-noise ratio estimation}
\label{sec:simu_snr_estimation}

The main scientific driver for the HiRISE project is the detailed characterization of known companions at small separations, where the detection and subsequent characterization of faint companions are typically limited by the stellar halo and by the high level of residuals after post-processing \citep[e.g.,][]{Cantalloube:2015}. We first estimate the performance of the combination of SPHERE and CRIRES+ in this specific regime by simulating known exoplanetary systems and evaluating the S/N on particular species expected in their atmospheres.

\begin{table*}
    \centering
    \caption{Planetary system parameters used as input for simulations.}
    \label{tab:targets}
    \begin{tabular}{l|cccc|ccccc|c}
    \hline\hline
                 & \multicolumn{4}{c}{Star}                & \multicolumn{5}{c}{Companion}  \\
    \hline
    System       & $H$  & $K_s$ & \Teff\tablefootmark{a}   & \logg\tablefootmark{a}      & Sep.  & $\Delta H$  & $\Delta K$  & \Teff\tablefootmark{a}       & \logg\tablefootmark{a} & References \\ 
                 &      &       & [K]         & [dex]      & [mas] & [mag]       & [mag]       & [K]         & [dex]      & \\
    \hline
    $\beta$\,Pic & 3.54 & 3.53  & 8052 [8100] & 4.15 [4.0] & 300   & 10.0        &  9.2        & 1650 [1700] & <4.7 [4.0] & 1, 2, 3 \\
    HIP\,65426   & 6.85 & 6.77  & 8840 [8800] & 4.23 [4.0] & 830   & 11.1        & 10.0        & 1450 [1500] & <4.5 [4.0] & 4, 5 \\
    51\,Eri      & 4.77 & 4.54  & 7256 [7200] & 4.13 [4.0] & 452   & 14.8        & 12.9        &  760 [ 700] & ~~4.3 [4.0] & 6, 7, 8 \\
    PDS\,70      & 8.82 & 8.54  & 3972 [4000] & 3.90 [4.0] & 199   & 9.12        &  7.8        & 1200 [1200] & ~~3.9 [4.0] & 9, 10\\ 
    \hline
    \end{tabular}
    \tablefoot{\tablefoottext{a}{The \Teff and \logg values from the literature are reported on the left side of the columns, while the values used for the models in the simulation are reported between brackets on the right side of the columns.}}
    \tablebib{(1) \citet{Bonnefoy:2013}; (2) \citet{Lagrange:2019}; (3) \citet{Chilcote:2017}; (4) \citet{Chauvin:2017}; (5) \citet{Cheetham:2019}; (6) \citet{Macintosh:2015}; (7) \citet{Samland:2017}; (8) \citet{Rajan:2017}; (9) \citet{Keppler:2018}; (10) \citet{Mueller:2018}.}
\end{table*}

The known systems that we evaluate in this section are HIP\,65426 \citep{Chauvin:2017a}, $\beta$\,Pictoris \citep{Lagrange:2009}, PDS\,70 \citep{Keppler:2018}, and 51\,Eridani \citep{Macintosh:2015}. These targets are planets (in contrast to brown dwarfs), relatively close in angular separation to their star and visible from the VLT. Their assumed physical and observational properties are provided in Table~\ref{tab:targets}. To avoid interpolations in the grids of stellar and sub-stellar atmospheric models,  for the stars and planets we use the models with the closest available \Teff and \logg. We simulate the realistic spectra for each of the targets with a varying integration time. Then we perform a cross-correlation analysis of the simulated total spectra, which includes all molecular contributions and noise, against both the input planetary spectrum and the spectral contribution of individual molecular species ($\mathrm{H}_2\mathrm{O}$, CO, and $\mathrm{CH}_4$). 

The simulations are performed for several instrumental configurations. First we consider CRIRES+ in standalone mode (i.e., without the HiRISE coupling). In this case only the telescope, atmosphere, and CRIRES+ transmission from Table~\ref{tab:trans} are considered in the simulation of the spectra. Then, for HiRISE we consider three different setups that include different types of coronagraphs: no coronagraph, APLC, and CLC. For the HiRISE simulations, the simulation of the spectra include the additional contributions in transmission from SPHERE, the FIM optics, the fiber coupling, and finally the FEM optics.

For this analysis we select the wavelength range available to each CRIRES+ grating setting, which we  extracted from the \texttt{CRIFORS} simulator. Both the observed and comparison spectra are preprocessed by removing the low-order (continuum) through a high-pass filter using a fast Fourier transform, as commonly done. This filtering step can be skipped if the full SPHERE/HiRISE/CRIRES+ combination can be spectrally calibrated. The S/N of the cross-correlation is then calculated using the matched filter approach \citep{Ruffio:2017} where the maximum likelihood S/N is defined by
\begin{equation}
    \mathrm{S/N}_v = \frac{\sum_{i}^{k}d_i\,m_{v,i} / \sigma_i^2}{\sqrt{\sum_{i}^{k} m_{v,i}^2 / \sigma_i^2}},
\end{equation}

\noindent where $d_i$ is the high-pass filtered observed data, $m_{v,i}$ is the high-pass filtered model shifted at radial velocity (RV) $v$ and resampled at the same wavelengths as $d_i$, and $\sigma_i$ is the estimated noise of the data $d_i$ (before the high-pass filter). The sums from $i=0$ to $k$ are over each spectral data point. The calculation is repeated for a range of radial velocities $v$ and the final S/N is evaluated at the input RV, which is equal to zero in the case of these simulations.

The noise in the data is derived by quadratically adding up all the known contributions of noise. In the case of the simulation the exact noise contributions are perfectly known, but  the case with real data will have to be handled differently. In HIRISE, in addition to the science fiber where the planet signal will be injected, we foresee the addition of reference fibers that will sample the stellar speckle field at locations around that of the planet. These reference signals will then be used in the data analysis for subtracting part of the stellar contribution and estimating the noise in the data.

As detailed in Sect.~\ref{sec:crires+}, CRIRES+ offers several grating settings that   cover a full band in up to four distinct observations. An initial set of cross-correlations on all $H$- and $K$-band settings separately revealed that the S/N values do not vary much between grating settings for the models that we use, even when focusing on individual atmospheric species,  presumably because the amount and the strength of the molecular lines that are covered do not change significantly between grating settings within the same band.  Hereafter we therefore only consider \texttt{H\_1\_4} and \texttt{K\_1\_4} settings to accelerate calculations.

\subsection{Performance as a function of exposure time}
\label{sec:exptime}

\begin{figure*}
  \centering
    \includegraphics[width=1.0\textwidth]{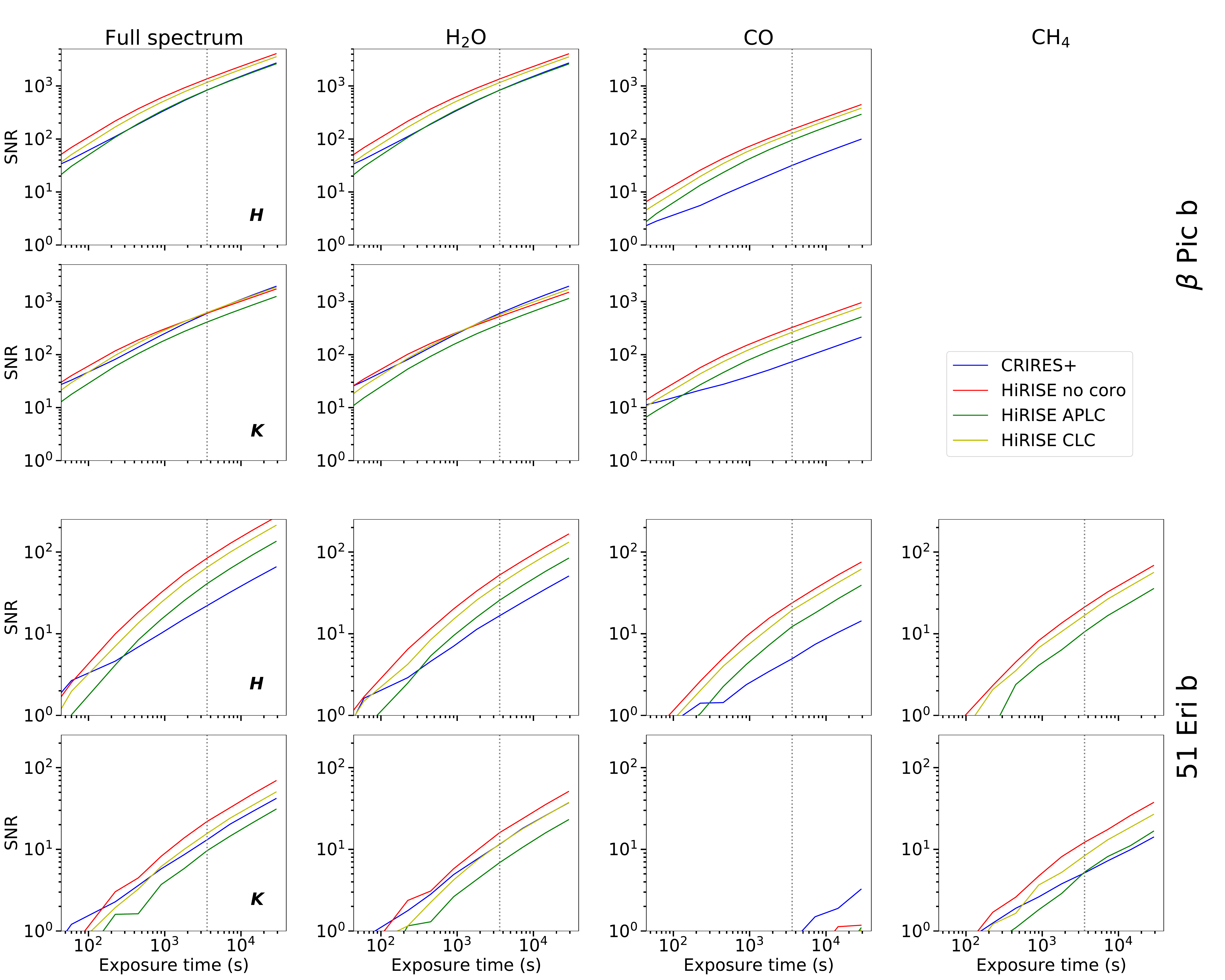}    \caption{Estimation of the S/N as a function of exposure time for the observation in $H$ and $K$ band of $\beta$\,Pictoris\,b (top rows) and 51\,Eridani\,b (bottom rows) with CRIRES+ (blue lines) and HiRISE in different coronagraphic configurations: no coronagraph (red lines), APLC (green lines), and CLC (yellow lines). The S/N is computed with a matched filtering approach (see Sect.~\ref{sec:simu_snr_estimation}) comparing the simulated spectra with the noiseless input planet spectrum (Full spectrum column) and with noiseless spectral templates for molecules (H$_2$O, CO and CH$_4$). The vertical dotted lines indicate 1 hour integration time.}
    \label{fig:molecule}
\end{figure*}

For a quantitative comparison of CRIRES+ standalone with HiRISE, we first evaluate the expected S/N as a function of the exposure time for the brightest and faintest planets in our sample, namely $\beta$\,Pictoris\,b and 51\,Eridani\,b. The simulated spectra are compared to both the noiseless input planet spectrum and to individual spectral templates for molecules expected in the atmospheres of the two planets. The results are plotted in Fig.~\ref{fig:molecule} for the molecules that are expected a priori  to have strong features, which are H$_2$O and CO for $\beta$\,Pic\,b and H$_2$O, CO, and CH$_4$ for 51\,Eridani\,b.

We first look at the S/N computed when comparing the simulated spectra with the input noiseless planet spectrum. We see a clear increase in S/N as a function of integration time in all instrumental configurations. Indeed, $\beta$\,Pictoris\,b and 51\,Eridani\,b have relatively bright host stars, $H=3.54$ and $H=4.77$, and are at very close angular separation from their host, 300\,mas and 452\,mas, respectively. Therefore, for both systems the noise is strongly dominated by the photon noise of the stellar halo, resulting in a slow increase of the S/N with exposure time. As we show in Sect.~\ref{sec:noise_breakdown}, the noise regime has a strong impact on performance.

In this comparison we also see that HiRISE without coronagraph almost always provides a gain of at least a factor $\gtrsim$1.4 in S/N with respect to CRIRES+. Assuming that the regime is indeed photon noise limited, this roughly translates to a gain of a factor of $\gtrsim$2 in exposure time to reach the same S/N. It is only at short exposure times (typically <100\,s) that in some cases CRIRES+ provides a better S/N than HiRISE as the short integration times can put the systems in the read noise limited regime where higher transmission is more important. We note that for the $\beta$\,Pic host star, HiRISE without coronagraph can be surpassed by HiRISE with coronagraph. We analyze the effect of the coronagraph in HiRISE in more detail in Sect.~\ref{sec:effect_coronagraph}.

The same results are observed when looking at individual molecules. We see that it is possible with HiRISE without  coronagraph to reach the same S/N  in a fraction of the observing time as CRIRES+ reaches in one hour. For H$_2$O in $H$ band we see that the same S/N achieved by CRIRES+ in 1 hour of integration time is reached in only 1560 seconds with HiRISE, corresponding to a factor of 2.3 gain, while for CO the same value as CRIRES+ is reached in only 300 seconds, corresponding to an even higher gain (a factor of 12). In $K$ band, HiRISE provides gains of approximately a factor of 11 for CO and no clear improvement for H$_2$O.

CH$_4$ is barely detected in the atmosphere of 51\,Eridani\,b with CRIRES+ in the $H$ band, while it appears to be easily detectable with HiRISE provided typical exposure times of longer than  a few hundred seconds. In $K$ band the detection appears easier for CRIRES+, but even longer exposure times than in $H$ band are required.

Interestingly, the S/N values reached for H$_2$O are very close to the values reached when comparing to the full input spectrum. H$_2$O is the dominant contributor in terms of spectral lines in the atmospheres of these planets,  in $H$ and in $K$ band, so it has a higher weight in the matched filtering and therefore drives the final value of the S/N. We also note that CO appears barely detectable in $K$ band for 51\,Eridani\,b, which is surprising considering the strong CO overtones starting at 2.29\,\mic. This absence of detection is likely due to the instruments starting at 1.9--2.0\,\mic and to the general faintness of the planet in this band ($K=17.4$).

We conclude that even though the global transmission of HiRISE is low, it provides a significant gain of observing time with respect to CRIRES+. The exact values provided in this section are model dependent and some molecules might be more pronounced in theory than seen in reality around these targets, but the order of magnitudes will hold whatever the model, as will the relative gains between HiRISE and CRIRES+.

\subsection{Noise breakdown}
\label{sec:noise_breakdown}

\begin{figure*}
  \centering
    \includegraphics[width=1\textwidth]{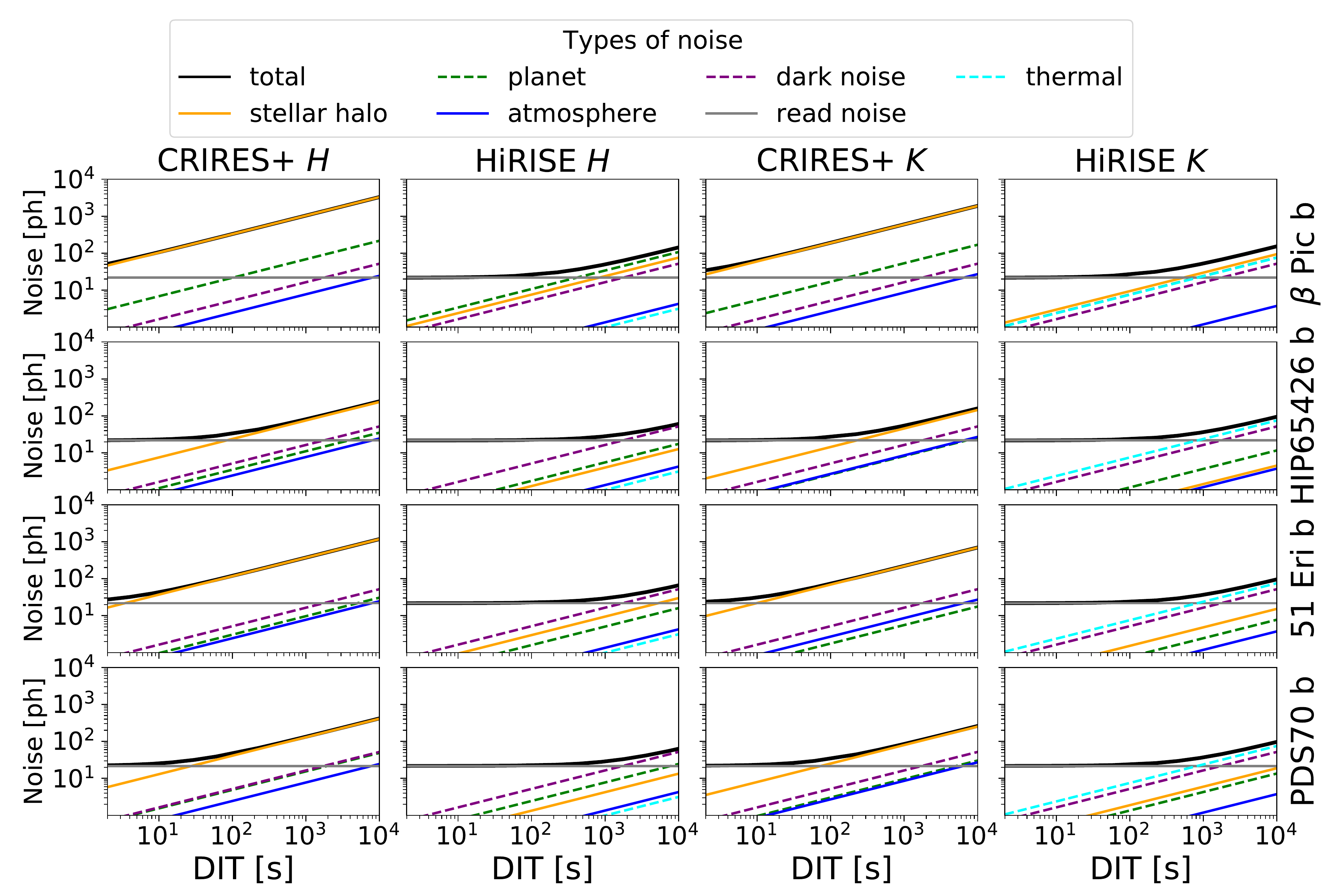}
    \caption{Breakdown of the noise into individual sources vs. detector integration time for four planets observed using CRIRES+ and HiRISE without coronagraph. The noise is effectively the average noise per resolution element within a certain band and is expressed in photons.}
    \label{fig:noise_breakdown}
\end{figure*}

The performance of HiRISE is strongly dependent on the noise regime in which the system is used, which itself depends on the brightness of the target that is observed. To illustrate the change in noise regime when using HiRISE, we show the square root of the mean of the variance within a certain band (the average noise per resolution element). In this simulation we consider the four planets presented in Sect.~\ref{sec:simu_snr_estimation}, which span a wide range of visual magnitudes in $H$ and $K$ band: $\beta$\,Pictoris\,b ($H=13.5$, $K=12.7$), PDS\,70\,b ($H=17.9$, $K=16.3$), HIP\,65426\,b ($H=18$, $K=16.8$), and 51\,Eridani\,b ($H=19.6$, $K=17.4$). We generate spectra for the four targets and we extract the noise for each of the noise sources.

Figure \ref{fig:noise_breakdown} shows the average noise per resolution element for each noise contribution in the simulated spectra for CRIRES+ and HiRISE without coronagraph as a function of time. The figure demonstrates that for each planet observed with CRIRES+ one of the strongest dominating noise sources is the Poisson noise on the stellar halo, which is therefore the performance limiting factor, and any suppression of this noise will directly provide a gain in S/N.

When using SPHERE in combination with CRIRES+ the noise regime changes drastically. In most cases, even without a coronagraph, HiRISE suppresses the stellar halo to a point where, due to the reduced transmission, the noise terms that originate in the sensor become relatively more important and the stellar halo noise is no longer the dominant component. In $H$ band the dominant noise source becomes either dark noise, read noise or noise on the planet itself depending on the target brightness, separation, planet to star contrast and integration time. For the brightest stars like $\beta$\,Pictoris, the stellar halo in addition remains an important noise contributor. In $K$ band the dominant noise sources are read noise, dark noise and thermal noise from SPHERE, with in addition the stellar halo for the very brightest stars. In general, after starlight suppression several noise terms are within one order of magnitude of each other. To avoid the read noise limited regime a detector integration time of at least $\sim$1000 seconds is needed for HiRISE. Detector integration times around this duration would also provide a reduced risk of saturation and less spectral blurring than much longer exposures. For planets at 100 mas around bright stars ($H=3$) the 62000 electrons full well capacity is reached in the science fiber without a coronagraph in 8000 seconds DIT for $H$ band (even when assuming the light is concentrated into 3 pixels per resolution element). For 50 mas it is reached in 600 seconds with the stellar halo as limiting factor. In $K$ band these limits are respectively $4000$ and $250$ seconds. Taking the 42000 electron per pixel limit for the linear regime, this level is reached in a fractionally shorter time (68\%), which requires exposures that are as short as 3 minutes at 50 mas in $H$ band. Fainter host stars allow for longer DITs. 

This noise breakdown illustrates the two main factors that drive the performance of high-contrast systems in general, and is especially applicable for systems coupled with high-resolution spectroscopy: the reduction of the stellar contribution at the location of the planet using ExAO and coronagraphy, and the global transmission of the system. Without the former, the performance is largely driven by the photon noise from the star, the situation in which any ideal system should be, but the noise level will be high compared to the faint signal of the planetary companions that we seek. Conversely, without the latter the other systematic effects in the system will start to become limiting factors compared to the photon noise.

\section{Discovery potential}
\label{sec:discovery_potential}

We anticipate the discovery of more directly imaged (or imageable) planets in the near future. The current SPHERE GTO survey SHINE \citep{Chauvin:2017} has discovered several potential planetary candidates at small angular separations that are difficult to confirm with current imaging capabilities (SHINE consortium, private communication). Additionally, the ESA/Gaia mission \citep{Gaia2016} is expected to release thousands of exoplanet candidates discovered through the astrometric method out to separations of 5\,AU and distances of 500\,pc \citep{Perryman:2014} in its final data release. Both categories of planetary candidates are found in regimes that are pushing  the limits of what the current generation of direct imagers can do, but provide a crucial view on an unprobed range of parameter space (i.e.,\,planets closer to their star, and older host stars). 

It is therefore interesting to look at the expected performance of HiRISE as a planet discovery or confirmation instrument. With a sparse field of view of a few single resolution elements, HiRISE certainly cannot be used to blindly search for planetary companions. However, with prior knowledge of the location of a candidate, HiRISE can become a powerful confirmation instrument and provide important high spectral resolution data on confirmed companions.

In this section we compare the detection performance of HiRISE and CRIRES+ (Sect.~\ref{sec:detection_performance}) for unseen companions, in particular as a function of the stellar host brightness (Sect.~\ref{sec:hmag}). Then we study in more detail  for HiRISE the impact of the coronagraph choice (Sect.~\ref{sec:effect_coronagraph}) and of the overall transmission of the system (Sect.~\ref{sec:effect_transmission}).

\subsection{Detection performance}
\label{sec:detection_performance}

We determine detection limits for HiRISE and CRIRES+ standalone by simulating observed spectra for a grid of planets in a range of separations ($20 \le r \le 1000$\,mas) and contrasts (delta magnitudes of $7.5 \le \Delta m \le 16$), and located around host stars with $H$-band magnitudes from 3 to 9. We set the integration time to 2 hours and calculate the S/N through a matched filter approach with the noiseless input planet spectrum. Although optimistic, this comparison with the input spectrum enables us to investigate the ultimate performance of the system. 

\begin{figure*}
  \centering
    \includegraphics[width=1.0\textwidth]{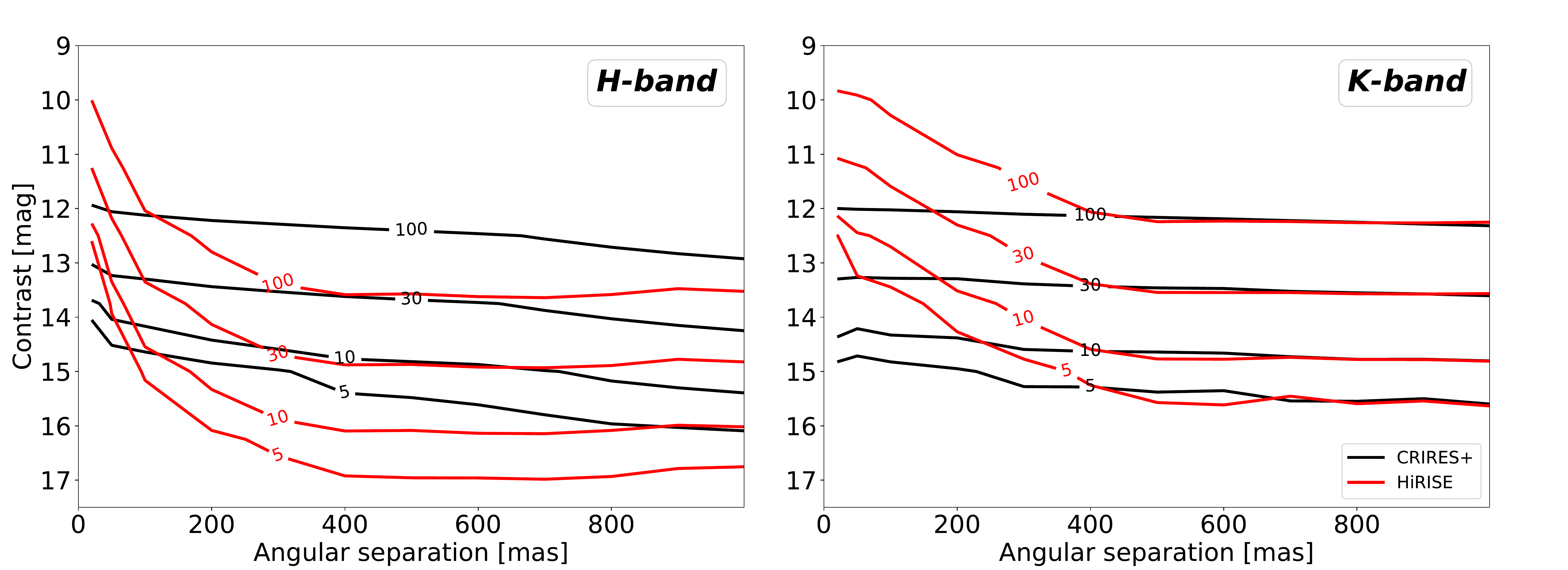}
    \caption{Signal-to-noise ratio as a function of contrast ($\Delta m $) and separation for HiRISE without a coronagraph (red contour lines) and CRIRES+ standalone (black contour lines). The simulation is performed for a $\beta$\,Pictoris-like host star with a 1200\,K planet and 2 hours of integration time. The S/N  is computed with a matched filtering approach (see Sect.~\ref{sec:simu_snr_estimation}) comparing the simulated spectra with the noiseless input planet spectrum. For this combination of parameters the S/N inward of $400-500$ mas is dominated by the noise on the stellar halo. Outside of $400-500$ mas it is limited by dark and read noise.}
    \label{fig:potential}
\end{figure*}

The results are presented in Fig.~\ref{fig:potential} with contours at fixed values of S/N for a bright host similar to $\beta$\,Pictoris (A5V, $H=3.5$) and a temperature of 1200 K. Additional results for a fainter host star similar to HIP\,65426 (A2V, $H=6.9$) are provided in Appendix~\ref{sec:contrast_curves}. In this figure only the HiRISE non-coronagraphic setup is shown in comparison to CRIRES+ as it is the mode that provides the best performance and best inner working angle (see Sect.~\ref{sec:effect_coronagraph}). As the noise breakdown has demonstrated, the S/N of CRIRES+ is dominated by the stellar halo at most separations. At separations larger than 1000\,mas where the noise for HiRISE is dominated by read, dark and instrumental thermal noise (depending on the band), CRIRES+ logically provides better performance than HiRISE. Likewise, at separations below 50 mas where the stellar halo starts to dominate even for HiRISE and where the planet and source are effectively unresolved, the additional transmission of CRIRES+ beats the HiRISE non-coronagraphic mode, although in this case the RV signal of the planet needs to be probed in time to disentangle star and planet light and the exposure time sufficiently short to avoid saturation on the star. The behavior at small separations is similar for the faint $H=6.8$ host (see Fig. \ref{fig:potential_hip}). 

Even so, in the intermediate range from 50\,mas up to several hundred mas (800\,mas for both bright and faint) HiRISE provides a substantial gain in sensitivity. At 200\,mas in the $H$ band, there is a 1.3 magnitude gain in contrast at which a S/N of 5 is reached. Conversely, looking at the S/N at a contrast $\Delta m=15$ and separation of 200\,mas, we can get an increase in S/N of a factor of $\sim$4, equivalent to a gain in observing efficiency of about 16 times. For the faint host star in $H$ band we see a gain of 1.0 magnitudes in contrast at 200 mas, and at $\Delta m=13.75$  and a separation of 200 mas we see an increase in S/N of a factor 2.4, which translates into a gain in observing efficiency of a factor 5.8.

In $K$ band around bright hosts, HiRISE narrowly outperforms CRIRES+ in a window between 400 and 650\,mas, with a gain of 0.2 magnitudes at 600\,mas for a S/N of 5. For fainter host stars, the window disappears for planets at 1200\,K. Despite the use of high-throughput fibers, the performance of HiRISE is clearly limited in $K$ band with respect to CRIRES+ in standalone. At low overall throughput, the contribution of the thermal noise stays an important limitation, which cannot be compensated by the gain from ExAO and spatial filtering by the fiber.

With a planet temperature of 1700\,K the highest gains seen in $H$ and $K$ band are 1.2 magnitudes at 300 mas and 0.6 mags at 500 mas respectively for host stars as bright as $\beta$\,Pictoris. For fainter stars like HIP\,65426 the highest gains at 200 mas are respectively 0.8 and 0.1 mags. The corresponding plots for these values are not shown but the numbers are provided as a reference.

\subsection{Relation with apparent magnitude of host}
\label{sec:hmag}

\begin{figure*}
    \centering
    \includegraphics[width=1.0\textwidth]{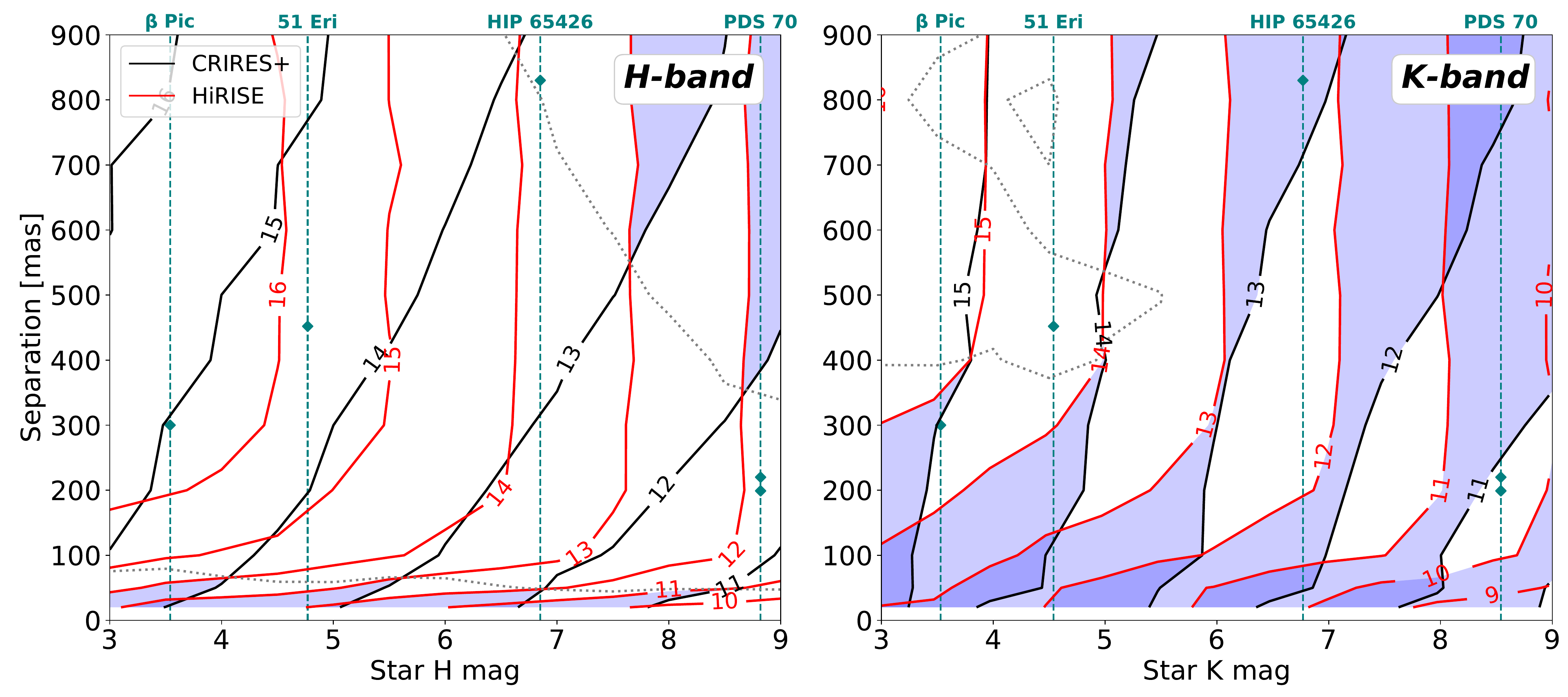}
    \caption{Contours showing the contrast ($\Delta m$) where an S/N of 5 is reached as a function of $H$ (left) and $K$ (right) magnitudes of the host star and separation of a 1200\,K planet in 2 hours of observing time. The black and red contours show the limits for CRIRES+ and HiRISE without coronagraph, respectively. The gray dotted contour shows where HiRISE and CRIRES+ have equal performance, and the blue shaded regions indicate where CRIRES+ outperforms HiRISE for the provided contour lines. The magnitudes of known planet hosting stars are indicated with vertical green dashed lines, and the separations of their known companions with diamond markers.}    \label{fig:potential_hmag}
\end{figure*}

The dependence of the performance on the host magnitude is an important parameter. To analyze how the performance varies with the host NIR magnitudes, we repeat the same S/N calculations as before, but vary the magnitude of the A5V host star in a range from 3 to 9. Figure \ref{fig:potential_hmag} shows the contrast at which a S/N of 5 is reached for a two-hour integration time and a 1200\,K companion, as a function of angular separation and host star magnitude. Again this comparison is between CRIRES+ and HiRISE without a coronagraph. This figure shows the trends previously identified, with HiRISE giving enhanced performance from separations of 50\,mas to at least 400\,mas in $H$ band, even for hosts as faint as $H=8.5$. For stars that are brighter than the above-mentioned magnitudes the outer limit extends beyond 400\,mas, while the inner limit remains similar. For $K$ band this region spans from about 400\,mas to 700\,mas for stars with $K=3.5$, but even in this region the gain provided by HiRISE is extremely limited. At $K=4.5$ the region becomes very narrow and the contrast improvement is only marginal. The gray dotted line shows where the performance of both instruments is equal and provides an overview of the regime that HiRISE should operate in (for the given simulation parameters), which  is inward of this line.

For HIP\,65426\,b we see that CRIRES+ has a small advantage over HiRISE in $H$ band. With a separation of 830\,mas and a moderately faint host star, the noise is initially dominated by the stellar halo but not as strongly as with candidates closer to the star (see Sect.~\ref{sec:noise_breakdown}). This means that suppressing the stellar noise with SPHERE gives a small improvement, but with the additional transmission losses of HiRISE this unfortunately does not give a greater improvement in S/N or contrast after a given amount of exposure time. For PDS\,70\,b, which has an even fainter host star, this similarly means that while the noise is dominated by the stellar halo for CRIRES+ due to the small separation, the reduction of throughput gives only a small improvement in S/N.

These results still need a few words of caution regarding the adaptive optics performance. In the HiRISE simulations, the $R$ magnitude of the host star is not varied, so the same $R=5$ value is assumed for all simulations. For fainter targets than $R=10$, the ExAO correction decreases, which necessarily impacts the fiber injection efficiency, reducing even further the overall transmission of the system. However, the situation is probably even worse for CRIRES+, which features only a low-order AO system with a wavefront sensor in the visible and which starts losing performance around $R=12$. While it is beyond the scope of this paper to analyze the full AO performance of the two instruments, it should be remembered that the performance of both HiRISE and CRIRES+ is decreased for faint host stars due to their respective AO systems.

\subsection{Effect of the coronagraph}
\label{sec:effect_coronagraph}

We showed in Sect.~\ref{sec:exptime} that the best results for HiRISE are obtained without a coronagraph. This is somewhat counter-intuitive because, as we highlighted in Sect.~\ref{sec:noise_breakdown}, the importance of reducing the stellar halo and diffraction pattern is exactly the purpose of a coronagraph. However, this simple statement does not factor in the reduced transmission induced by a coronagraph or its effect on the shape of the PSF.

\begin{figure*}
  \centering
    \includegraphics[width=1.0\textwidth]{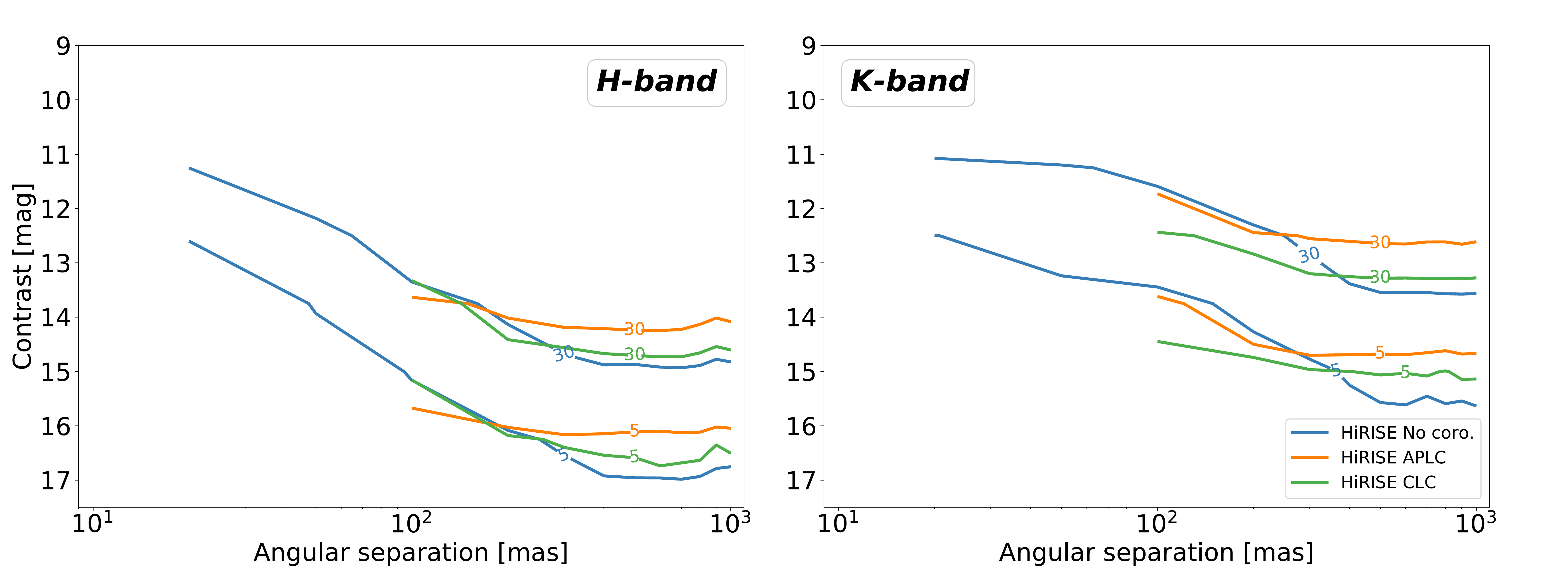}
    \caption{Signal-to-noise ratio as a function of contrast ($\Delta m $) and log-scaled separation for HiRISE without a coronagraph (blue lines), with an APLC (orange lines), and with a CLC (green lines). The simulation is performed for a $\beta$\,Pictoris-like host star with a 1200\,K planet and 2 hours of integration time. S/N values below the inner working angle radius of 92.5\,mas have been suppressed for the two modes using the focal plane mask. The S/N is computed with a matched filtering approach (see Sect.~\ref{sec:simu_snr_estimation}) comparing the simulated spectra with the noiseless input planet spectrum.}
    \label{fig:coronagraph}
\end{figure*}

In Fig.~\ref{fig:coronagraph} we compare the HiRISE detection limits at 5$\sigma$ and 30$\sigma$ for the different coronagraphic modes, again for a two-hour integration time on a 1200\,K planet around a $\beta$\,Pictoris-like host star. Additional results for a fainter host star similar to HIP\,65426 (A2V, $H=6.9$) are provided in Appendix~\ref{sec:contrast_curves}. Far away from the star, where dark noise, readout noise, and atmospheric noise dominate, the highest S/N is reached for the configuration with the highest transmission (i.e., without any type of coronagraph). Closer to the star, the optimal mode strongly depends on the brightness of the host. For a bright host the stellar halo remains a significant noise component close to the star, and therefore the extra starlight suppression provided by coronagraphs can further improve the S/N. Between 100 and 250\,mas of separation, the CLC slightly outperforms the non-coronagraphic mode, although for the faintest planets with a contrast of 16 magnitudes the APLC has a small advantage below 200\,mas. Interestingly, the gain provided by the coronagraph is higher in $K$ band, probably because of the improved stellar light rejection in this band thanks to a better correction of the turbulence and the stronger contribution of the halo due to the effective decrease in separation when expressed in $\lambda/D$. For a fainter host close to the star (Fig.~\ref{fig:coronagraph_hip}), the reduced transmission that goes hand in hand with the improved coronagraphic performance leads to a sub-optimal result and the S/N is highest in $H$ band with the non-coronagraphic mode of HiRISE.

We note that the APLC and the CLC both provide an improvement only at very small angular separations, typically below 200--250\,mas where the Airy ring intensity is highest. Considering that the focal plane mask has a radius of $\sim$90\,mas, this leaves only a very narrow window where the use of a coronagraph with HiRISE provides a meaningful gain, although this is a region close to where we expect to find currently undiscovered planets. Moreover, the CLC almost always outperforms the APLC despite its much reduced diffraction suppression at small inner-working angles. This is largely due to the significant loss in transmission induced by the pupil apodizer that blocks 57\% of the incoming light. The Gaussianization of the PSF induced by the apodizer in the APLC configuration leads to a slightly improved injection efficiency in the science fiber, but this gain is far outweighed by the loss of transmission, also caused by the apodizer. While the coronagraphic improvement is only seen in a limited range of angular separations, it enables us to choose an optimal coronagraph (among the types currently available in SPHERE) in an observing campaign and thereby to maximize the S/N for each target.

\subsection{Effect of transmission}
\label{sec:effect_transmission}

As previously hinted at in Fig. \ref{fig:noise_breakdown}, the enhanced starlight suppression only makes an impact when the noise of the stellar halo is the dominant noise factor, in which case it improves the S/N by the inverse of the square root of the starlight suppression factor. Increasing the transmission will increase the S/N by the square root of the transmission as the signal increases linearly and the dominant noise source increases by a square root, unless the noise is dominated by readout or dark noise, in which case the S/N improvement will be linear. In any case, additional transmission allows us to be less limited by the other noise sources. It maximizes the effectiveness of the increased stellar suppression and is therefore a critical parameter to optimize. 

\begin{figure}
  \centering
    \includegraphics[width=1.0\columnwidth]{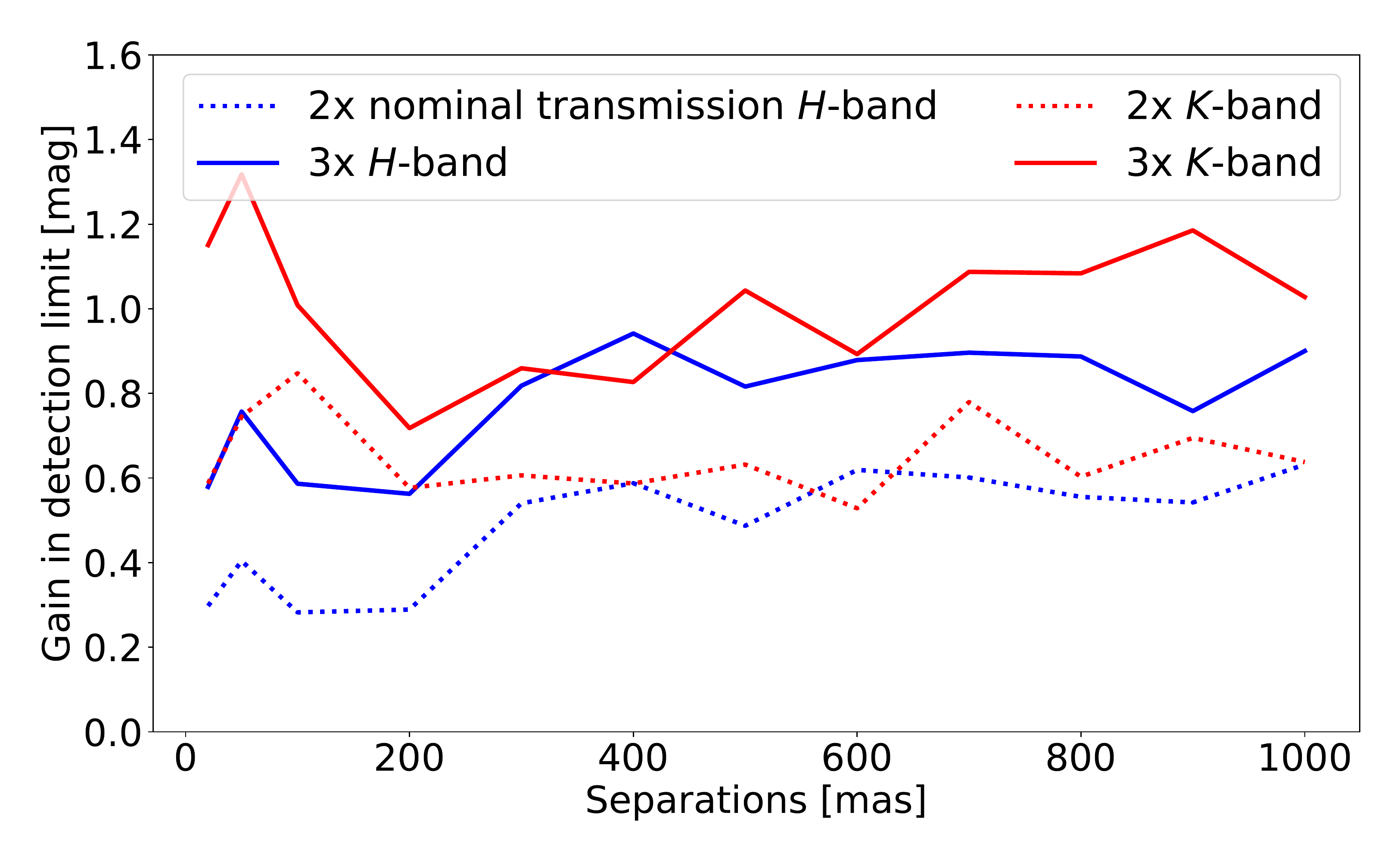}
    \caption{Gain in detection limit (in magnitude at $5 \sigma$ significance) with respect to nominal transmission as a function of separation for HiRISE without a coronagraph, evaluated for different transmission factors: two (dotted) and three (solid) times nominal transmission. The simulation is performed for a $\beta$\,Pictoris-like host star with a 1200\,K planet and 2 hours of integration time. The S/N is computed with a matched filtering approach (see Sect.~\ref{sec:simu_snr_estimation}) comparing the simulated spectra with the noiseless input planet spectrum.}
    \label{fig:fudge}
\end{figure}

To investigate the impact of transmission on HiRISE performance, we run the same simulations as in Sect.~\ref{sec:detection_performance} to obtain S/N as a function of contrast and separation, but in this case we increase the transmission of the spectrograph (here CRIRES+) by a multiplicative factor to simulate the effect of transmission enhancement measures. We compare the gain in detection limit expressed in magnitudes for a 5$\sigma$ detection for 2$\times$ and 3$\times$ transmission with respect to nominal transmission in Fig. \ref{fig:fudge}. The figure shows a step-wise improvement in contrast limit within 2 hours of integration with increased transmission of the spectrograph. Generally, we see that each increase in transmission increases the S/N by a factor of approximately 1.5 (close to $\sqrt{2}$). For the faintest objects where the dark and read noise are an important part of the total noise, a 2$\times$ increase in transmission gives an S/N increase of a factor 1.8, close to the factor 2 improvement that is expected when completely limited by read and dark noise. For both bright or faint hosts, in both $H$ and $K$ bands, there is an increase in contrast between 0.5 and 0.7 magnitudes at all separations when the throughput is doubled.

The overall transmission is clearly a driving parameter of the performance of HiRISE, and of any system coupling high-contrast imaging and high-resolution spectroscopy. While a specifically designed ExAO coronagraphic instrument with a fiber-fed high-resolution spectrograph would be able to achieve significant transmission improvements, it is interesting to look at options that could be implemented in existing instrumentation to gain in transmission. Whatever the coronagraph or spectrograph, small gains can be obtained by minimizing the number of optics and improving the efficiency of dichroic filters and anti-reflection coatings. Although these seem like minor contributors, they add up when considering many surfaces. For SPHERE, a more transmissive dichroic would give a 24\% and 58\% increase for $H$ and $K$ band respectively. If only the $H$ band is considered for science, a small transmission increase a ($\sim$10\%) gain can be achieved with respect to ZBLAN fibers by using standard telecom fibers made from fused silica, but in this case the $K$ band is totally lost due to the poor performance of such fibers beyond 1.7--1.8\,\mic. Additionally, reducing the amount of fiber couplers in HiRISE would increase throughput with a gain of 3.5\% per coupling.

From the coronagraphic and fiber injection point of view, the transmission can be significantly improved by reshaping the SPHERE PSF into a Gaussian using a PIAA-like beam shaper \citep{Jovanovic:2017}. Contrary to the APLC, which uses an amplitude mask that blocks part of the incoming photons, the PIAA reshapes the beam using phase only, which does not induce any loss of photons. In theory, the fiber injection efficiency can be increased up to almost 100\% with a PIAA. However, the PIAA requires very specific, hard-to-manufacture optics, which can be hard to retrofit into an existing instrument like SPHERE. An alternative option for benefitting from diffraction suppression without the use of loss-inducing amplitude masks is to use wavefront control to null any underlying speckle and create a dark hole at the known location of a companion. While this option will not increase the injection efficiency, it will largely decrease the level of stellar light at the location of the planet. Recent progress has been made in this direction in the SPHERE instrument with pair-wise probing and electric field conjugation to estimate and correct the remaining speckles \citep{Potier:2020}.

Finally, having a dedicated, compact, diffraction-limited, and fiber-fed spectrograph designed for high transmission would certainly be a significant gain. One step in this direction is the Virtual Image Phased Array (VIPA) spectrograph \citep{Bourdarot:2018}, which is a high-resolution spectrograph ($R=80\,000$)  based on a VIPA optic for spectral dispersion. Because it is optimized to be fed by a diffraction limited PSF, the design is compact ($\sim$40\,cm) and can achieve high spectral resolution. The transmission is higher than CRIRES+ by a factor of two due to its efficient optical design, as seen in Table 4 of \citet{Bourdarot:2018}. In the case of HiRISE, a VIPA-like spectrograph would enable a direct gain of a factor 2 in transmission from the spectrograph. Additionally, because of the compactness of the spectrograph, it could be directly installed next to the SPHERE instrument on the Nasmyth platform, therefore reducing the fiber length required to reach the spectrograph and all the associated losses.

\section{Conclusion}
\label{sec:conclusion}

The direct detection of exoplanets has stepped into a new era with dedicated high-contrast imagers like SPHERE, GPI, or SCExAO. However, their characterization is still in large part limited by the accessible spectral resolution \citep[e.g.,][]{Zurlo:2016,DeRosa:2016,Greenbaum:2018,Samland:2017,Rajan:2017,Cheetham:2019}, although recent progress has been made with VLTI/GRAVITY \citep{Lacour:2019,Nowak:2020} at $R=4000$. For another major step forward, access to much higher spectral resolutions is required, as was demonstrated by the seminal study on $\beta$\,Pictoris\,b by \citet{Snellen:2014}. The return of CRIRES+ at the VLT on the same Unit Telescope as SPHERE opens a major window of opportunity to combine both high-contrast imaging with high spectral resolution, thanks to the proposed HiRISE coupling between the two instruments \citep{Vigan:2018}.

In this work, we present detailed simulation of the expected performance for HiRISE in $H$ and $K$ band based on an end-to-end model of the instrument, and we compare them to the expected performance of CRIRES+ for the same science cases. Our simulations show that the exoplanet characterization performance of high-dispersion spectrograph CRIRES+ will be fundamentally limited by the stellar halo noise. HiRISE can overcome this limitation by significantly increasing the contrast through the extreme adaptive optics system and coronagraphs provided by SPHERE.

For the characterization of known companions in both $H$ and $K$ bands, we show that whatever the exposure time, HiRISE without coronagraph significantly outperforms CRIRES+ in terms of the time required to reach a given S/N. The observed gains are typically of factor of at least 2 in exposure time, but can reach a factor of more than 12 in some cases. While CRIRES+ observations are always limited by the stellar halo photon noise, the situation is very different for HiRISE once the diffraction has been suppressed with a coronagraph and because of the overall low transmission of the system. With HiRISE, most noise sources become very close to each other in terms of relative contribution to the complete noise budget. The final S/N reached by HiRISE is therefore very dependent on the noise regime in which the observations are done.

The detection limits that we derive demonstrate the full the potential of HiRISE for companions located at separations between 50 and 800\,mas around bright hosts. At 200\,mas, the 5$\sigma$ detection limit for HiRISE in $H$ band is 1.3~magnitudes deeper than for CRIRES+. Conversely, the S/N reached with HiRISE at a contrast of 15\,mag is $\sim$4 times higher than with CRIRES+, which roughly translates into a factor of 16 gain in observing efficiency in the stellar noise limited regime. In $K$ band the HiRISE gains are marginal at best without throughput increasing measures due to the sub-optimal transmission in that band. Based on these simulations the use of $K$ band cannot be realistically considered for implementation with the current instruments and would require profound changes to become attractive for HiRISE. We therefore conclude that the sweet spot for HiRISE is without any doubt the $H$ band where it can provide a substantial gain with respect to CRIRES+. In this band, the gain of ExAO, the spatial filtering of the fiber, and the low thermal emission of the sky and instruments nicely counterbalance the low throughput of the entire coupling, providing good performance for the detection and characterization of companions.

We also investigate the impact of the stellar host magnitude. Our simulations reveal the optimal parameter space for HiRISE to be relatively wide in $H$ band. In this band, CRIRES+ only starts to be competitive for faint hosts stars and at separations outward of $\sim$400\,mas. For brighter hosts and small separations, HiRISE is the best option, except inward of 50\,mas. At such small separations, the two instruments are limited by the stellar photon noise, but CRIRES+ starts outperforming HiRISE because of its much higher transmission. Separating starlight from planet light would have to be done based on the planet's RV diversity caused by the orbital motion. In $K$ band, at a planet temperature of 1200K the optimal parameter space for HiRISE is strongly reduced and only the brightest stars show a gain, again largely due to the low transmission.

Finally, we investigate the effect of the choice of coronagraph for HiRISE and the expected impact of increased transmission. These two aspects are tightly linked because the addition of a coronagraph can have a strong impact on the transmission. Generally, we find that the highest performance for HiRISE is reached without a coronagraph. Although the APLC makes the PSF slightly more Gaussian, which increases the coupling efficiency, this effect is counterbalanced by the loss of more than 50\% of the incoming photons due to the pupil amplitude apodizer. In the end, the whole performance comes down to the overall transmission of the system. We simulate the impact of increasing the transmission of the high-resolution spectrograph, CRIRES+ in our case, by factors of 2 and 3. This change directly translates into a gain in S/N at all angular separations, both in $H$ and $K$ band. This gain factor in the S/N is equal to $\sim$1.5 for a gain of a factor of 2 in transmission of just the spectrograph. This highlights the importance of maximizing the transmission throughout the whole system by optimizing the number of optics, the coatings, the coronagraph, the type of fibers and their lengths, and of course the final high-resolution spectrograph.

\begin{figure}
  \centering
    \includegraphics[width=1.0\columnwidth]{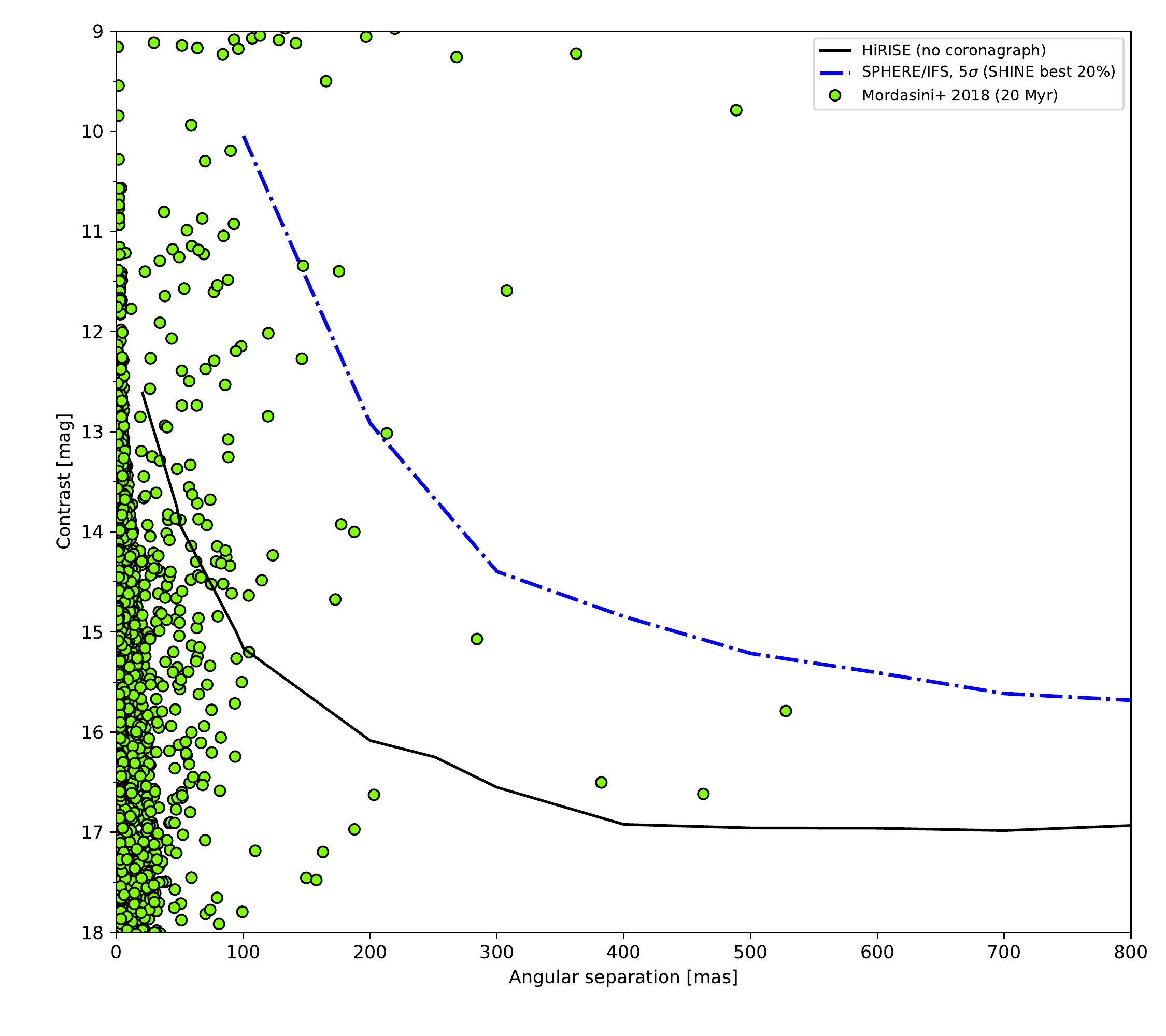}
    \caption{Detection limits of 5\,$\sigma$  for HiRISE  derived in the present work for a bright nearby young star ($H=3.5$, 19\,pc, 20\,Myr), compared to the 20\% best SPHERE/SHINE detection limits. We overplot state-of-the-art population synthesis models based on the core accretion formation scenario \citep{Mordasini:2018,Emsenhuber:2020A,Emsenhuber:2020B}.}
    \label{fig:sphere}
\end{figure}

An important goal for current direct imaging instruments is to study giant exoplanet formation in young systems to better understand the early stages and evolution of planetary systems. Two formation mechanisms are typically considered for giant planets: gravitational instability \citep[GI,][]{Kuiper:51,Boss:1997}, and core accretion \citep[CA,][]{Perri:74,Mizuno:78,Bodenheimer:86}. GI is a binary star-like framework where planets form very quickly in the outer parts of disks from clumps that detach from the rest of the disk, become gravitationally bound, and contract into a giant planet, while CA starts with a smaller Earth-sized core and forms a giant planet through the rapid accumulation of gas from the disk onto this core. It is expected that the two formation mechanisms will lead to different planet occurrence distributions. Recent observational results from large direct imaging surveys \citep[e.g.,][]{SHINEPaperIII} tend to corroborate this expectation, but the very small number of detections makes it impossible to draw strong conclusions just from the statistical point of view. However, it is also expected that GI and CA may lead to different chemical compositions (e.g.,\,metallicity [Fe/H] or C/O ratio) which can be derived from spectra \citep{Oberg:2011,Piso:2015a,Mordasini:2016}. This is why the detailed characterization of more directly detected planets at high spectral resolution (e.g.,\,with HiRISE) is crucial.

Thanks to the powerful combination of ExAO and spatial filtering with single-mode fibers, HiRISE offers very deep detection limits at short angular separations, in particular for bright stars where the readout noise from the spectrograph detector is not the main limitation. We illustrate this in Fig.~\ref{fig:sphere} by comparing the HiRISE detection limits around a bright nearby young star similar to $\beta$\,Pictoris ($H=3.5$, 19\,pc, 20\,Myr) with the 20\% best detection limits from the SPHERE/SHINE direct imaging survey performed using coronagraphic imaging in the $H$ band and modern post-processing techniques \citep{SHINEPaperI,SHINEPaperII,SHINEPaperIII}. The gain brought by ExAO, spatial-filtering, and high spectral resolution is clearly visible in the 50--500\,mas range, where HiRISE outperforms SPHERE by several magnitudes. It is  not a completely fair comparison, however,  because HiRISE cannot be used  a priori for planet searches since it samples only a few sparse spatial resolution elements. It nonetheless illustrates that there is a major potential for the characterization of companions that cannot be seen directly with SPHERE, but could potentially be detected with other techniques.

In this context, it makes sense to compare the detection limits with the outputs of state-of-the-art population synthesis models that predict the expected population of planets that could be formed around nearby stars. We overplot in Fig.~\ref{fig:sphere} the output of population \texttt{NG76} from the new generation planetary population synthesis (NGPPS) model from Bern \citep{Mordasini:2018,Emsenhuber:2020A,Emsenhuber:2020B}. In this plot we assume a $\beta$\,Pictoris analog (A5, 19\,pc, 20\,Myr). The brightness of the planets in the $H$ band is directly computed by the model and translated into contrast assuming a $H=3.5$ magnitude for the star. These synthetic populations only provide a statistical vision of the population of giant exoplanets expected around nearby stars, but they are interesting because they can  assess a discovery potential. Assuming favorable inclinations for the systems, we see in Fig.~\ref{fig:sphere} that some planets would clearly be within reach of HiRISE for detailed spectral characterization. Unless the SPHERE instrument undergoes a major upgrade \citep[e.g.,][]{Boccaletti:2020}, these planets are currently not directly detectable. However, some of them could potentially be detectable by the ESA/Gaia survey \citep{Gaia2016,Gaia2018} in its ultimate data release a few years from now. The final detection limits of Gaia are not yet known and will strongly dependent on its final astrometric accuracy \citep[e.g.,][]{Lindegren:2018}, but it is reasonable to assume that a few detections can be expected around young nearby stars. Although astrometry by itself is not sufficient to know exactly where the planet is located, and therefore where to place the HiRISE fiber, a combination of RV with astrometry would easily break any remaining degeneracy for giant planets \citep[e.g.,][]{Brandt:2019}. This would open a huge potential for confirmation and characterization with HiRISE of objects that would otherwise remain  unreachable by SPHERE.

\begin{acknowledgements}
    This project has received funding from the European Research Council (ERC) under the European Union’s Horizon 2020 research and innovation programme, grant agreements No. 757561 (HiRISE) \& 678777 (ICARUS).     
    SPHERE is an instrument designed and built by a consortium consisting of IPAG (Grenoble, France), MPIA (Heidelberg, Germany), LAM (Marseille, France), LESIA (Paris, France), Laboratoire Lagrange (Nice, France), INAF-Osservatorio di Padova (Italy), Observatoire de Gen\`{e}ve (Switzerland), ETH Zurich (Switzerland), NOVA (Netherlands), ONERA (France) and ASTRON (Netherlands) in collaboration with ESO. SPHERE was funded by ESO, with additional contributions from CNRS (France), MPIA (Germany), INAF (Italy), FINES (Switzerland) and NOVA (Netherlands). SPHERE also received funding from the European Commission Sixth and Seventh Framework Programmes as part of the Optical Infrared Coordination Network for Astronomy (OPTICON) under grant number RII3-Ct-2004-001566 for FP6 (2004-2008), grant number 226604 for FP7 (2009-2012) and grant number 312430 for FP7 (2013-2016).
    
    This publication makes use of data products from the Two Micron All Sky Survey, which is a joint project of the University of Massachusetts and the Infrared Processing and Analysis Center/California Institute of Technology, funded by the National Aeronautics and Space Administration and the National Science Foundation.
    
    This publication makes use of VOSA, developed under the Spanish Virtual Observatory project supported by the Spanish MINECO through grant AyA2017-84089. 
    
    This research made use of \texttt{PyAstronomy}, \texttt{Numpy} \citep{Numpy}, \texttt{SciPy} \citep{SciPy}.
    
    We thank Jean-Charles Lambert at LAM for his HPC support. We thank Dimitri Mawet, Nem Jovanovic, Jacques-Robert Delorme and Sebastiaan Haffert for the discussions on fiber injection, the impact of transmission and optimal cross-correlations.
    We thank the anonymous referee for providing constructive comments that have helped improve the paper.
\end{acknowledgements}

\bibliographystyle{aa}
\bibliography{otten}

\begin{thebibliography}{103}
\expandafter\ifx\csname natexlab\endcsname\relax\def\natexlab#1{#1}\fi

\bibitem[{{Antichi} {et~al.}(2009){Antichi}, {Dohlen}, {Gratton}, {Mesa},
  {Claudi}, {Giro}, {Boccaletti}, {Mouillet}, {Puget}, \&
  {Beuzit}}]{Antichi:2009}
{Antichi}, J., {Dohlen}, K., {Gratton}, R.~G., {et~al.} 2009, \apj, 695, 1042

\bibitem[{{Arsenault} {et~al.}(2003){Arsenault}, {Alonso}, {Bonnet}, {Brynnel},
  {Delabre}, {Donaldson}, {Dupuy}, {Fedrigo}, {Farinato}, {Hubin}, {Ivanescu},
  {Kasper}, {Paufique}, {Rossi}, {Tordo}, {Stroebele}, {Lizon}, {Gigan},
  {Delplancke}, {Silber}, {Quattri}, \& {Reiss}}]{Arsenault:2003}
{Arsenault}, R., {Alonso}, J., {Bonnet}, H., {et~al.} 2003, Society of
  Photo-Optical Instrumentation Engineers (SPIE) Conference Series, Vol. 4839,
  {MACAO-VLTI: An Adaptive Optics system for the ESO VLT interferometer}, ed.
  P.~L. {Wizinowich} \& D.~{Bonaccini}, 174--185

\bibitem[{{Barman} {et~al.}(2015){Barman}, {Konopacky}, {Macintosh}, \&
  {Marois}}]{Barman:2015}
{Barman}, T.~S., {Konopacky}, Q.~M., {Macintosh}, B., \& {Marois}, C. 2015,
  \apj, 804, 61

\bibitem[{Baudrand \& Walker(2001)}]{Baudrand:2001}
Baudrand, J. \& Walker, G.~A.~H. 2001, Publications of the Astronomical Society
  of the Pacific, 113, 851

\bibitem[{{Bayo} {et~al.}(2008){Bayo}, {Rodrigo}, {Barrado Y Navascu{\'e}s},
  {Solano}, {Guti{\'e}rrez}, {Morales-Calder{\'o}n}, \& {Allard}}]{Bayo:2008}
{Bayo}, A., {Rodrigo}, C., {Barrado Y Navascu{\'e}s}, D., {et~al.} 2008, \aap,
  492, 277

\bibitem[{{Beuzit} {et~al.}(2019){Beuzit}, {Vigan}, {Mouillet}, {Dohlen},
  {Gratton}, {Boccaletti}, {Sauvage}, {Schmid}, {Langlois}, {Petit},
  {Baruffolo}, {Feldt}, {Milli}, {Wahhaj}, {Abe}, {Anselmi}, {Antichi},
  {Barette}, {Baudrand}, {Baudoz}, {Bazzon}, {Bernardi}, {Blanchard}, {Brast},
  {Bruno}, {Buey}, {Carbillet}, {Carle}, {Cascone}, {Chapron}, {Charton},
  {Chauvin}, {Claudi}, {Costille}, {De Caprio}, {de Boer}, {Delboulb{\'e}},
  {Desidera}, {Dominik}, {Downing}, {Dupuis}, {Fabron}, {Fantinel}, {Farisato},
  {Feautrier}, {Fedrigo}, {Fusco}, {Gigan}, {Ginski}, {Girard}, {Giro},
  {Gisler}, {Gluck}, {Gry}, {Henning}, {Hubin}, {Hugot}, {Incorvaia}, {Jaquet},
  {Kasper}, {Lagadec}, {Lagrange}, {Le Coroller}, {Le Mignant}, {Le Ruyet},
  {Lessio}, {Lizon}, {Llored}, {Lundin}, {Madec}, {Magnard}, {Marteaud},
  {Martinez}, {Maurel}, {M{\'e}nard}, {Mesa}, {M{\"o}ller-Nilsson}, {Moulin},
  {Moutou}, {Orign{\'e}}, {Parisot}, {Pavlov}, {Perret}, {Pragt}, {Puget},
  {Rabou}, {Ramos}, {Reess}, {Rigal}, {Rochat}, {Roelfsema}, {Rousset}, {Roux},
  {Saisse}, {Salasnich}, {Santambrogio}, {Scuderi}, {Segransan}, {Sevin},
  {Siebenmorgen}, {Soenke}, {Stadler}, {Suarez}, {Tiph{\`e}ne}, {Turatto},
  {Udry}, {Vakili}, {Waters}, {Weber}, {Wildi}, {Zins}, \&
  {Zurlo}}]{Beuzit:2019}
{Beuzit}, J.~L., {Vigan}, A., {Mouillet}, D., {et~al.} 2019, \aap, 631, A155

\bibitem[{{Birkby} {et~al.}(2013){Birkby}, {de Kok}, {Brogi}, {de Mooij},
  {Schwarz}, {Albrecht}, \& {Snellen}}]{Birkby:2013}
{Birkby}, J.~L., {de Kok}, R.~J., {Brogi}, M., {et~al.} 2013, \mnras, 436, L35

\bibitem[{{Boccaletti} {et~al.}(2020){Boccaletti}, {Chauvin}, {Mouillet},
  {Absil}, {Allard}, {Antoniucci}, {Augereau}, {Barge}, {Baruffolo}, {Baudino},
  {Baudoz}, {Beaulieu}, {Benisty}, {Beuzit}, {Bianco}, {Biller}, {Bonavita},
  {Bonnefoy}, {Bos}, {Bouret}, {Brandner}, {Buchschache}, {Carry},
  {Cantalloube}, {Cascone}, {Carlotti}, {Charnay}, {Chiavassa}, {Choquet},
  {Clenet}, {Crida}, {De Boer}, {De Caprio}, {Desidera}, {Desert}, {Delisle},
  {Delorme}, {Dohlen}, {Doelman}, {Dominik}, {Orazi}, {Dougados}, {Doute},
  {Fedele}, {Feldt}, {Ferreira}, {Fontanive}, {Fusco}, {Galicher}, {Garufi},
  {Gendron}, {Ghedina}, {Ginski}, {Gonzalez}, {Gratadour}, {Gratton},
  {Guillot}, {Haffert}, {Hagelberg}, {Henning}, {Huby}, {Janson}, {Kamp},
  {Keller}, {Kenworthy}, {Kervella}, {Kral}, {Kuhn}, {Lagadec}, {Laibe},
  {Langlois}, {Lagrange}, {Launhardt}, {Leboulleux}, {Le Coroller}, {Li Causi},
  {Loupias}, {Maire}, {Marleau}, {Martinache}, {Martinez}, {Mary}, {Mattioli},
  {Mazoyer}, {Meheut}, {Menard}, {Mesa}, {Meunier}, {Miguel}, {Milli}, {Min},
  {Molliere}, {Mordasini}, {Moretto}, {Mugnier}, {Muro Arena}, {Nardetto},
  {Diaye}, {Nesvadba}, {Pedichini}, {Pinilla}, {Por}, {Potier}, {Quanz},
  {Rameau}, {Roelfsema}, {Rouan}, {Rigliaco}, {Salasnich}, {Samland},
  {Sauvage}, {Schmid}, {Segransan}, {Snellen}, {Snik}, {Soulez}, {Stadler},
  {Stam}, {Tallon}, {Thebault}, {Thiebaut}, {Tschudi}, {Udry}, {van Holstein},
  {Vernazza}, {Vidal}, {Vigan}, {Waters}, {Wildi}, {Willson}, {Zanutta},
  {Zavagno}, \& {Zurlo}}]{Boccaletti:2020}
{Boccaletti}, A., {Chauvin}, G., {Mouillet}, D., {et~al.} 2020, arXiv e-prints,
  arXiv:2003.05714

\bibitem[{{Bodenheimer} \& {Pollack}(1986)}]{Bodenheimer:86}
{Bodenheimer}, P. \& {Pollack}, J.~B. 1986, \icarus, 67, 391

\bibitem[{{Bonnefoy} {et~al.}(2013){Bonnefoy}, {Boccaletti}, {Lagrange},
  {Allard}, {Mordasini}, {Beust}, {Chauvin}, {Girard}, {Homeier}, {Apai},
  {Lacour}, \& {Rouan}}]{Bonnefoy:2013}
{Bonnefoy}, M., {Boccaletti}, A., {Lagrange}, A.~M., {et~al.} 2013, \aap, 555,
  A107

\bibitem[{Boss(1997)}]{Boss:1997}
Boss, A.~P. 1997, Science, 276, 1836

\bibitem[{{Bourdarot} {et~al.}(2018){Bourdarot}, {Le Coarer}, {Mouillet},
  {Correia}, {Jocou}, {Rabou}, {Carlotti}, {Bonfils}, {Artigau}, {Vallee},
  {Doyon}, {Forveille}, {Stadler}, {Magnard}, \& {Vigan}}]{Bourdarot:2018}
{Bourdarot}, G., {Le Coarer}, E., {Mouillet}, D., {et~al.} 2018, in Society of
  Photo-Optical Instrumentation Engineers (SPIE) Conference Series, Vol. 10702,
  \procspie, 107025Y

\bibitem[{{Brandt} {et~al.}(2019){Brandt}, {Dupuy}, \& {Bowler}}]{Brandt:2019}
{Brandt}, T.~D., {Dupuy}, T.~J., \& {Bowler}, B.~P. 2019, \aj, 158, 140

\bibitem[{{Brogi} {et~al.}(2012){Brogi}, {Snellen}, {de Kok}, {Albrecht},
  {Birkby}, \& {de Mooij}}]{Brogi:2012}
{Brogi}, M., {Snellen}, I. A.~G., {de Kok}, R.~J., {et~al.} 2012, \nat, 486,
  502

\bibitem[{{Bryan} {et~al.}(2018){Bryan}, {Benneke}, {Knutson}, {Batygin}, \&
  {Bowler}}]{Bryan:2018}
{Bryan}, M.~L., {Benneke}, B., {Knutson}, H.~A., {Batygin}, K., \& {Bowler},
  B.~P. 2018, Nature Astronomy, 2, 138

\bibitem[{{Cantalloube} {et~al.}(2015){Cantalloube}, {Mouillet}, {Mugnier},
  {Milli}, {Absil}, {Gomez Gonzalez}, {Chauvin}, {Beuzit}, \&
  {Cornia}}]{Cantalloube:2015}
{Cantalloube}, F., {Mouillet}, D., {Mugnier}, L.~M., {et~al.} 2015, \aap, 582,
  A89

\bibitem[{{Carbillet} {et~al.}(2011){Carbillet}, {Bendjoya}, {Abe}, {Guerri},
  {Boccaletti}, {Daban}, {Dohlen}, {Ferrari}, {Robbe-Dubois}, {Douet}, \&
  {Vakili}}]{Carbillet:2011}
{Carbillet}, M., {Bendjoya}, P., {Abe}, L., {et~al.} 2011, Experimental
  Astronomy, 30, 39

\bibitem[{{Chauvin} {et~al.}(2017{\natexlab{a}}){Chauvin}, {Desidera},
  {Lagrange}, {Vigan}, {Feldt}, {Gratton}, {Langlois}, {Cheetham}, {Bonnefoy},
  \& {Meyer}}]{Chauvin:2017}
{Chauvin}, G., {Desidera}, S., {Lagrange}, A.-M., {et~al.} 2017{\natexlab{a}},
  in SF2A-2017: Proceedings of the Annual meeting of the French Society of
  Astronomy and Astrophysics, ed. C.~{Reyl{\'e}}, P.~{Di Matteo}, F.~{Herpin},
  E.~{Lagadec}, A.~{Lan{\c c}on}, Z.~{Meliani}, \& F.~{Royer}, 331--335

\bibitem[{{Chauvin} {et~al.}(2017{\natexlab{b}}){Chauvin}, {Desidera},
  {Lagrange}, {Vigan}, {Gratton}, {Langlois}, {Bonnefoy}, {Beuzit}, {Feldt},
  {Mouillet}, {Meyer}, {Cheetham}, {Biller}, {Boccaletti}, {D'Orazi},
  {Galicher}, {Hagelberg}, {Maire}, {Mesa}, {Olofsson}, {Samland}, {Schmidt},
  {Sissa}, {Bonavita}, {Charnay}, {Cudel}, {Daemgen}, {Delorme},
  {Janin-Potiron}, {Janson}, {Keppler}, {Le Coroller}, {Ligi}, {Marleau},
  {Messina}, {Molli{\`e}re}, {Mordasini}, {M{\"u}ller}, {Peretti}, {Perrot},
  {Rodet}, {Rouan}, {Zurlo}, {Dominik}, {Henning}, {Menard}, {Schmid},
  {Turatto}, {Udry}, {Vakili}, {Abe}, {Antichi}, {Baruffolo}, {Baudoz},
  {Baudrand}, {Blanchard}, {Bazzon}, {Buey}, {Carbillet}, {Carle}, {Charton},
  {Cascone}, {Claudi}, {Costille}, {Deboulbe}, {De Caprio}, {Dohlen},
  {Fantinel}, {Feautrier}, {Fusco}, {Gigan}, {Giro}, {Gisler}, {Gluck},
  {Hubin}, {Hugot}, {Jaquet}, {Kasper}, {Madec}, {Magnard}, {Martinez},
  {Maurel}, {Le Mignant}, {M{\"o}ller-Nilsson}, {Llored}, {Moulin},
  {Orign{\'e}}, {Pavlov}, {Perret}, {Petit}, {Pragt}, {Puget}, {Rabou},
  {Ramos}, {Rigal}, {Rochat}, {Roelfsema}, {Rousset}, {Roux}, {Salasnich},
  {Sauvage}, {Sevin}, {Soenke}, {Stadler}, {Suarez}, {Weber}, {Wildi},
  {Antoniucci}, {Augereau}, {Baudino}, {Brandner}, {Engler}, {Girard}, {Gry},
  {Kral}, {Kopytova}, {Lagadec}, {Milli}, {Moutou}, {Schlieder},
  {Szul{\'a}gyi}, {Thalmann}, \& {Wahhaj}}]{Chauvin:2017a}
{Chauvin}, G., {Desidera}, S., {Lagrange}, A.~M., {et~al.} 2017{\natexlab{b}},
  \aap, 605, L9

\bibitem[{{Chauvin} {et~al.}(2005){Chauvin}, {Lagrange}, {Dumas}, {Zuckerman},
  {Mouillet}, {Song}, {Beuzit}, \& {Lowrance}}]{Chauvin:2005}
{Chauvin}, G., {Lagrange}, A.~M., {Dumas}, C., {et~al.} 2005, \aap, 438, L25

\bibitem[{{Cheetham} {et~al.}(2019){Cheetham}, {Samland}, {Brems}, {Launhardt},
  {Chauvin}, {S{\'e}gransan}, {Henning}, {Quirrenbach}, {Avenhaus}, {Cugno},
  {Girard}, {Godoy}, {Kennedy}, {Maire}, {Metchev}, {M{\"u}ller}, {Musso
  Barcucci}, {Olofsson}, {Pepe}, {Quanz}, {Queloz}, {Reffert}, {Rickman}, {van
  Boekel}, {Boccaletti}, {Bonnefoy}, {Cantalloube}, {Charnay}, {Delorme},
  {Janson}, {Keppler}, {Lagrange}, {Langlois}, {Lazzoni}, {Menard}, {Mesa},
  {Meyer}, {Schmidt}, {Sissa}, {Udry}, \& {Zurlo}}]{Cheetham:2019}
{Cheetham}, A.~C., {Samland}, M., {Brems}, S.~S., {et~al.} 2019, \aap, 622, A80

\bibitem[{{Chilcote} {et~al.}(2017){Chilcote}, {Pueyo}, {De Rosa}, {Vargas},
  {Macintosh}, {Bailey}, {Barman}, {Bauman}, {Bruzzone}, {Bulger}, {Burrows},
  {Cardwell}, {Chen}, {Cotten}, {Dillon}, {Doyon}, {Draper}, {Duch{\^e}ne},
  {Dunn}, {Erikson}, {Fitzgerald}, {Follette}, {Gavel}, {Goodsell}, {Graham},
  {Greenbaum}, {Hartung}, {Hibon}, {Hung}, {Ingraham}, {Kalas}, {Konopacky},
  {Larkin}, {Maire}, {Marchis}, {Marley}, {Marois}, {Metchev},
  {Millar-Blanchaer}, {Morzinski}, {Nielsen}, {Norton}, {Oppenheimer},
  {Palmer}, {Patience}, {Perrin}, {Poyneer}, {Rajan}, {Rameau},
  {Rantakyr{\"o}}, {Sadakuni}, {Saddlemyer}, {Savransky}, {Schneider}, {Serio},
  {Sivaramakrishnan}, {Song}, {Soummer}, {Thomas}, {Wallace}, {Wang},
  {Ward-Duong}, {Wiktorowicz}, \& {Wolff}}]{Chilcote:2017}
{Chilcote}, J., {Pueyo}, L., {De Rosa}, R.~J., {et~al.} 2017, \aj, 153, 182

\bibitem[{{De Rosa} {et~al.}(2016){De Rosa}, {Rameau}, {Patience}, {Graham},
  {Doyon}, {Lafreni{\`e}re}, {Macintosh}, {Pueyo}, {Rajan}, {Wang},
  {Ward-Duong}, {Hung}, {Maire}, {Nielsen}, {Ammons}, {Bulger}, {Cardwell},
  {Chilcote}, {Galvez}, {Gerard}, {Goodsell}, {Hartung}, {Hibon}, {Ingraham},
  {Johnson-Groh}, {Kalas}, {Konopacky}, {Marchis}, {Marois}, {Metchev},
  {Morzinski}, {Oppenheimer}, {Perrin}, {Rantakyr{\"o}}, {Savransky}, \&
  {Thomas}}]{DeRosa:2016}
{De Rosa}, R.~J., {Rameau}, J., {Patience}, J., {et~al.} 2016, \apj, 824, 121

\bibitem[{{Desidera} {et~al.}({submitted}){Desidera}, {Chauvin}, {Bonavita}, \&
  {SHINE consortium}}]{SHINEPaperI}
{Desidera}, S., {Chauvin}, G., {Bonavita}, M., \& {SHINE consortium}.
  {submitted}, \aap

\bibitem[{{Dohlen} {et~al.}(2016){Dohlen}, {Vigan}, {Mouillet}, {Wildi},
  {Sauvage}, {Fusco}, {Beuzit}, {Puget}, {Le Mignant}, {Roelfsema}, {Pragt},
  {Schmid}, {Gratton}, {Mesa}, {Claudi}, {Langlois}, {Costille}, {Hugot},
  {O'Neil}, {Guerra}, {N'Diaye}, {Girard}, {Mawet}, \& {Zins}}]{Dohlen:2016}
{Dohlen}, K., {Vigan}, A., {Mouillet}, D., {et~al.} 2016, Society of
  Photo-Optical Instrumentation Engineers (SPIE) Conference Series, Vol. 9908,
  {SPHERE on-sky performance compared with budget predictions}, 99083D

\bibitem[{{Dorn} {et~al.}(2016){Dorn}, {Follert}, {Bristow}, {Cumani},
  {Eschbaumer}, {Grunhut}, {Haimerl}, {Hatzes}, {Heiter}, {Hinterschuster},
  {Ives}, {Jung}, {Kerber}, {Klein}, {Lavail}, {Lizon}, {L{\"o}winger},
  {Molina-Conde}, {Nicholson}, {Marquart}, {Oliva}, {Origlia}, {Pasquini},
  {Paufique}, {Piskunov}, {Reiners}, {Seemann}, {Stegmeier}, {Stempels}, \&
  {Tordo}}]{Dorn:2016}
{Dorn}, R.~J., {Follert}, R., {Bristow}, P., {et~al.} 2016, Society of
  Photo-Optical Instrumentation Engineers (SPIE) Conference Series, Vol. 9908,
  {The ``+'' for CRIRES: enabling better science at infrared wavelength and
  high spectral resolution at the ESO VLT}, 99080I

\bibitem[{{Emsenhuber} {et~al.}(2020{\natexlab{a}}){Emsenhuber}, {Mordasini},
  {Burn}, {Alibert}, {Benz}, \& {Asphaug}}]{Emsenhuber:2020A}
{Emsenhuber}, A., {Mordasini}, C., {Burn}, R., {et~al.} 2020{\natexlab{a}},
  arXiv e-prints, arXiv:2007.05561

\bibitem[{{Emsenhuber} {et~al.}(2020{\natexlab{b}}){Emsenhuber}, {Mordasini},
  {Burn}, {Alibert}, {Benz}, \& {Asphaug}}]{Emsenhuber:2020B}
{Emsenhuber}, A., {Mordasini}, C., {Burn}, R., {et~al.} 2020{\natexlab{b}},
  arXiv e-prints, arXiv:2007.05562

\bibitem[{{Fusco} {et~al.}(2006){Fusco}, {Rousset}, {Sauvage}, {Petit},
  {Beuzit}, {Dohlen}, {Mouillet}, {Charton}, {Nicolle}, {Kasper}, {Baudoz}, \&
  {Puget}}]{Fusco:2006}
{Fusco}, T., {Rousset}, G., {Sauvage}, J.~F., {et~al.} 2006, Optics Express,
  14, 7515

\bibitem[{{Gaia Collaboration} {et~al.}(2018){Gaia Collaboration}, {Brown},
  {Vallenari}, {Prusti}, {de Bruijne}, {Babusiaux}, {Bailer-Jones}, {Biermann},
  {Evans}, {Eyer}, {Jansen}, {Jordi}, {Klioner}, {Lammers}, {Lindegren},
  {Luri}, {Mignard}, {Panem}, {Pourbaix}, {Randich}, {Sartoretti}, {Siddiqui},
  {Soubiran}, {van Leeuwen}, {Walton}, {Arenou}, {Bastian}, {Cropper},
  {Drimmel}, {Katz}, {Lattanzi}, {Bakker}, {Cacciari}, {Casta{\~n}eda},
  {Chaoul}, {Cheek}, {De Angeli}, {Fabricius}, {Guerra}, {Holl}, {Masana},
  {Messineo}, {Mowlavi}, {Nienartowicz}, {Panuzzo}, {Portell}, {Riello},
  {Seabroke}, {Tanga}, {Th{\'e}venin}, {Gracia-Abril}, {Comoretto},
  {Garcia-Reinaldos}, {Teyssier}, {Altmann}, {Andrae}, {Audard},
  {Bellas-Velidis}, {Benson}, {Berthier}, {Blomme}, {Burgess}, {Busso},
  {Carry}, {Cellino}, {Clementini}, {Clotet}, {Creevey}, {Davidson}, {De
  Ridder}, {Delchambre}, {Dell'Oro}, {Ducourant},
  {Fern{\'a}ndez-Hern{\'a}ndez}, {Fouesneau}, {Fr{\'e}mat}, {Galluccio},
  {Garc{\'\i}a-Torres}, {Gonz{\'a}lez-N{\'u}{\~n}ez}, {Gonz{\'a}lez-Vidal},
  {Gosset}, {Guy}, {Halbwachs}, {Hambly}, {Harrison}, {Hern{\'a}ndez},
  {Hestroffer}, {Hodgkin}, {Hutton}, {Jasniewicz}, {Jean-Antoine-Piccolo},
  {Jordan}, {Korn}, {Krone-Martins}, {Lanzafame}, {Lebzelter}, {L{\"o}ffler},
  {Manteiga}, {Marrese}, {Mart{\'\i}n-Fleitas}, {Moitinho}, {Mora}, {Muinonen},
  {Osinde}, {Pancino}, {Pauwels}, {Petit}, {Recio-Blanco}, {Richards},
  {Rimoldini}, {Robin}, {Sarro}, {Siopis}, {Smith}, {Sozzetti}, {S{\"u}veges},
  {Torra}, {van Reeven}, {Abbas}, {Abreu Aramburu}, {Accart}, {Aerts},
  {Altavilla}, {{\'A}lvarez}, {Alvarez}, {Alves}, {Anderson}, {Andrei},
  {Anglada Varela}, {Antiche}, {Antoja}, {Arcay}, {Astraatmadja}, {Bach},
  {Baker}, {Balaguer-N{\'u}{\~n}ez}, {Balm}, {Barache}, {Barata}, {Barbato},
  {Barblan}, {Barklem}, {Barrado}, {Barros}, {Barstow}, {Bartholom{\'e}
  Mu{\~n}oz}, {Bassilana}, {Becciani}, {Bellazzini}, {Berihuete}, {Bertone},
  {Bianchi}, {Bienaym{\'e}}, {Blanco-Cuaresma}, {Boch}, {Boeche}, {Bombrun},
  {Borrachero}, {Bossini}, {Bouquillon}, {Bourda}, {Bragaglia}, {Bramante},
  {Breddels}, {Bressan}, {Brouillet}, {Br{\"u}semeister}, {Brugaletta},
  {Bucciarelli}, {Burlacu}, {Busonero}, {Butkevich}, {Buzzi}, {Caffau},
  {Cancelliere}, {Cannizzaro}, {Cantat-Gaudin}, {Carballo}, {Carlucci},
  {Carrasco}, {Casamiquela}, {Castellani}, {Castro-Ginard}, {Charlot},
  {Chemin}, {Chiavassa}, {Cocozza}, {Costigan}, {Cowell}, {Crifo}, {Crosta},
  {Crowley}, {Cuypers}, {Dafonte}, {Damerdji}, {Dapergolas}, {David}, {David},
  {de Laverny}, {De Luise}, {De March}, {de Martino}, {de Souza}, {de Torres},
  {Debosscher}, {del Pozo}, {Delbo}, {Delgado}, {Delgado}, {Di Matteo},
  {Diakite}, {Diener}, {Distefano}, {Dolding}, {Drazinos}, {Dur{\'a}n},
  {Edvardsson}, {Enke}, {Eriksson}, {Esquej}, {Eynard Bontemps}, {Fabre},
  {Fabrizio}, {Faigler}, {Falc{\~a}o}, {Farr{\`a}s Casas}, {Federici},
  {Fedorets}, {Fernique}, {Figueras}, {Filippi}, {Findeisen}, {Fonti},
  {Fraile}, {Fraser}, {Fr{\'e}zouls}, {Gai}, {Galleti}, {Garabato},
  {Garc{\'\i}a-Sedano}, {Garofalo}, {Garralda}, {Gavel}, {Gavras}, {Gerssen},
  {Geyer}, {Giacobbe}, {Gilmore}, {Girona}, {Giuffrida}, {Glass}, {Gomes},
  {Granvik}, {Gueguen}, {Guerrier}, {Guiraud}, {Guti{\'e}rrez-S{\'a}nchez},
  {Haigron}, {Hatzidimitriou}, {Hauser}, {Haywood}, {Heiter}, {Helmi}, {Heu},
  {Hilger}, {Hobbs}, {Hofmann}, {Holland}, {Huckle}, {Hypki}, {Icardi},
  {Jan{\ss}en}, {Jevardat de Fombelle}, {Jonker}, {Juh{\'a}sz}, {Julbe},
  {Karampelas}, {Kewley}, {Klar}, {Kochoska}, {Kohley}, {Kolenberg},
  {Kontizas}, {Kontizas}, {Koposov}, {Kordopatis}, {Kostrzewa-Rutkowska},
  {Koubsky}, {Lambert}, {Lanza}, {Lasne}, {Lavigne}, {Le Fustec}, {Le
  Poncin-Lafitte}, {Lebreton}, {Leccia}, {Leclerc}, {Lecoeur-Taibi},
  {Lenhardt}, {Leroux}, {Liao}, {Licata}, {Lindstr{\o}m}, {Lister}, {Livanou},
  {Lobel}, {L{\'o}pez}, {Managau}, {Mann}, {Mantelet}, {Marchal}, {Marchant},
  {Marconi}, {Marinoni}, {Marschalk{\'o}}, {Marshall}, {Martino}, {Marton},
  {Mary}, {Massari}, {Matijevi{\v{c}}}, {Mazeh}, {McMillan}, {Messina},
  {Michalik}, {Millar}, {Molina}, {Molinaro}, {Moln{\'a}r}, {Montegriffo},
  {Mor}, {Morbidelli}, {Morel}, {Morris}, {Mulone}, {Muraveva}, {Musella},
  {Nelemans}, {Nicastro}, {Noval}, {O'Mullane}, {Ord{\'e}novic},
  {Ord{\'o}{\~n}ez-Blanco}, {Osborne}, {Pagani}, {Pagano}, {Pailler},
  {Palacin}, {Palaversa}, {Panahi}, {Pawlak}, {Piersimoni}, {Pineau}, {Plachy},
  {Plum}, {Poggio}, {Poujoulet}, {Pr{\v{s}}a}, {Pulone}, {Racero}, {Ragaini},
  {Rambaux}, {Ramos-Lerate}, {Regibo}, {Reyl{\'e}}, {Riclet}, {Ripepi}, {Riva},
  {Rivard}, {Rixon}, {Roegiers}, {Roelens}, {Romero-G{\'o}mez}, {Rowell},
  {Royer}, {Ruiz-Dern}, {Sadowski}, {Sagrist{\`a} Sell{\'e}s}, {Sahlmann},
  {Salgado}, {Salguero}, {Sanna}, {Santana-Ros}, {Sarasso}, {Savietto},
  {Schultheis}, {Sciacca}, {Segol}, {Segovia}, {S{\'e}gransan}, {Shih},
  {Siltala}, {Silva}, {Smart}, {Smith}, {Solano}, {Solitro}, {Sordo}, {Soria
  Nieto}, {Souchay}, {Spagna}, {Spoto}, {Stampa}, {Steele},
  {Steidelm{\"u}ller}, {Stephenson}, {Stoev}, {Suess}, {Surdej}, {Szabados},
  {Szegedi-Elek}, {Tapiador}, {Taris}, {Tauran}, {Taylor}, {Teixeira},
  {Terrett}, {Teyssand ier}, {Thuillot}, {Titarenko}, {Torra Clotet}, {Turon},
  {Ulla}, {Utrilla}, {Uzzi}, {Vaillant}, {Valentini}, {Valette}, {van Elteren},
  {Van Hemelryck}, {van Leeuwen}, {Vaschetto}, {Vecchiato}, {Veljanoski},
  {Viala}, {Vicente}, {Vogt}, {von Essen}, {Voss}, {Votruba}, {Voutsinas},
  {Walmsley}, {Weiler}, {Wertz}, {Wevers}, {Wyrzykowski}, {Yoldas},
  {{\v{Z}}erjal}, {Ziaeepour}, {Zorec}, {Zschocke}, {Zucker}, {Zurbach}, \&
  {Zwitter}}]{Gaia2018}
{Gaia Collaboration}, {Brown}, A.~G.~A., {Vallenari}, A., {et~al.} 2018, \aap,
  616, A1

\bibitem[{{Gaia Collaboration} {et~al.}(2016){Gaia Collaboration}, {Prusti},
  {de Bruijne}, {Brown}, {Vallenari}, {Babusiaux}, {Bailer-Jones}, {Bastian},
  {Biermann}, {Evans}, \& et~al.}]{Gaia2016}
{Gaia Collaboration}, {Prusti}, T., {de Bruijne}, J.~H.~J., {et~al.} 2016,
  \aap, 595, A1

\bibitem[{Ge {et~al.}(1998)Ge, Angel, \& Shelton}]{Ge:1998}
Ge, J., Angel, J. R.~P., \& Shelton, J.~C. 1998, in Optical Astronomical
  Instrumentation, ed. S.~D'Odorico, Vol. 3355, International Society for
  Optics and Photonics (SPIE), 253 -- 263

\bibitem[{{Goyal} {et~al.}(2018){Goyal}, {Mayne}, {Sing}, {Drummond},
  {Tremblin}, {Amundsen}, {Evans}, {Carter}, {Spake}, {Baraffe}, {Nikolov},
  {Manners}, {Chabrier}, \& {Hebrard}}]{Goyal:2018}
{Goyal}, J.~M., {Mayne}, N., {Sing}, D.~K., {et~al.} 2018, Monthly Notices of
  the Royal Astronomical Society, 474, 5158

\bibitem[{{Gravity Collaboration} {et~al.}(2019){Gravity Collaboration},
  {Lacour}, {Nowak}, {Wang}, {Pfuhl}, {Eisenhauer}, {Abuter}, {Amorim},
  {Anugu}, {Benisty}, {Berger}, {Beust}, {Blind}, {Bonnefoy}, {Bonnet},
  {Bourget}, {Brandner}, {Buron}, {Collin}, {Charnay}, {Chapron}, {Cl{\'e}net},
  {Coud{\'e} Du Foresto}, {de Zeeuw}, {Deen}, {Dembet}, {Dexter}, {Duvert},
  {Eckart}, {F{\"o}rster Schreiber}, {F{\'e}dou}, {Garcia}, {Garcia Lopez},
  {Gao}, {Gendron}, {Genzel}, {Gillessen}, {Gordo}, {Greenbaum}, {Habibi},
  {Haubois}, {Hau{\ss}mann}, {Henning}, {Hippler}, {Horrobin}, {Hubert},
  {Jimenez Rosales}, {Jocou}, {Kendrew}, {Kervella}, {Kolb}, {Lagrange},
  {Lapeyr{\`e}re}, {Le Bouquin}, {L{\'e}na}, {Lippa}, {Lenzen}, {Maire},
  {Molli{\`e}re}, {Ott}, {Paumard}, {Perraut}, {Perrin}, {Pueyo}, {Rabien},
  {Ram{\'\i}rez}, {Rau}, {Rodr{\'\i}guez-Coira}, {Rousset}, {Sanchez-Bermudez},
  {Scheithauer}, {Schuhler}, {Straub}, {Straubmeier}, {Sturm}, {Tacconi},
  {Vincent}, {van Dishoeck}, {von Fellenberg}, {Wank}, {Waisberg}, {Widmann},
  {Wieprecht}, {Wiest}, {Wiezorrek}, {Woillez}, {Yazici}, {Ziegler}, \&
  {Zins}}]{Lacour:2019}
{Gravity Collaboration}, {Lacour}, S., {Nowak}, M., {et~al.} 2019, \aap, 623,
  L11

\bibitem[{{Gravity Collaboration} {et~al.}(2020){Gravity Collaboration},
  {Nowak}, {Lacour}, {Molli{\`e}re}, {Wang}, {Charnay}, {van Dishoeck},
  {Abuter}, {Amorim}, {Berger}, {Beust}, {Bonnefoy}, {Bonnet}, {Brandner},
  {Buron}, {Cantalloube}, {Collin}, {Chapron}, {Cl{\'e}net}, {Coud{\'e} Du
  Foresto}, {de Zeeuw}, {Dembet}, {Dexter}, {Duvert}, {Eckart}, {Eisenhauer},
  {F{\"o}rster Schreiber}, {F{\'e}dou}, {Garcia Lopez}, {Gao}, {Gendron},
  {Genzel}, {Gillessen}, {Hau{\ss}mann}, {Henning}, {Hippler}, {Hubert},
  {Jocou}, {Kervella}, {Lagrange}, {Lapeyr{\`e}re}, {Le Bouquin}, {L{\'e}na},
  {Maire}, {Ott}, {Paumard}, {Paladini}, {Perraut}, {Perrin}, {Pueyo}, {Pfuhl},
  {Rabien}, {Rau}, {Rodr{\'\i}guez-Coira}, {Rousset}, {Scheithauer},
  {Shangguan}, {Straub}, {Straubmeier}, {Sturm}, {Tacconi}, {Vincent},
  {Widmann}, {Wieprecht}, {Wiezorrek}, {Woillez}, {Yazici}, \&
  {Ziegler}}]{Nowak:2020}
{Gravity Collaboration}, {Nowak}, M., {Lacour}, S., {et~al.} 2020, \aap, 633,
  A110

\bibitem[{{Greenbaum} {et~al.}(2018){Greenbaum}, {Pueyo}, {Ruffio}, {Wang}, {De
  Rosa}, {Aguilar}, {Rameau}, {Barman}, {Marois}, {Marley}, {Konopacky},
  {Rajan}, {Macintosh}, {Ansdell}, {Arriaga}, {Bailey}, {Bulger}, {Burrows},
  {Chilcote}, {Cotten}, {Doyon}, {Duch{\^e}ne}, {Fitzgerald}, {Follette},
  {Gerard}, {Goodsell}, {Graham}, {Hibon}, {Hung}, {Ingraham}, {Kalas},
  {Larkin}, {Maire}, {Marchis}, {Metchev}, {Millar-Blanchaer}, {Nielsen},
  {Norton}, {Oppenheimer}, {Palmer}, {Patience}, {Perrin}, {Poyneer},
  {Rantakyr{\"o}}, {Savransky}, {Schneider}, {Sivaramakrishnan}, {Song},
  {Soummer}, {Thomas}, {Wallace}, {Ward-Duong}, {Wiktorowicz}, \&
  {Wolff}}]{Greenbaum:2018}
{Greenbaum}, A.~Z., {Pueyo}, L., {Ruffio}, J.-B., {et~al.} 2018, \aj, 155, 226

\bibitem[{{Guerri} {et~al.}(2011){Guerri}, {Daban}, {Robbe-Dubois}, {Douet},
  {Abe}, {Baudrand }, {Carbillet}, {Boccaletti}, {Bendjoya}, {Gouvret}, \&
  {Vakili}}]{Guerri:2011}
{Guerri}, G., {Daban}, J.-B., {Robbe-Dubois}, S., {et~al.} 2011, Experimental
  Astronomy, 30, 59

\bibitem[{{Haffert} {et~al.}(2020){Haffert}, {Por}, {Keller}, {Kenworthy},
  {Doelman}, {Snik}, \& {Escuti}}]{Haffert:2018}
{Haffert}, S.~Y., {Por}, E.~H., {Keller}, C.~U., {et~al.} 2020, A\&A, 635, A56

\bibitem[{{Hoeijmakers} {et~al.}(2018){Hoeijmakers}, {Schwarz}, {Snellen}, {de
  Kok}, {Bonnefoy}, {Chauvin}, {Lagrange}, \& {Girard}}]{Hoeijmakers:2018}
{Hoeijmakers}, H.~J., {Schwarz}, H., {Snellen}, I.~A.~G., {et~al.} 2018, \aap,
  617, A144

\bibitem[{{Husser} {et~al.}(2013){Husser}, {Wende-von Berg}, {Dreizler},
  {Homeier}, {Reiners}, {Barman}, \& {Hauschildt}}]{Husser:2013}
{Husser}, T.~O., {Wende-von Berg}, S., {Dreizler}, S., {et~al.} 2013, \aap,
  553, A6

\bibitem[{{Jones} {et~al.}(2013){Jones}, {Noll}, {Kausch}, {Szyszka}, \&
  {Kimeswenger}}]{Jones:2013}
{Jones}, A., {Noll}, S., {Kausch}, W., {Szyszka}, C., \& {Kimeswenger}, S.
  2013, \aap, 560, A91

\bibitem[{Jones {et~al.}(2001)Jones, Oliphant, Peterson, {et~al.}}]{SciPy}
Jones, E., Oliphant, T., Peterson, P., {et~al.} 2001, {SciPy}: Open source
  scientific tools for {Python}, [Online; accessed 2019]

\bibitem[{{Jovanovic} {et~al.}(2015){Jovanovic}, {Martinache}, {Guyon},
  {Clergeon}, {Singh}, {Kudo}, {Garrel}, {Newman}, {Doughty}, {Lozi}, {Males},
  {Minowa}, {Hayano}, {Takato}, {Morino}, {Kuhn}, {Serabyn}, {Norris},
  {Tuthill}, {Schworer}, {Stewart}, {Close}, {Huby}, {Perrin}, {Lacour},
  {Gauchet}, {Vievard}, {Murakami}, {Oshiyama}, {Baba}, {Matsuo}, {Nishikawa},
  {Tamura}, {Lai}, {Marchis}, {Duchene}, {Kotani}, \&
  {Woillez}}]{Jovanovic:2015}
{Jovanovic}, N., {Martinache}, F., {Guyon}, O., {et~al.} 2015, \pasp, 127, 890

\bibitem[{{Jovanovic} {et~al.}(2017){Jovanovic}, {Schwab}, {Guyon}, {Lozi},
  {Cvetojevic}, {Martinache}, {Leon-Saval}, {Norris}, {Gross}, {Doughty},
  {Currie}, \& {Takato}}]{Jovanovic:2017}
{Jovanovic}, N., {Schwab}, C., {Guyon}, O., {et~al.} 2017, \aap, 604, A122

\bibitem[{{Kaeufl} {et~al.}(2004){Kaeufl}, {Ballester}, {Biereichel},
  {Delabre}, {Donaldson}, {Dorn}, {Fedrigo}, {Finger}, {Fischer}, {Franza},
  {Gojak}, {Huster}, {Jung}, {Lizon}, {Mehrgan}, {Meyer}, {Moorwood}, {Pirard},
  {Paufique}, {Pozna}, {Siebenmorgen}, {Silber}, {Stegmeier}, \&
  {Wegerer}}]{Kaeufl:2004}
{Kaeufl}, H.-U., {Ballester}, P., {Biereichel}, P., {et~al.} 2004, Society of
  Photo-Optical Instrumentation Engineers (SPIE) Conference Series, Vol. 5492,
  {CRIRES: a high-resolution infrared spectrograph for ESO's VLT}, ed. A.~F.~M.
  {Moorwood} \& M.~{Iye}, 1218--1227

\bibitem[{{Kawahara} \& {Hirano}(2014)}]{Kawahara:2014}
{Kawahara}, H. \& {Hirano}, T. 2014, arXiv e-prints, arXiv:1409.5740

\bibitem[{{Kawahara} {et~al.}(2014){Kawahara}, {Murakami}, {Matsuo}, \&
  {Kotani}}]{Kawahara:2014a}
{Kawahara}, H., {Murakami}, N., {Matsuo}, T., \& {Kotani}, T. 2014, \apjs, 212,
  27

\bibitem[{{Keppler} {et~al.}(2018){Keppler}, {Benisty}, {M{\"u}ller},
  {Henning}, {van Boekel}, {Cantalloube}, {Ginski}, {van Holstein}, {Maire},
  {Pohl}, {Samland }, {Avenhaus}, {Baudino}, {Boccaletti}, {de Boer},
  {Bonnefoy}, {Chauvin}, {Desidera}, {Langlois}, {Lazzoni}, {Marleau},
  {Mordasini}, {Pawellek}, {Stolker}, {Vigan}, {Zurlo}, {Birnstiel},
  {Brandner}, {Feldt}, {Flock}, {Girard}, {Gratton}, {Hagelberg}, {Isella},
  {Janson}, {Juhasz}, {Kemmer}, {Kral}, {Lagrange}, {Launhardt}, {Matter},
  {M{\'e}nard}, {Milli}, {Molli{\`e}re}, {Olofsson}, {P{\'e}rez}, {Pinilla},
  {Pinte}, {Quanz}, {Schmidt}, {Udry}, {Wahhaj}, {Williams}, {Buenzli},
  {Cudel}, {Dominik}, {Galicher}, {Kasper}, {Lannier}, {Mesa}, {Mouillet},
  {Peretti}, {Perrot}, {Salter}, {Sissa}, {Wildi}, {Abe}, {Antichi},
  {Augereau}, {Baruffolo}, {Baudoz}, {Bazzon}, {Beuzit}, {Blanchard}, {Brems},
  {Buey}, {De Caprio}, {Carbillet}, {Carle}, {Cascone}, {Cheetham}, {Claudi},
  {Costille}, {Delboulb{\'e}}, {Dohlen}, {Fantinel}, {Feautrier}, {Fusco},
  {Giro}, {Gluck}, {Gry}, {Hubin}, {Hugot}, {Jaquet}, {Le Mignant}, {Llored},
  {Madec}, {Magnard}, {Martinez}, {Maurel}, {Meyer}, {M{\"o}ller-Nilsson},
  {Moulin}, {Mugnier}, {Orign{\'e}}, {Pavlov}, {Perret}, {Petit}, {Pragt},
  {Puget}, {Rabou}, {Ramos}, {Rigal}, {Rochat}, {Roelfsema}, {Rousset}, {Roux},
  {Salasnich}, {Sauvage}, {Sevin}, {Soenke}, {Stadler}, {Suarez}, {Turatto}, \&
  {Weber}}]{Keppler:2018}
{Keppler}, M., {Benisty}, M., {M{\"u}ller}, A., {et~al.} 2018, \aap, 617, A44

\bibitem[{{Konopacky} {et~al.}(2013){Konopacky}, {Barman}, {Macintosh}, \&
  {Marois}}]{Konopacky:2013}
{Konopacky}, Q.~M., {Barman}, T.~S., {Macintosh}, B.~A., \& {Marois}, C. 2013,
  Science, 339, 1398

\bibitem[{{Kotani} {et~al.}(2018){Kotani}, {Tamura}, {Nishikawa}, {Ueda},
  {Kuzuhara}, {Omiya}, {Hashimoto}, {Ishizuka}, {Hirano}, {Suto}, {Kurokawa},
  {Kokubo}, {Mori}, {Tanaka}, {Kashiwagi}, {Konishi}, {Kudo}, {Sato},
  {Jacobson}, {Hodapp}, {Hall}, {Aoki}, {Usuda}, {Nishiyama}, {Nakajima},
  {Ikeda}, {Yamamuro}, {Morino}, {Baba}, {Hosokawa}, {Ishikawa}, {Narita},
  {Kokubo}, {Hayano}, {Izumiura}, {Kambe}, {Kusakabe}, {Kwon}, {Ikoma}, {Hori},
  {Genda}, {Fukui}, {Fujii}, {Kawahara}, {Olivier}, {Jovanovic}, {Harakawa},
  {Hayashi}, {Hidai}, {Machida}, {Matsuo}, {Nagata}, {Ogihara}, {Takami},
  {Takato}, {Terada}, \& {Oh}}]{Kotani:2018}
{Kotani}, T., {Tamura}, M., {Nishikawa}, J., {et~al.} 2018, in Society of
  Photo-Optical Instrumentation Engineers (SPIE) Conference Series, Vol. 10702,
  \procspie, 1070211

\bibitem[{Kuiper(1951)}]{Kuiper:51}
Kuiper, G.~P. 1951, Proceedings of the National Academy of Sciences, 37, 1

\bibitem[{{Lagrange} {et~al.}(2019){Lagrange}, {Boccaletti}, {Langlois},
  {Chauvin}, {Gratton}, {Beust}, {Desidera}, {Milli}, {Bonnefoy}, {Cheetham},
  {Feldt}, {Meyer}, {Vigan}, {Biller}, {Bonavita}, {Baudino}, {Cantalloube},
  {Cudel}, {Daemgen}, {Delorme}, {D'Orazi}, {Girard}, {Fontanive}, {Hagelberg},
  {Janson}, {Keppler}, {Koypitova}, {Galicher}, {Lannier}, {Le Coroller},
  {Ligi}, {Maire}, {Mesa}, {Messina}, {M{\"u}eller}, {Peretti}, {Perrot},
  {Rouan}, {Salter}, {Samland}, {Schmidt}, {Sissa}, {Zurlo}, {Beuzit},
  {Mouillet}, {Dominik}, {Henning}, {Lagadec}, {M{\'e}nard}, {Schmid},
  {Turatto}, {Udry}, {Bohn}, {Charnay}, {Gomez Gonzales}, {Gry}, {Kenworthy},
  {Kral}, {Mordasini}, {Moutou}, {van der Plas}, {Schlieder}, {Abe}, {Antichi},
  {Baruffolo}, {Baudoz}, {Baudrand}, {Blanchard}, {Bazzon}, {Buey},
  {Carbillet}, {Carle}, {Charton}, {Cascone}, {Claudi}, {Costille}, {Deboulbe},
  {De Caprio}, {Dohlen}, {Fantinel}, {Feautrier}, {Fusco}, {Gigan}, {Giro},
  {Gisler}, {Gluck}, {Hubin}, {Hugot}, {Jaquet}, {Kasper}, {Madec}, {Magnard},
  {Martinez}, {Maurel}, {Le Mignant}, {M{\"o}ller-Nilsson}, {Llored}, {Moulin},
  {Orign{\'e}}, {Pavlov}, {Perret}, {Petit}, {Pragt}, {Szulagyi}, \&
  {Wildi}}]{Lagrange:2019}
{Lagrange}, A.~M., {Boccaletti}, A., {Langlois}, M., {et~al.} 2019, \aap, 621,
  L8

\bibitem[{{Lagrange} {et~al.}(2009){Lagrange}, {Gratadour}, {Chauvin}, {Fusco},
  {Ehrenreich}, {Mouillet}, {Rousset}, {Rouan}, {Allard}, {Gendron}, {Charton},
  {Mugnier}, {Rabou}, {Montri}, \& {Lacombe}}]{Lagrange:2009}
{Lagrange}, A.~M., {Gratadour}, D., {Chauvin}, G., {et~al.} 2009, \aap, 493,
  L21

\bibitem[{{Langlois} {et~al.}({in prep.}){Langlois}, {Gratton}, {Lagrange},
  {Delorme}, {Boccaletti}, {Bonnefoy}, {Maire}, {Mesa}, {Chauvin}, {Desidera},
  {Vigan}, \& {SHINE consortium}}]{SHINEPaperII}
{Langlois}, M., {Gratton}, R., {Lagrange}, A.-M., {et~al.} {in prep.}, \aap

\bibitem[{{Lindegren} {et~al.}(2018){Lindegren}, {Hern{\'a}ndez}, {Bombrun},
  {Klioner}, {Bastian}, {Ramos-Lerate}, {de Torres}, {Steidelm{\"u}ller},
  {Stephenson}, {Hobbs}, {Lammers}, {Biermann}, {Geyer}, {Hilger}, {Michalik},
  {Stampa}, {McMillan}, {Casta{\~n}eda}, {Clotet}, {Comoretto}, {Davidson},
  {Fabricius}, {Gracia}, {Hambly}, {Hutton}, {Mora}, {Portell}, {van Leeuwen},
  {Abbas}, {Abreu}, {Altmann}, {Andrei}, {Anglada}, {Balaguer-N{\'u}{\~n}ez},
  {Barache}, {Becciani}, {Bertone}, {Bianchi}, {Bouquillon}, {Bourda},
  {Br{\"u}semeister}, {Bucciarelli}, {Busonero}, {Buzzi}, {Cancelliere},
  {Carlucci}, {Charlot}, {Cheek}, {Crosta}, {Crowley}, {de Bruijne}, {de
  Felice}, {Drimmel}, {Esquej}, {Fienga}, {Fraile}, {Gai}, {Garralda},
  {Gonz{\'a}lez-Vidal}, {Guerra}, {Hauser}, {Hofmann}, {Holl}, {Jordan},
  {Lattanzi}, {Lenhardt}, {Liao}, {Licata}, {Lister}, {L{\"o}ffler},
  {Marchant}, {Martin-Fleitas}, {Messineo}, {Mignard}, {Morbidelli}, {Poggio},
  {Riva}, {Rowell}, {Salguero}, {Sarasso}, {Sciacca}, {Siddiqui}, {Smart},
  {Spagna}, {Steele}, {Taris}, {Torra}, {van Elteren}, {van Reeven}, \&
  {Vecchiato}}]{Lindegren:2018}
{Lindegren}, L., {Hern{\'a}ndez}, J., {Bombrun}, A., {et~al.} 2018, \aap, 616,
  A2

\bibitem[{{Lovis} {et~al.}(2017){Lovis}, {Snellen}, {Mouillet}, {Pepe},
  {Wildi}, {Astudillo-Defru}, {Beuzit}, {Bonfils}, {Cheetham}, {Conod},
  {Delfosse}, {Ehrenreich}, {Figueira}, {Forveille}, {Martins}, {Quanz},
  {Santos}, {Schmid}, {S{\'e}gransan}, \& {Udry}}]{Lovis:2017}
{Lovis}, C., {Snellen}, I., {Mouillet}, D., {et~al.} 2017, \aap, 599, A16

\bibitem[{{Macintosh} {et~al.}(2015){Macintosh}, {Graham}, {Barman}, {De Rosa},
  {Konopacky}, {Marley}, {Marois}, {Nielsen}, {Pueyo}, {Rajan}, {Rameau},
  {Saumon}, {Wang}, {Patience}, {Ammons}, {Arriaga}, {Artigau}, {Beckwith},
  {Brewster}, {Bruzzone}, {Bulger}, {Burningham}, {Burrows}, {Chen}, {Chiang},
  {Chilcote}, {Dawson}, {Dong}, {Doyon}, {Draper}, {Duch{\^e}ne}, {Esposito},
  {Fabrycky}, {Fitzgerald}, {Follette}, {Fortney}, {Gerard}, {Goodsell},
  {Greenbaum}, {Hibon}, {Hinkley}, {Cotten}, {Hung}, {Ingraham},
  {Johnson-Groh}, {Kalas}, {Lafreniere}, {Larkin}, {Lee}, {Line}, {Long},
  {Maire}, {Marchis}, {Matthews}, {Max}, {Metchev}, {Millar-Blanchaer},
  {Mittal}, {Morley}, {Morzinski}, {Murray-Clay}, {Oppenheimer}, {Palmer},
  {Patel}, {Perrin}, {Poyneer}, {Rafikov}, {Rantakyr{\"o}}, {Rice}, {Rojo},
  {Rudy}, {Ruffio}, {Ruiz}, {Sadakuni}, {Saddlemyer}, {Salama}, {Savransky},
  {Schneider}, {Sivaramakrishnan}, {Song}, {Soummer}, {Thomas}, {Vasisht},
  {Wallace}, {Ward-Duong}, {Wiktorowicz}, {Wolff}, \&
  {Zuckerman}}]{Macintosh:2015}
{Macintosh}, B., {Graham}, J.~R., {Barman}, T., {et~al.} 2015, Science, 350, 64

\bibitem[{{Macintosh} {et~al.}(2014){Macintosh}, {Graham}, {Ingraham},
  {Konopacky}, {Marois}, {Perrin}, {Poyneer}, {Bauman}, {Barman}, {Burrows},
  {Cardwell}, {Chilcote}, {De Rosa}, {Dillon}, {Doyon}, {Dunn}, {Erikson},
  {Fitzgerald}, {Gavel}, {Goodsell}, {Hartung}, {Hibon}, {Kalas}, {Larkin},
  {Maire}, {Marchis}, {Marley}, {McBride}, {Millar-Blanchaer}, {Morzinski},
  {Norton}, {Oppenheimer}, {Palmer}, {Patience}, {Pueyo}, {Rantakyro},
  {Sadakuni}, {Saddlemyer}, {Savransky}, {Serio}, {Soummer},
  {Sivaramakrishnan}, {Song}, {Thomas}, {Wallace}, {Wiktorowicz}, \&
  {Wolff}}]{Macintosh:2014}
{Macintosh}, B., {Graham}, J.~R., {Ingraham}, P., {et~al.} 2014, Proceedings of
  the National Academy of Science, 111, 12661

\bibitem[{{Marois} {et~al.}(2008){Marois}, {Macintosh}, {Barman}, {Zuckerman},
  {Song}, {Patience}, {Lafreni{\`e}re}, \& {Doyon}}]{Marois:2008}
{Marois}, C., {Macintosh}, B., {Barman}, T., {et~al.} 2008, Science, 322, 1348

\bibitem[{{Mawet} {et~al.}(2017){Mawet}, {Ruane}, {Xuan}, {Echeverri},
  {Klimovich}, {Randolph}, {Fucik}, {Wallace}, {Wang}, {Vasisht}, {Dekany},
  {Mennesson}, {Choquet}, {Delorme}, \& {Serabyn}}]{Mawet:2017}
{Mawet}, D., {Ruane}, G., {Xuan}, W., {et~al.} 2017, \apj, 838, 92

\bibitem[{{Mawet} {et~al.}(2016){Mawet}, {Wizinowich}, {Dekany}, {Chun},
  {Hall}, {Cetre}, {Guyon}, {Wallace}, {Bowler}, {Liu}, {Ruane}, {Serabyn},
  {Bartos}, {Wang}, {Vasisht}, {Fitzgerald}, {Skemer}, {Ireland}, {Fucik},
  {Fortney}, {Crossfield}, {Hu}, \& {Benneke}}]{Mawet:2016}
{Mawet}, D., {Wizinowich}, P., {Dekany}, R., {et~al.} 2016, Society of
  Photo-Optical Instrumentation Engineers (SPIE) Conference Series, Vol. 9909,
  {Keck Planet Imager and Characterizer: concept and phased implementation},
  99090D

\bibitem[{Milli {et~al.}(2017)Milli, Mouillet, Fusco, Girard, Masciadri, Pena,
  Sauvage, Reyes, Dohlen, Beuzit, \& et~al.}]{Milli:2017}
Milli, J., Mouillet, D., Fusco, T., {et~al.} 2017, Proceedings of the Adaptive
  Optics for Extremely Large Telescopes 5

\bibitem[{{Mizuno} {et~al.}(1978){Mizuno}, {Nakazawa}, \&
  {Hayashi}}]{Mizuno:78}
{Mizuno}, H., {Nakazawa}, K., \& {Hayashi}, C. 1978, Progress of Theoretical
  Physics, 60, 699

\bibitem[{{Mordasini}(2018)}]{Mordasini:2018}
{Mordasini}, C. 2018, {Planetary Population Synthesis}, 143

\bibitem[{{Mordasini} {et~al.}(2016){Mordasini}, {van Boekel}, {Molli{\`e}re},
  {Henning}, \& {Benneke}}]{Mordasini:2016}
{Mordasini}, C., {van Boekel}, R., {Molli{\`e}re}, P., {Henning}, T., \&
  {Benneke}, B. 2016, \apj, 832, 41

\bibitem[{{M{\"u}ller} {et~al.}(2018){M{\"u}ller}, {Keppler}, {Henning},
  {Samland}, {Chauvin}, {Beust}, {Maire}, {Molaverdikhani}, {van Boekel},
  {Benisty}, {Boccaletti}, {Bonnefoy}, {Cantalloube}, {Charnay}, {Baudino},
  {Gennaro}, {Long}, {Cheetham}, {Desidera}, {Feldt}, {Fusco}, {Girard},
  {Gratton}, {Hagelberg}, {Janson}, {Lagrange}, {Langlois}, {Lazzoni}, {Ligi},
  {M{\'e}nard}, {Mesa}, {Meyer}, {Molli{\`e}re}, {Mordasini}, {Moulin},
  {Pavlov}, {Pawellek}, {Quanz}, {Ramos}, {Rouan}, {Sissa}, {Stadler}, {Vigan},
  {Wahhaj}, {Weber}, \& {Zurlo}}]{Mueller:2018}
{M{\"u}ller}, A., {Keppler}, M., {Henning}, T., {et~al.} 2018, \aap, 617, L2

\bibitem[{{Noll} {et~al.}(2012){Noll}, {Kausch}, {Barden}, {Jones}, {Szyszka},
  {Kimeswenger}, \& {Vinther}}]{Noll:2012}
{Noll}, S., {Kausch}, W., {Barden}, M., {et~al.} 2012, \aap, 543, A92

\bibitem[{{{\"O}berg} {et~al.}(2011){{\"O}berg}, {Murray-Clay}, \&
  {Bergin}}]{Oberg:2011}
{{\"O}berg}, K.~I., {Murray-Clay}, R., \& {Bergin}, E.~A. 2011, \apjl, 743, L16

\bibitem[{{Pasquini} {et~al.}(2002){Pasquini}, {Avila}, {Blecha}, {Cacciari},
  {Cayatte}, {Colless}, {Damiani}, {de Propris}, {Dekker}, {di Marcantonio},
  {Farrell}, {Gillingham}, {Guinouard}, {Hammer}, {Kaufer}, {Hill}, {Marteaud},
  {Modigliani}, {Mulas}, {North}, {Popovic}, {Rossetti}, {Royer}, {Santin},
  {Schmutzer}, {Simond}, {Vola}, {Waller}, \& {Zoccali}}]{Pasquini:2002}
{Pasquini}, L., {Avila}, G., {Blecha}, A., {et~al.} 2002, The Messenger, 110, 1

\bibitem[{Paufique {et~al.}(2006)Paufique, Biereichel, Delabre, Donaldson,
  Esteves, Fedrigo, Gigan, Gojak, Hubin, Kasper, Käufl, Lizon, Marchetti,
  Oberti, Pirard, Pozna, Santos, Stroebele, \& Tordo}]{Paufique:2006}
Paufique, J., Biereichel, P., Delabre, B., {et~al.} 2006, in Advances in
  Adaptive Optics II, ed. B.~L. Ellerbroek \& D.~B. Calia, Vol. 6272,
  International Society for Optics and Photonics (SPIE), 379 -- 390

\bibitem[{{Perri} \& {Cameron}(1974)}]{Perri:74}
{Perri}, F. \& {Cameron}, A.~G.~W. 1974, \icarus, 22, 416

\bibitem[{{Perryman} {et~al.}(2014){Perryman}, {Hartman}, {Bakos}, \&
  {Lindegren}}]{Perryman:2014}
{Perryman}, M., {Hartman}, J., {Bakos}, G.~{\'A}., \& {Lindegren}, L. 2014,
  \apj, 797, 14

\bibitem[{{Petit} {et~al.}(2014){Petit}, {Sauvage}, {Fusco}, {Sevin}, {Suarez},
  {Costille}, {Vigan}, {Soenke}, {Perret}, {Rochat}, {Barrufolo}, {Salasnich},
  {Beuzit}, {Dohlen}, {Mouillet}, {Puget}, {Wildi}, {Kasper}, {Conan},
  {Kulcs{\'a}r}, \& {Raynaud}}]{Petit:2014}
{Petit}, C., {Sauvage}, J.~F., {Fusco}, T., {et~al.} 2014, in Adaptive Optics
  Systems IV, Vol. 9148, 91480O

\bibitem[{{Phillips} {et~al.}(2020){Phillips}, {Tremblin}, {Baraffe},
  {Chabrier}, {Allard}, {Spiegelman}, {Goyal}, {Drummond}, \&
  {H{\'e}brard}}]{Phillips:2020}
{Phillips}, M.~W., {Tremblin}, P., {Baraffe}, I., {et~al.} 2020, \aap, 637, A38

\bibitem[{{Piso} {et~al.}(2015){Piso}, {{\"O}berg}, {Birnstiel}, \&
  {Murray-Clay}}]{Piso:2015a}
{Piso}, A.-M.~A., {{\"O}berg}, K.~I., {Birnstiel}, T., \& {Murray-Clay}, R.~A.
  2015, \apj, 815, 109

\bibitem[{{Plavchan} {et~al.}(2013){Plavchan}, {Bottom}, {Gao}, {Wallace},
  {Mennesson}, {Ciardi}, {Crawford}, {Lin}, {Beichman}, {Brinkworth},
  {Johnson}, {Davison}, {White}, {Anglada-Escude}, {von Braun}, {Vasisht},
  {Prato}, {Kane}, {Tanner}, {Walp}, \& {Mills}}]{Plavchan:2013}
{Plavchan}, P.~P., {Bottom}, M., {Gao}, P., {et~al.} 2013, Society of
  Photo-Optical Instrumentation Engineers (SPIE) Conference Series, Vol. 8864,
  {Precision near-infrared radial velocity instrumentation II: noncircular core
  fiber scrambler}, 88640G

\bibitem[{{Por} \& {Haffert}(2020)}]{Por:2018}
{Por}, E.~H. \& {Haffert}, S.~Y. 2020, A\&A, 635, A55

\bibitem[{{Potier, A.} {et~al.}(2020){Potier, A.}, {Galicher, R.}, {Baudoz,
  P.}, {Huby, E.}, {Milli, J.}, {Wahhaj, Z.}, {Boccaletti, A.}, {Vigan, A.},
  {N\'{}Diaye, M.}, \& {Sauvage, J.-F.}}]{Potier:2020}
{Potier, A.}, {Galicher, R.}, {Baudoz, P.}, {et~al.} 2020, A\&A, 638, A117

\bibitem[{{Rajan} {et~al.}(2017){Rajan}, {Rameau}, {De Rosa}, {Marley},
  {Graham}, {Macintosh}, {Marois}, {Morley}, {Patience}, {Pueyo}, {Saumon},
  {Ward-Duong}, {Ammons}, {Arriaga}, {Bailey}, {Barman}, {Bulger}, {Burrows},
  {Chilcote}, {Cotten}, {Czekala}, {Doyon}, {Duch{\^e}ne}, {Esposito},
  {Fitzgerald}, {Follette}, {Fortney}, {Goodsell}, {Greenbaum}, {Hibon},
  {Hung}, {Ingraham}, {Johnson-Groh}, {Kalas}, {Konopacky}, {Lafreni{\`e}re},
  {Larkin}, {Maire}, {Marchis}, {Metchev}, {Millar-Blanchaer}, {Morzinski},
  {Nielsen}, {Oppenheimer}, {Palmer}, {Patel}, {Perrin}, {Poyneer},
  {Rantakyr{\"o}}, {Ruffio}, {Savransky}, {Schneider}, {Sivaramakrishnan},
  {Song}, {Soummer}, {Thomas}, {Vasisht}, {Wallace}, {Wang}, {Wiktorowicz}, \&
  {Wolff}}]{Rajan:2017}
{Rajan}, A., {Rameau}, J., {De Rosa}, R.~J., {et~al.} 2017, \aj, 154, 10

\bibitem[{{Riaud} \& {Schneider}(2007)}]{Riaud:2007}
{Riaud}, P. \& {Schneider}, J. 2007, \aap, 469, 355

\bibitem[{{Ruffio} {et~al.}(2017){Ruffio}, {Macintosh}, {Wang}, {Pueyo},
  {Nielsen}, {De Rosa}, {Czekala}, {Marley}, {Arriaga}, {Bailey}, {Barman},
  {Bulger}, {Chilcote}, {Cotten}, {Doyon}, {Duch{\^e}ne}, {Fitzgerald},
  {Follette}, {Gerard}, {Goodsell}, {Graham}, {Greenbaum}, {Hibon}, {Hung},
  {Ingraham}, {Kalas}, {Konopacky}, {Larkin}, {Maire}, {Marchis}, {Marois},
  {Metchev}, {Millar-Blanchaer}, {Morzinski}, {Oppenheimer}, {Palmer},
  {Patience}, {Perrin}, {Poyneer}, {Rajan}, {Rameau}, {Rantakyr{\"o}},
  {Savransky}, {Schneider}, {Sivaramakrishnan}, {Song}, {Soummer}, {Thomas},
  {Wallace}, {Ward-Duong}, {Wiktorowicz}, \& {Wolff}}]{Ruffio:2017}
{Ruffio}, J.-B., {Macintosh}, B., {Wang}, J.~J., {et~al.} 2017, \apj, 842, 14

\bibitem[{{Ruilier}(1998)}]{Ruilier:98}
{Ruilier}, C. 1998, in Astronomical Interferometry, ed. R.~D. Reasenberg, Vol.
  3350, International Society for Optics and Photonics (SPIE), 319 -- 329

\bibitem[{{Samland} {et~al.}(2017){Samland}, {Molli{\`e}re}, {Bonnefoy},
  {Maire}, {Cantalloube}, {Cheetham}, {Mesa}, {Gratton}, {Biller}, {Wahhaj},
  {Bouwman}, {Brandner}, {Melnick}, {Carson}, {Janson}, {Henning}, {Homeier},
  {Mordasini}, {Langlois}, {Quanz}, {van Boekel}, {Zurlo}, {Schlieder},
  {Avenhaus}, {Beuzit}, {Boccaletti}, {Bonavita}, {Chauvin}, {Claudi}, {Cudel},
  {Desidera}, {Feldt}, {Fusco}, {Galicher}, {Kopytova}, {Lagrange}, {Le
  Coroller}, {Martinez}, {Moeller-Nilsson}, {Mouillet}, {Mugnier}, {Perrot},
  {Sevin}, {Sissa}, {Vigan}, \& {Weber}}]{Samland:2017}
{Samland}, M., {Molli{\`e}re}, P., {Bonnefoy}, M., {et~al.} 2017, \aap, 603,
  A57

\bibitem[{{Sauvage} {et~al.}(2014){Sauvage}, {Fusco}, {Petit}, {Mouillet},
  {Dohlen}, {Costille}, {Beuzit}, {Baruffolo}, {Kasper}, {Suarez Valles},
  {Downing}, {Feautrier}, {Mugnier}, \& {Baudoz}}]{Sauvage:2014}
{Sauvage}, J.-F., {Fusco}, T., {Petit}, C., {et~al.} 2014, in SPIE Conference
  Series, Vol. 9148

\bibitem[{{Schwarz} {et~al.}(2016){Schwarz}, {Ginski}, {de Kok}, {Snellen},
  {Brogi}, \& {Birkby}}]{Schwarz:2016}
{Schwarz}, H., {Ginski}, C., {de Kok}, R.~J., {et~al.} 2016, \aap, 593, A74

\bibitem[{{Skrutskie} {et~al.}(2006){Skrutskie}, {Cutri}, {Stiening},
  {Weinberg}, {Schneider}, {Carpenter}, {Beichman}, {Capps}, {Chester},
  {Elias}, {Huchra}, {Liebert}, {Lonsdale}, {Monet}, {Price}, {Seitzer},
  {Jarrett}, {Kirkpatrick}, {Gizis}, {Howard}, {Evans}, {Fowler}, {Fullmer},
  {Hurt}, {Light}, {Kopan}, {Marsh}, {McCallon}, {Tam}, {Van Dyk}, \&
  {Wheelock}}]{Skrutskie:2006}
{Skrutskie}, M.~F., {Cutri}, R.~M., {Stiening}, R., {et~al.} 2006, \aj, 131,
  1163

\bibitem[{{Snellen} {et~al.}(2015){Snellen}, {de Kok}, {Birkby}, {Brandl},
  {Brogi}, {Keller}, {Kenworthy}, {Schwarz}, \& {Stuik}}]{Snellen:2015}
{Snellen}, I., {de Kok}, R., {Birkby}, J.~L., {et~al.} 2015, \aap, 576, A59

\bibitem[{{Snellen} {et~al.}(2014){Snellen}, {Brandl}, {de Kok}, {Brogi},
  {Birkby}, \& {Schwarz}}]{Snellen:2014}
{Snellen}, I. A.~G., {Brandl}, B.~R., {de Kok}, R.~J., {et~al.} 2014, \nat,
  509, 63

\bibitem[{{Snellen} {et~al.}(2010){Snellen}, {de Kok}, {de Mooij}, \&
  {Albrecht}}]{Snellen:2010}
{Snellen}, I. A.~G., {de Kok}, R.~J., {de Mooij}, E. J.~W., \& {Albrecht}, S.
  2010, \nat, 465, 1049

\bibitem[{{Snellen} {et~al.}(2013){Snellen}, {de Kok}, {le Poole}, {Brogi}, \&
  {Birkby}}]{Snellen:2013}
{Snellen}, I.~A.~G., {de Kok}, R.~J., {le Poole}, R., {Brogi}, M., \& {Birkby},
  J. 2013, \apj, 764, 182

\bibitem[{{Soummer}(2005)}]{Soummer:2005}
{Soummer}, R. 2005, \apjl, 618, L161

\bibitem[{{Sparks} \& {Ford}(2002)}]{Sparks:2002}
{Sparks}, W.~B. \& {Ford}, H.~C. 2002, \apj, 578, 543

\bibitem[{{Tennyson} {et~al.}(2016){Tennyson}, {Yurchenko}, {Al-Refaie},
  {Barton}, {Chubb}, {Coles}, {Diamantopoulou}, {Gorman}, {Hill}, {Lam},
  {Lodi}, {McKemmish}, {Na}, {Owens}, {Polyansky}, {Rivlin}, {Sousa-Silva},
  {Underwood}, {Yachmenev}, \& {Zak}}]{Tennyson:2016}
{Tennyson}, J., {Yurchenko}, S.~N., {Al-Refaie}, A.~F., {et~al.} 2016, Journal
  of Molecular Spectroscopy, 327, 73

\bibitem[{{Thatte} {et~al.}(2007){Thatte}, {Abuter}, {Tecza}, {Nielsen},
  {Clarke}, \& {Close}}]{Thatte:2007}
{Thatte}, N., {Abuter}, R., {Tecza}, M., {et~al.} 2007, \mnras, 378, 1229

\bibitem[{{Vigan}(2019)}]{Vigan:2019a}
{Vigan}, A. 2019, in The Very Large Telescope in 2030, 39

\bibitem[{{Vigan} {et~al.}(2020){Vigan}, {Fontanive}, {Meyer}, {Biller},
  {Bonavita}, {Feldt}, {Desidera}, {Marleau}, {Emsenhuber}, {Galicher}, {Rice},
  {Forgan}, {Mordasini}, {Gratton}, {Le Coroller}, {Maire}, {Cantalloube},
  {Chauvin}, {Cheetham}, {Hagelberg}, {Lagrange}, {Langlois}, {Bonnefoy},
  {Beuzit}, {Boccaletti}, {D'Orazi}, {Delorme}, {Dominik}, {Henning}, {Janson},
  {Lagadec}, {Lazzoni}, {Ligi}, {Menard}, {Mesa}, {Messina}, {Moutou},
  {M{\"u}ller}, {Perrot}, {Samland}, {Schmid}, {Schmidt}, {Sissa}, {Turatto},
  {Udry}, {Zurlo}, {Abe}, {Antichi}, {Asensio-Torres}, {Baruffolo}, {Baudoz},
  {Baudrand}, {Bazzon}, {Blanchard}, {Bohn}, {Brown Sevilla}, {Carbillet},
  {Carle}, {Cascone}, {Charton}, {Claudi}, {Costille}, {De Caprio},
  {Delboulb{\'e}}, {Dohlen}, {Engler}, {Fantinel}, {Feautrier}, {Fusco},
  {Gigan}, {Girard}, {Giro}, {Gisler}, {Gluck}, {Gry}, {Hubin}, {Hugot},
  {Jaquet}, {Kasper}, {Le Mignant}, {Llored}, {Madec}, {Magnard}, {Martinez},
  {Maurel}, {M{\"o}ller-Nilsson}, {Mouillet}, {Moulin}, {Orign{\'e}}, {Pavlov},
  {Perret}, {Petit}, {Pragt}, {Puget}, {Rabou}, {Ramos}, {Rickman}, {Rigal},
  {Rochat}, {Roelfsema}, {Rousset}, {Roux}, {Salasnich}, {Sauvage}, {Sevin},
  {Soenke}, {Stadler}, {Suarez}, {Wahhaj}, {Weber}, \& {Wildi}}]{SHINEPaperIII}
{Vigan}, A., {Fontanive}, C., {Meyer}, M., {et~al.} 2020, arXiv e-prints,
  arXiv:2007.06573

\bibitem[{{Vigan} {et~al.}(2008){Vigan}, {Langlois}, {Moutou}, \&
  {Dohlen}}]{Vigan:2008}
{Vigan}, A., {Langlois}, M., {Moutou}, C., \& {Dohlen}, K. 2008, \aap, 489,
  1345

\bibitem[{{Vigan} {et~al.}(2019){Vigan}, {N'Diaye}, {Dohlen}, {Sauvage},
  {Milli}, {Zins}, {Petit}, {Wahhaj}, {Cantalloube}, {Caillat}, {Costille}, {Le
  Merrer}, {Carlotti}, {Beuzit}, \& {Mouillet}}]{Vigan:2019}
{Vigan}, A., {N'Diaye}, M., {Dohlen}, K., {et~al.} 2019, \aap, 629, A11

\bibitem[{{Vigan} {et~al.}(2018){Vigan}, {Otten}, {Muslimov}, {Dohlen},
  {Philipps}, {Seemann}, {Beuzit}, {Dorn}, {Kasper}, {Mouillet}, {Baraffe}, \&
  {Reiners}}]{Vigan:2018}
{Vigan}, A., {Otten}, G.~P.~P.~L., {Muslimov}, E., {et~al.} 2018, in Society of
  Photo-Optical Instrumentation Engineers (SPIE) Conference Series, Vol. 10702,
  \procspie, 1070236

\bibitem[{{Wagner} \& {Tomlinson}(1982)}]{Wagner:1982}
{Wagner}, R.~E. \& {Tomlinson}, W.~J. 1982, \ao, 21, 2671

\bibitem[{Walt {et~al.}(2011)Walt, Colbert, \& Varoquaux}]{Numpy}
Walt, S. v.~d., Colbert, S.~C., \& Varoquaux, G. 2011, Computing in Science \&
  Engineering, 13, 22

\bibitem[{{Wang} {et~al.}(2017){Wang}, {Mawet}, {Ruane}, {Hu}, \&
  {Benneke}}]{Wang:2017}
{Wang}, J., {Mawet}, D., {Ruane}, G., {Hu}, R., \& {Benneke}, B. 2017, \aj,
  153, 183

\bibitem[{{Zurlo} {et~al.}(2016){Zurlo}, {Vigan}, {Galicher}, {Maire}, {Mesa},
  {Gratton}, {Chauvin}, {Kasper}, {Moutou}, {Bonnefoy}, {Desidera}, {Abe},
  {Apai}, {Baruffolo}, {Baudoz}, {Baudrand}, {Beuzit}, {Blancard},
  {Boccaletti}, {Cantalloube}, {Carle}, {Cascone}, {Charton}, {Claudi},
  {Costille}, {de Caprio}, {Dohlen}, {Dominik}, {Fantinel}, {Feautrier},
  {Feldt}, {Fusco}, {Gigan}, {Girard}, {Gisler}, {Gluck}, {Gry}, {Henning},
  {Hugot}, {Janson}, {Jaquet}, {Lagrange}, {Langlois}, {Llored}, {Madec},
  {Magnard}, {Martinez}, {Maurel}, {Mawet}, {Meyer}, {Milli},
  {Moeller-Nilsson}, {Mouillet}, {Orign{\'e}}, {Pavlov}, {Petit}, {Puget},
  {Quanz}, {Rabou}, {Ramos}, {Rousset}, {Roux}, {Salasnich}, {Salter},
  {Sauvage}, {Schmid}, {Soenke}, {Stadler}, {Suarez}, {Turatto}, {Udry},
  {Vakili}, {Wahhaj}, {Wildi}, \& {Antichi}}]{Zurlo:2016}
{Zurlo}, A., {Vigan}, A., {Galicher}, R., {et~al.} 2016, \aap, 587, A57

\end{thebibliography}

\appendix

\onecolumn

\section{Contrast}
\label{sec:rawcontrast}

\begin{figure}
  \centering
    \includegraphics[width=1.0\columnwidth]{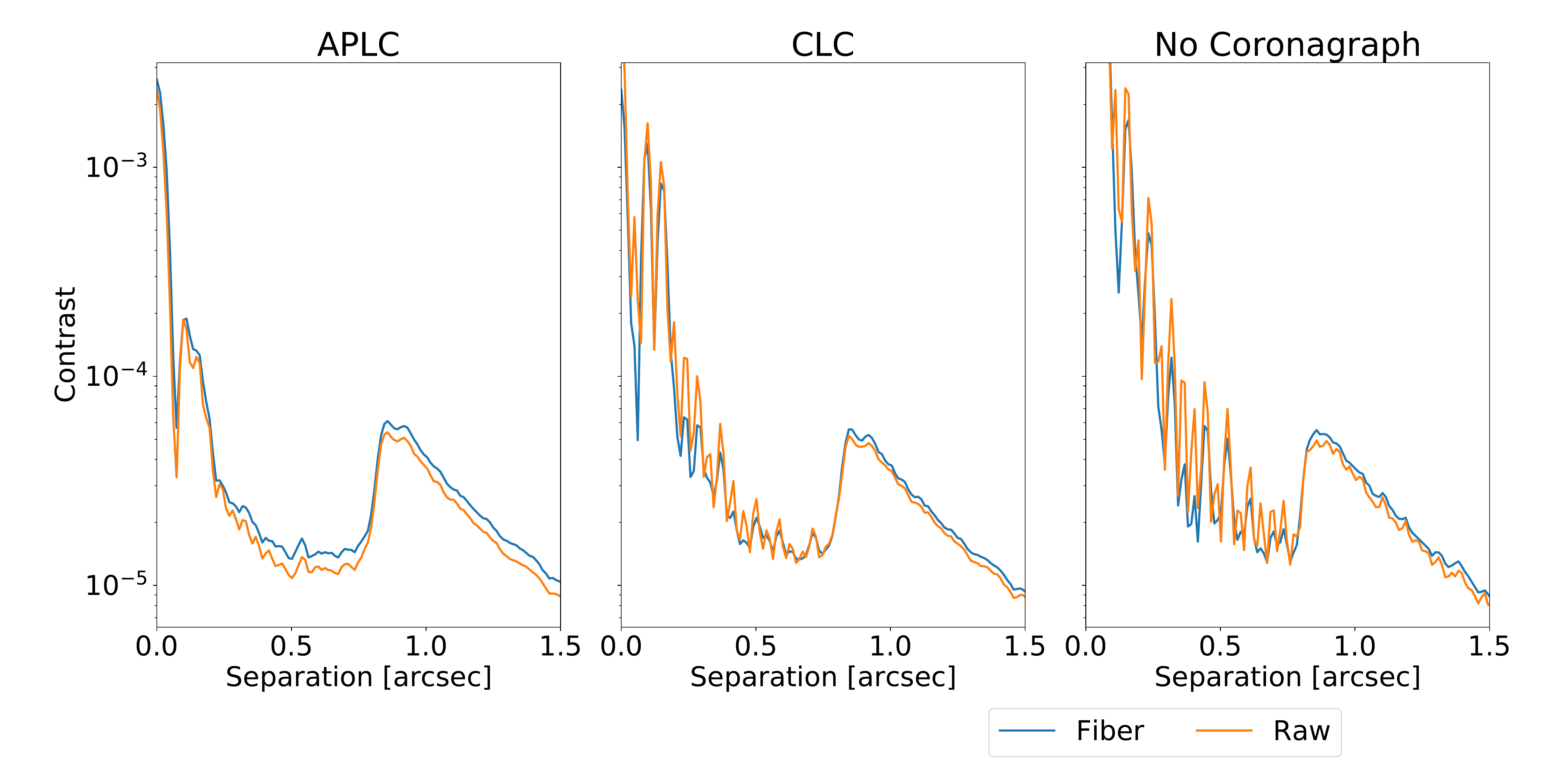}
    \caption{Radial profile of the coronagraphic PSF normalized to the peak flux of the PSF   obtained when the coronagraphic focal plane mask is removed (raw). Overplotted is the radial profile of the coronagraphic fiber efficiency normalized to the maximum fiber efficiency without the coronagraphic mask (fiber). Evaluated at a wavelength of 1.6 \mic.}
    \label{fig:rawcontrast}
\end{figure}

In Fig. \ref{fig:rawcontrast} we show the raw contrast plotted together with a rescaled version of the fiber injection efficiency. We divided the fiber injection efficiency as calculated in Sect. \ref{sec:fibereff} by the peak efficiency without the focal plane mask. This scaling allows us to directly compare the two methods. Without a coronagraph we see a small improvement in the contrast at the location of the airy rings. With the LC this is approximately equal and with the APLC we see a slight reduction in contrast with respect to the raw PSF. 

\section{Fiber injection efficiency tolerances}
\label{app:fiberinjection}

The fiber injection efficiency strongly depends on the offset and tilt of the PSF with respect to the Gaussian mode of the fiber. Therefore, we simulate the effect of a tilt and a shift separately to put tolerances on these values. The lateral shift of the PSF is simulated by translating the Gaussian and redoing the calculation described in Sect. \ref{sec:fibereff}. The tilt is simulated by adding a phase gradient on top of the PSF. A pupil shift of one pupil diameter $D$ has the effect of creating a tilt of $2\pi$, or one wave over 1\,$\lambda/D$. With the $F$-number used by our setup, this value translates into a tilt of the fiber of $\tan^{-1}\left(1/3.5\right)=15.95^{\circ}$ away from the normal.

\begin{figure}[ht]
    \centering
    \includegraphics[width=0.5\columnwidth]{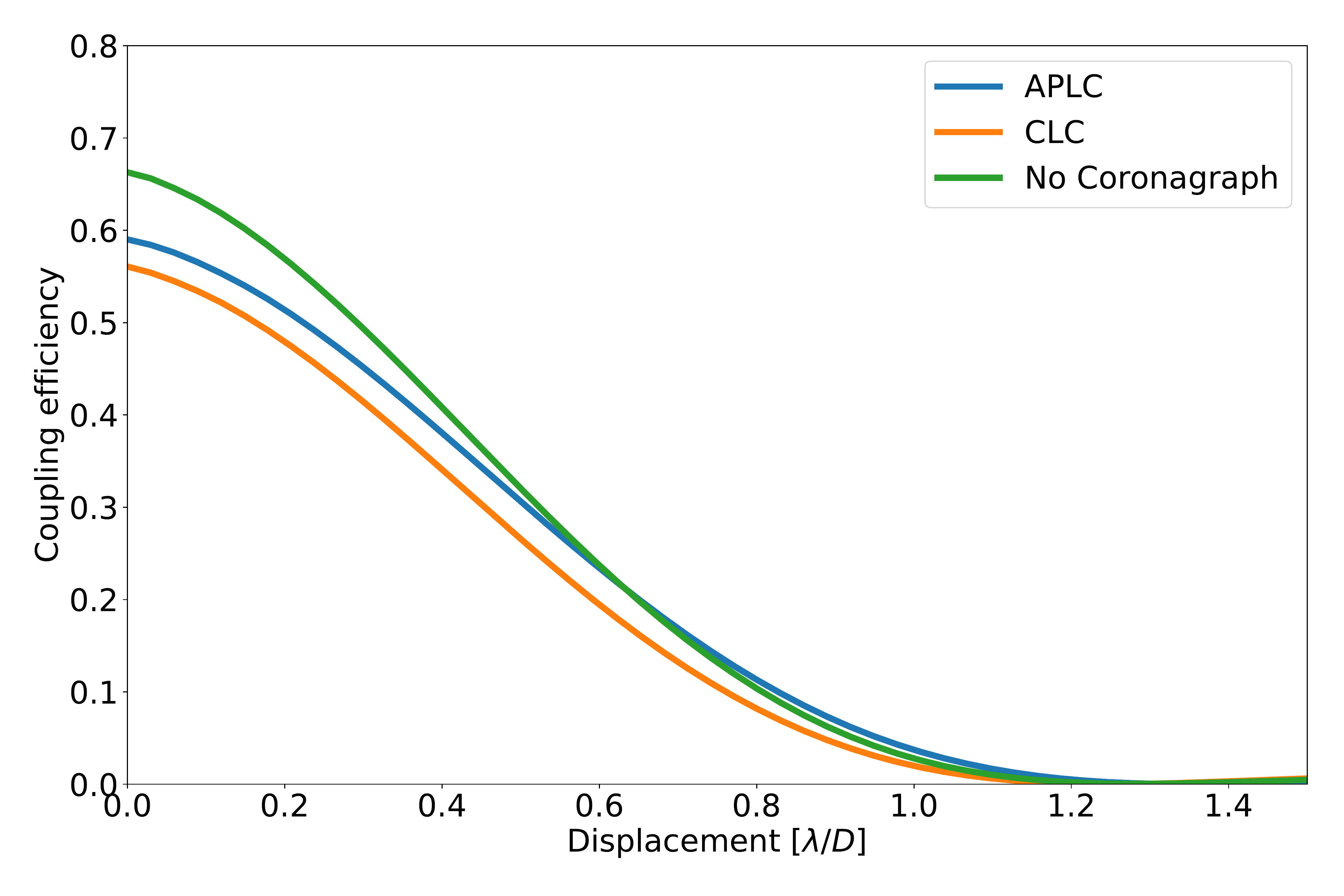}
    \caption{Coupling efficiency as a function of displacement between the PSF core and the Gaussian mode of the fiber. Evaluated at a wavelength of 1.6\,\mic for three different coronagraph options.}
    \label{fig:couplingshift}
\end{figure}

\begin{figure}[ht]
    \centering
    \includegraphics[width=0.5\columnwidth]{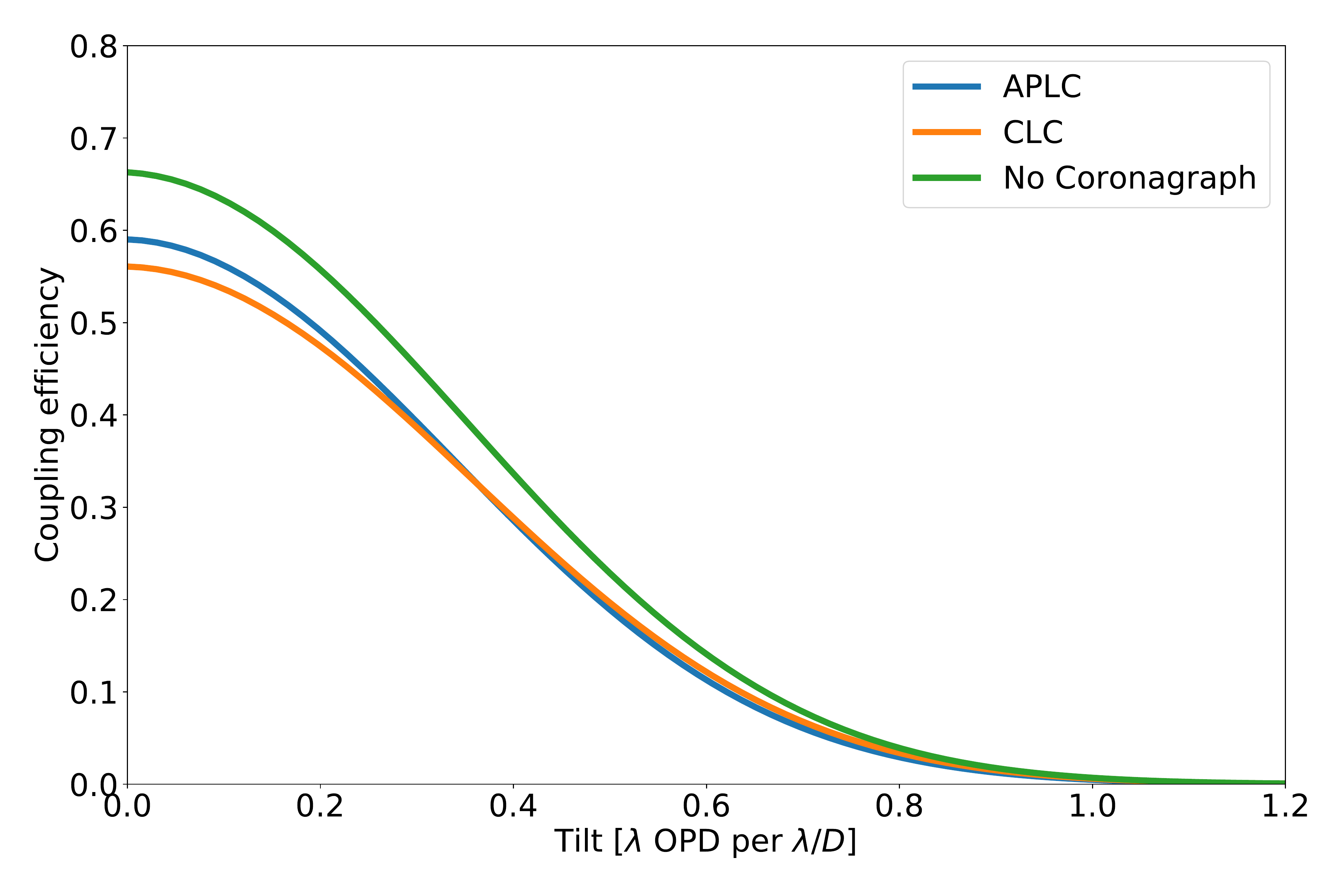}
    \caption{Coupling efficiency as a function of focal plane wavefront tilt between the PSF and the Gaussian mode of the fiber. Evaluated at a wavelength of 1.6\,\mic for three different coronagraph options. A tilt of one $\lambda$ of optical path difference (OPD) per $\lambda/D$ equals $\tan^{-1}\left(1/3.5\right)=15.95^{\circ}$.}
    \label{fig:couplingtilt}
\end{figure}

We also note that for ExAO guide-star magnitudes beyond $R=10$, the increase in residual ExAO wavefront error (WFE) leads to a related decrease in the coupling efficiency that is, by approximation, linearly proportional to the Strehl ratio of the AO correction ($\sim e^{-\sigma_\mathrm{WFE}^2}$), where $\sigma_\mathrm{WFE}$ is the standard deviation of the residual WFE.

\section{Contrast curves}
\label{sec:contrast_curves}

The plots in this section show the same detection limits presented in Sect.~\ref{sec:detection_performance} and Sect.~\ref{sec:effect_coronagraph}, but for a fainter host star similar to HIP\,65426 ($H=18$, $K=16.8$).

\begin{figure*}
  \centering
    \includegraphics[width=1.0\textwidth]{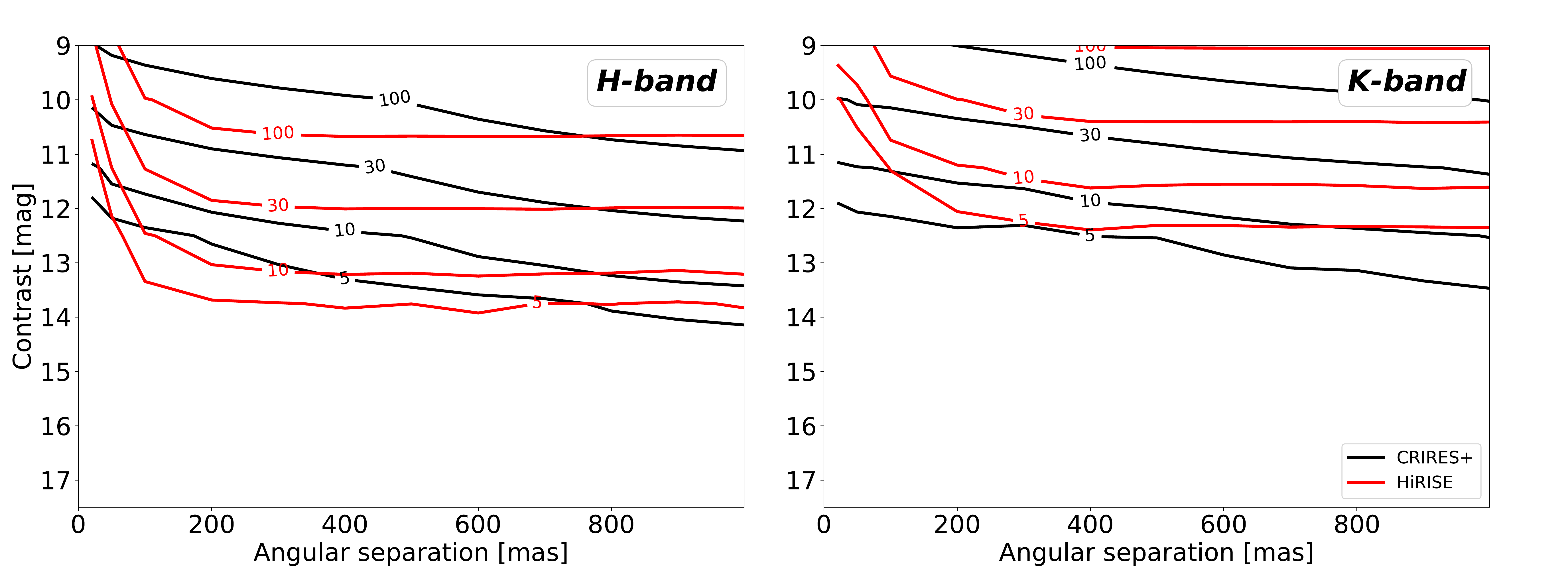}
    \caption{Signal-to-noise ratio as a function of contrast ($\Delta m $) and separation for HiRISE without a coronagraph (red contour lines) and CRIRES+ standalone (black contour lines). The simulation is performed for a HIP\,65426-like host star with a 1200\,K planet and 2 hours of integration time. The S/N is computed with a matched filtering approach (see Sect.~\ref{sec:simu_snr_estimation}) comparing the simulated spectra with the noiseless input planet spectrum. For this combination of parameters the S/N inward of $200-300$ mas is dominated by the noise on the stellar halo. Outside of $200-300$ mas it is limited by dark and read noise.}
    \label{fig:potential_hip}
\end{figure*}

\begin{figure*}
  \centering
    \includegraphics[width=1.0\textwidth]{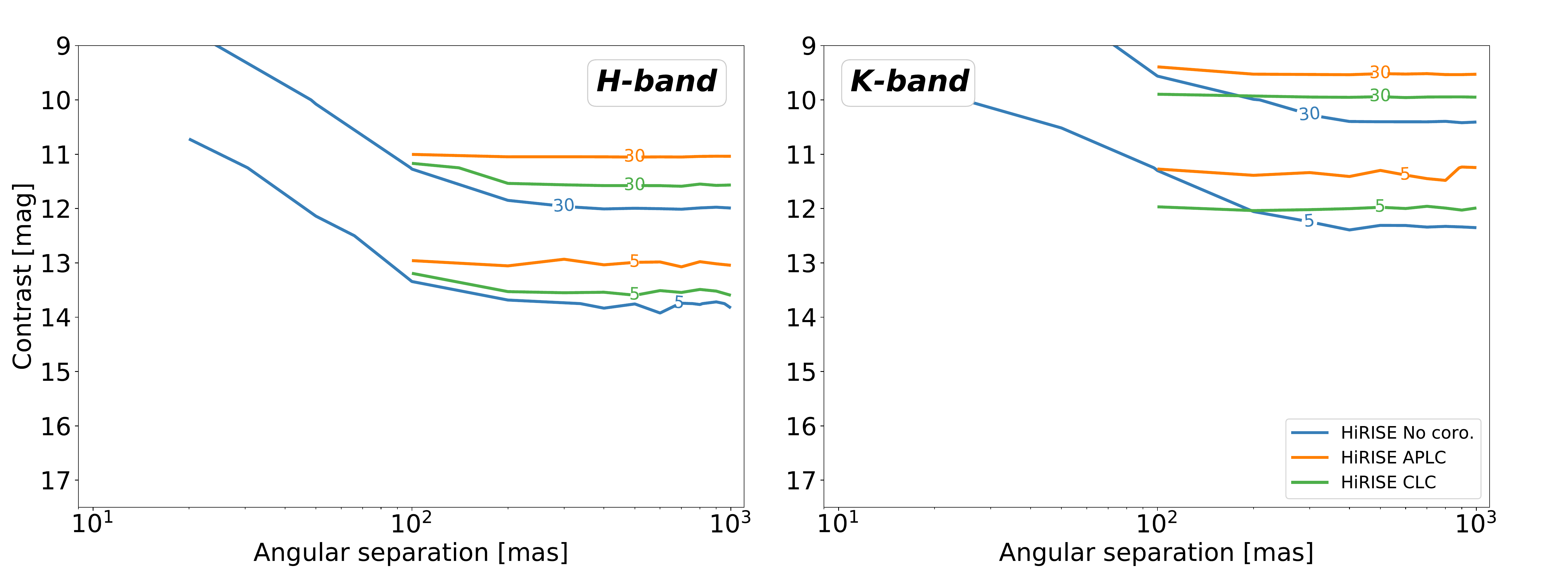}
    \caption{Signal-to-noise ratio as a function of contrast ($\Delta m $) and log-scaled separation for HiRISE without a coronagraph (blue lines), with an APLC (orange lines), and with a CLC (green lines). The simulation is performed for a HIP\,65426-like host star with a 1200\,K planet and 2 hours of integration time. S/N values below the inner working angle radius of 92.5\,mas have been suppressed for the two modes using the focal plane mask. The S/N is computed with a matched filtering approach (see Sect.~\ref{sec:simu_snr_estimation}) comparing the simulated spectra with the noiseless input planet spectrum.}
    \label{fig:coronagraph_hip}
\end{figure*}

\end{document}